\documentclass[12pt,a4paper]{article}
\newcount\driver
\newcount\bozza

\font\cs=cmcsc10 scaled\magstep1%
\font\ottorm=cmr8 scaled\magstep1%
\font\euftw=eufm10 scaled\magstep1%
\font\euftwww=eufm7 scaled\magstep1%
\font\msytw=msbm10 scaled\magstep1%
\font\msytwww=msbm7 scaled\magstep1%
\font\indbf=cmbx10 scaled\magstep2%

{\count255=\time\divide\count255 by 60 \xdef\hourmin{\number\count255}
        \multiply\count255 by-60\advance\count255 by\time
   \xdef\hourmin{\hourmin:\ifnum\count255<10 0\fi\the\count255}}

\let\a=\alpha \let\b=\beta  \let\g=\gamma     \let\d=\delta  \let\e=\varepsilon
\let\z=\zeta  \let\h=\eta    \def\th{\theta}
   \let\l=\lambda
\let\m=\mu    \let\n=\nu    \let\x=\xi        \let\p=\pi      \let\r=\rho
\let\s=\sigma \let\t=\tau        \let\c=\chi
\let\ps=\psi   \let\o=\omega 
      \let\L=\Lambda  
    \let\Si=\Sigma       
\let\O=\Omega 

\def\\{\hfill\break} \let\==\equiv

\let\io=\infty 

\let\0=\noindent

\def\ie{\hbox{\it i.e.\ }}

\let\dpr=\partial 
\let\bs=\backslash

\def\tende#1{\,\vtop{\ialign{##\crcr\rightarrowfill\crcr
 \noalign{\kern-1pt\nointerlineskip}
 \hskip3.pt${\scriptstyle #1}$\hskip3.pt\crcr}}\,}
\def\otto{\,{\kern-1.truept\leftarrow\kern-5.truept\to\kern-1.truept}\,}

\def\PP{{\cal P}}\def\EE{{\cal E}}\def\MM{{\cal M}}\def\VV{{\cal V}}
\def\FF{{\cal F}}\def\HH{{\cal H}}\def\WW{{\cal W}}
\def\TT{{\cal T}}\def\NN{{\cal N}}\def\MM{{\cal M}}
\def\RR{{\cal R}}\def\LL{{\cal L}}
\def\DD{{\cal D}}\def\SS{{\cal S}}
\def\OO{{\cal O}}

\def\indica{\leaders \hbox to 0.5cm{\hss.\hss}\hfill}
\def\guida{\leaders\hbox to 1em{\hss.\hss}\hfill}
\mathchardef\oo= "0521

\def\pp{{\bf p}}\def\xx{{\bf x}}
\def\yy{{\bf y}}\def\kk{{\bf k}}\def\nn{{\bf n}}
\def\uu{{\bf u}}\def\vv{{\bf v}}
 \def\bP{{\bf P}}\def\rr{{\bf r}}
\def\tt{{\bf t}} \def\bO{{\bf O}} \def\bT{{\bf T}}

\def\V#1{{\vec #1}}
\def\vt{\V \t}  \def\grad{{\vec\nabla}} \def\vbb{{\V b}}
\def\vpp{{\vec p}}\def\vqq{{\vec q}}\def\vxx{{\vec x}}
\def\vyy{{\vec y}}\def\vkk{{\vec k}}\def\vnn{{\vec n}}

\def\vrr{{\vec r}}\def\vee{{\vec e}}

\def\Halmos{\hfill\vrule height6pt width4pt depth2pt \par\hbox to \hsize{}}
\def\virg{\quad,\quad}
\def\proof{\0{\cs Proof - } }

\def\ss{{\underline \sigma}}\def\oo{{\underline \omega}}
 \def\ux{{\underline x}} \def\uk{{\underline k}}
\def\xxx{{\underline\xx}} \def\vxxx{{\underline\vxx}}
\def\kkk{{\underline\kk}} \def\vkkk{{\underline\vkk}}

\def\qed{\raise1pt\hbox{\vrule height5pt width5pt depth0pt}}

\def\indic{\hbox{\raise-2pt \hbox{\indbf 1}}}

\def\RRR{\hbox{\msytw R}} 
\def\rrr{\hbox{\msytwww R}}

 \def\ZZZ{\hbox{\msytw Z}}
 
\def\TTT{\hbox{\msytw T}} 

\def\SSS{\hbox{\euftw S}}
\def\SSSS{\hbox{\euftwww S}}
\def\FFF{\hbox{\euftw F}} \def\MMM{\hbox{\euftw M}}

\def\ins#1#2#3{\vbox to0pt{\kern-#2 \hbox{\kern#1 #3}\vss}\nointerlineskip}
\newdimen\xshift \newdimen\xwidth \newdimen\yshift
\newcount\griglia

\def\insertplot#1#2#3#4#5#6{%
\begin{figure}[h]
\begin{center}
\vspace{#2pt}
\begin{minipage}{#1pt}
#3
\ifnum\driver=1
\griglia=#6
\ifnum\griglia=1
\openout13=griglia.ps
\write13{gsave .2 setlinewidth}
\write13{0 10 #1 {dup 0 moveto #2 lineto } for}
\write13{0 10 #2 {dup 0 exch moveto #1 exch lineto } for}
\write13{stroke}
\write13{.5 setlinewidth}
\write13{0 50 #1 {dup 0 moveto #2 lineto } for}
\write13{0 50 #2 {dup 0 exch moveto #1 exch lineto } for}
\write13{stroke grestore}
\closeout13
\includegraphics{griglia.ps}\fi
\includegraphics{#4.pst}\fi
\ifnum\driver=2
\fi
\end{minipage}
\end{center}
\caption{#5}
\end{figure}
}

\newdimen\shift \shift=-1truecm
\def\lb#1{%
\ifnum\bozza=1
\label{#1}\rlap{\kern\shift{$\scriptstyle#1$}}
\else\label{#1}
\fi}

\def\be{\begin{equation}}
\def\ee{\end{equation}}
\def\bea{\begin{eqnarray}}\def\eea{\end{eqnarray}}
\def\bean{\begin{eqnarray*}}\def\eean{\end{eqnarray*}}
\def\bfr{\begin{flushright}}\def\efr{\end{flushright}}
\def\bc{\begin{center}}\def\ec{\end{center}}
\def\ba#1{\begin{array}{#1}} \def\ea{\end{array}}
\def\bd{\begin{description}}\def\ed{\end{description}}
\def\bv{\begin{verbatim}}\def\ev{\end{verbatim}}
\def\nn{\nonumber}
\def\Halmos{\hfill\vrule height10pt width4pt depth2pt \par\hbox to \hsize{}}
\def\pref#1{(\ref{#1})}
 
\def\virg{\quad,\quad}

\newtheorem{lemma}{Lemma}[section]
\newtheorem{theorem}{Theorem}[section]

\begin{document}
\title{{Low temperature analysis of two dimensional Fermi
systems with symmetric Fermi surface}}

\author{
G. Benfatto\thanks{Supported by M.I.U.R. - Dipartimento di
Matematica, Universit\`a di Roma ``Tor Vergata'', Via della
Ricerca Scientifica, I-00133, Roma}
\\e-mail: benfatto@mat.uniroma2.it,
\and A. Giuliani\thanks{Dipartimento di Fisica, Universit\`a di
Roma ``La Sapienza'', Piazzale Aldo Moro 5, I-00185, Roma}
\\e-mail: alessandro.giuliani@roma1.infn.it
\and V. Mastropietro{${}^\ast$}
\\e-mail: mastropi@mat.uniroma2.it
}

\date{5 July 2002}
\maketitle

\begin{abstract}
We prove the convergence of the perturbative expansion, based on
Renormalization Group, of the two point Schwinger function of a
system of weakly interacting fermions in $d=2$, with symmetric
Fermi surface and up to exponentially small temperatures, close to
the expected onset of superconductivity.
\end{abstract}

\section{Introduction and main results}\lb{ss1}

\subsection{Motivations}\lb{ss1.1}

The unexpected properties of recently discovered materials,
showing high-$T_c$ superconductivity and significative deviations
from Fermi liquid behavior in their normal phase (\ie above $T_c$)
\cite{VLSAR}, provides the main physical motivation for the search
of well established results on models for interacting non
relativistic fermions, describing the conduction electrons in
metals. One can consider such models not only in $d=3$, but also
in $d=1,2$, to describe metals so anisotropic that the conduction
electrons move essentially on a chain or on a plane.
Renormalization Group (RG) methods provide a powerful technique
for studying such models. While in $d=1$ RG methods were applied
since long time \cite{So} and many rigorous results up to $T=0$
were established (see for instance \cite{BGPS}, \cite{BoM},
\cite{BM} and \cite{GM} for an updated review), in $d>1$ the
application of RG methods is much more recent and started in
\cite{BG}, \cite{FT}. At the moment RG methods seem unable to get
a rigorous control of such models in $d>1$ up to $T=0$, for the
generic presence of phase transitions (for instance to a
superconducting state) at low temperatures (unless such phase
transitions are forbidden by a careful choice of the dispersion
relation, see \cite{FKT}). On the other hand, RG methods seem well
suited to obtain rigorous information on the behavior of $d>1$
models at temperatures above $T_c$, and to clarify the microscopic
origin of Fermi or non Fermi liquid behavior in the normal phase.
One can write, in the weakly interacting case, an expansion for
the Schwinger functions based on RG ideas; the finite temperature
acts as an infrared cut-off so that each perturbative order is
trivially finite; the mathematical non trivial problem is to prove
that the expansion is convergent, and it turns out that such
problem is more and more difficult as the temperature of the
system decreases. Indeed, if $\l$ is the interaction strength, the
cancellations due to the anticommutativity properties of fermions
allow quite easily to prove convergence of naive perturbation
theory for $T\ge|\l|^{\a}$, for some constant $\a>0$. On the other
hand, the critical temperature in the weak coupling case at which
phase transitions are expected is $O(e^{-(a|\l|)^{-1}})$ where $a$
is a constant essentially given by the second order contributions
\cite{AGD} of the expansion, \ie it is exponentially small and so
quite smaller than $|\l|^{\a}$, if $\l$ is small enough. In
\cite{FMRT} and \cite{DR} the perturbative expansion convergence
was proved for the effective potential up to exponentially small
temperatures, in the $d=2$ {\it Jellium model}, describing
fermions in the continuum, with dispersion relation $\e(\vec
k)=|\vec k|^2/(2m)$ and a rotation invariant weak interaction. One
of the main difficulties of the proof is that non perturbative
bounds are naturally obtained in coordinate space, while one has
to exploit the geometric properties of the Fermi surface (\ie the
set of momenta $\vec k$ such that $\e(\vec k)=\mu$), which are
naturally investigated in momentum space. In \cite{FMRT} and
\cite{DR} an expansion based on RG is considered, such that only
the relevant (but not the marginal) terms are renormalized; this
has the effect that one has convergence for $T\ge e^{-{(c
|\l|)^{-1}}}$, with $c$ related to an all order bound, hence much
bigger than $a$. The proof uses in a crucial way the {\it rotation
invariance} of the Jellium model, an hypothesis which is indeed
quite unrealistic (it corresponds to completely neglecting the
effect of the lattice).

The aim of this paper is to prove convergence of the perturbative
expansion for the two point Schwinger function, in the case of an
interacting system of fermions in a lattice or in the continuum.
Since the interaction modifies the Fermi surface, we write the
dispersion relation $\e_0(\vkk)$ of the free model in the form
$\e_0(\vkk) = \e(\vkk) +\d\e(\vkk)$ and try to choose the {\it
counterterm} $\d\e(\vkk)$, which becomes part of the interaction,
as a suitable function of the original interaction, so that the
Fermi surface of the interacting system is the set $F=\{\vkk:
\e(\vkk)=\m \}$. We can face this problem if $\e(\vkk)$ satisfies
some conditions, implying mainly that $F$ is a smooth, convex
curve, symmetric with respect to the origin.

We prove convergence for weak coupling and up to temperatures
$T\ge \exp\{-(c_0|\l|)^{-1}\}$, where $c_0$ depends on a few terms
of the first and second order, implying a bound of the critical
temperature very close to the expected value. Note that, in order
to get this type of result, we consider an expansion in which both
the relevant and marginal terms are renormalized.

\subsection{The model}\lb{ss1.2}

There are two main classes of models of interacting fermions,
depending wether the Fermi operators space coordinates are
continuous or discrete variables. Our analysis deals with both
such possibilities, so we give the following definitions. \*

\0 1){\it Continuum models.} In such a case, given a square
$[0,L]^2\in\RRR^2$, the inverse temperature $\b$ and the (large)
integer $M$, we introduce in $\L=[0,L]^2\times[0,\b]$ a lattice
$\L_M$, whose sites are given by the {\it space-time points}
$\xx=(x_0, \vec x)=(n_0 a_0, n_1 a, n_2 a)$, $a=L/M$, $a_0=\b/M$,
$n_1,n_2,n_0=0,1,\ldots,M-1$. We also consider the set $\DD$ of
{\it space-time momenta} $\kk=(k_0, {\vec k})$, with ${\vec
k}={2\p{\vec n} \over L}$, ${\vec n}\in\ZZZ^2$, ${\vec
n}=(n_1,n_2)$, $-M\le n_i\le M-1$ and $k_0={2\pi\over
\b}(n_0+{1\over 2})$, $n_1,n_2,n_0=0,1,\ldots,M-1$. With each
$\kk\in\DD$ we associate four Grassmanian variables
$\hat\psi^\e_{\kk,\s}$, $\e,\s\in\{+,-\}$. The lattice $\L_M$ is
introduced only for technical reasons so that the number of
Grassmanian variables is finite, and eventually the (essentially
trivial) limit $M\to\io$ is taken.

\*

\0 2){\it Lattice models}. In such a case, given
$[0,L]^2\in\ZZZ^2$, the inverse temperature $\b$ and the (large)
integer $M$, we introduce in $\L=[0,L]^2\times[0,\b]$ a lattice
$\L_M$, whose sites are given by the {\it space-time points}
$\xx=(x_0, \vec x)=(n_0 a_0, n_1 , n_2), a_0=\b/M$,
$n_1,n_2=0,..,L-1$ and $n_0=0,1,\ldots,M-1$; this definition is
obtained from the previous one by defining $a=1$. In such a case
$\DD$ is a set of space-time momenta $\kk=(k_0, \vec k)$, with
$k_0={2\p \over \b}(n+{1\over 2})$, $n\in\ZZZ$, $-M\le n\le M-1$;
and ${\vec k}={2\p{\vec n} \over L}$, ${\vec n}\in\ZZZ^2$, ${\vec
n}=(n_1,n_2)$, $-[{L\over 2}]\le n_i\le [{(L-1)\over 2}]$. With
each $\kk\in\DD$ we associate four Grassmanian variables
$\hat\psi^\e_{\kk,\s}$, $\e,\s\in\{+,-\}$.

\*

All the models are defined by introducing a linear functional
$P(d\psi)$ on the Grassmanian algebra generated by the variables
$\hat\psi^\e_{\kk,\s}$, such that
\bea &&\int P(d \psi) \hat \psi^-_{\kk_1,\s_1}\hat
\psi^+_{\kk_2,\s_2} = L^2\b \d_{\s_1,\s_2}\d_{\kk_1,\kk_2} \hat
g(\kk_1)\;,\lb{2.1}\\
&& \qquad \hat g(\kk) = {\bar C_0^{-1}(\vec k)\over -ik_0+ \e(\vec
k)-\mu}\;,\nn\eea
where $\e(\vkk)$, the {\it dispersion relation} of the model, is a
function strictly positive for $\vkk\not=0$ and equal to $0$ for
$\vkk=0$, $\m$ is the {\it chemical potential} and $\bar
C_0^{-1}(\vec k)$ is the {\it ultraviolet cut-off}. In the case of
lattice models we choose $\bar C_0^{-1}(\vec k)=1$, while for
continuum models the function $\bar C_0^{-1}(\vec k)$ is defined
as $\bar C_0^{-1}(\vec k)=H\left(\e(\vec k)-\mu\right)$ where
$H(t) \in C^{\io}(\RRR)$ is a smooth function of compact support
such that, for example, $H(t) = 1$ for $t<1$ and $H(t)=0$ for
$t>2$.

We introduce the propagator in coordinate space:
\be g^{L,\b}(\xx-\yy) \= \lim_{M\to\io}
{1\over L^2\b} \sum_{\kk\in {\cal D}} \, e^{-i\kk\cdot(\xx-\yy)}
\, \hat g(\kk) = \lim_{M\to\io} \int P(d \psi) \psi^-_{\xx,\s}
\psi^+_{\yy,\s} \;,\lb{2.3}\ee
where the {\sl Grassmanian field} $ \psi^\e_\xx$ is defined by
\be \psi_{\xx,\s}^{\pm}= {1\over L^2\b}
\sum_{\kk\in {\cal D}}\hat \psi_{\kk,\s}^{\pm} e^{\pm
i\kk\cdot\xx}\; .\lb{2.4}\ee

The ``Gaussian measure'' $P(d \psi)$ has a simple representation
in terms of the ``Lebesgue Grassmanian measure''
\be \hbox{\sl D}\psi=\prod_{\kk\in\DD,\s=\pm\atop
C_0^{-1}(\vec k)>0}d\hat\psi_{\kk,\s}^+
d\hat\psi_{\kk,\s}^-,\lb{2.5}\ee
defined as the linear functional on the Grassmanian algebra, such
that, given a monomial $Q( \hat\psi^-, \hat\psi^+)$ in the
variables $\hat\psi_{\kk,\s}^-, \hat\psi_{\kk,\s}^+$, its value is
$0$, except in the case $Q( \hat\psi^-, \hat\psi^+)=
\prod_{\kk,\s} \hat\psi^-_{\kk,\s} \hat\psi^+_{\kk,\s}$, up to a
permutation of the variable, in which case its value is $1$. We
define
\be
P(d\psi) = N^{-1} \hbox{\sl D}\psi \cdot\;\exp \left\{-{1\over
L^2\b} \sum_{\kk\in\DD,\s=\pm\atop \bar C_0^{-1}(\vec k)>0 } \bar
C_{0}(\vec k) (-i k_0+ \e(\vec
k)-\mu)\hat\psi^{+}_{\kk,\s}\hat\psi^{-}_{\kk,\s}\right\}\;,
\lb{2.6}\ee
with $N=\prod_{\kk\in\DD,\s=\pm}[(L^2\b)^{-1}(-ik_0+\e(\vec
k)-\mu) \bar C_{0}(\vec k)]$.

The {\it Schwinger functions} are defined by the following {\it
Grassmanian functional integral}
\be S(\xx_1,\e_1,\s_1;..,\xx_n,\e_n,\s_n) = \lim_{L\to\io}\lim_{M\to\io}
{\int P(d\psi) e^{-\VV(\psi) -\NN(\psi)}
\psi^{\e_1}_{\xx_1,\s_1}...\psi^{\e_n,\s_n}_{\xx_n} \over \int
P(d\psi) e^{-\VV(\psi) -\NN(\psi)}}\;,\lb{2.9}\ee
where
\be \NN(\psi) = {1\over L^2\b} \sum_{\kk\in\DD, \s=\pm\atop C_0^{-1}(\vec k)>0 }
\hat\n(\vec k,\l)\psi^+_{\kk,\s} \psi^-_{\kk,\s},\lb{2.10}\ee
and, if we use $\int d\xx$ and $\d(x_0-y_0)$ as short-hands for
$\sum_{\xx\in \Lambda_M}a_0 a^2$ and $a_0^{-1}\d_{x_0,y_0}$,
\be \VV(\psi)=\l\sum_{\s,\s'}\int d\xx d\yy \d(x_0-y_0) v_{\s,\s'}(\vxx-\vyy)
\psi^+_{\xx,\s}\psi^-_{\xx,\s}
\psi^+_{\yy,\s'}\psi^-_{\yy,\s'}\;,\lb{2.8}\ee
$v_{\s,\s'}(\vxx)$ being smooth functions such that $\max_{\s,\s'}
\int d\vxx(1+|\vxx|^2) |v_{\s,\s'}(\vxx)|$ is bounded.

Note that $\hat\n(\vec k,\l)$ is related to the counterterm
$\d\e(\vkk)$ introduced in \S\ref{ss1.1} by the relation $\d\e(\vkk)=
\bar C_0^{-1}(\vec k) \hat\n(\vec k,\l)$

\*

In order to make more precise the model, we have to specify some
properties of the dispersion relation. We will assume that
$\e(\vec k)$ verifies the following properties (whose consequences
are discussed in \S\ref{ssA1}. From now on $c,c_1,c_2,\ldots$,
will denote suitable positive constants.

\begin{enumerate}

\item There exists $e_0$ such that, for $|e|\le e_0$, $\e(\vec
k)-\mu=e$ defines a regular $C^\io$ convex curve $\Sigma(e)$
encircling the origin, which can be represented in polar
coordinates as $\vpp=u(\th,e)\vee_r(\th)$ with
$\vee_r(\th)=(\cos\th,\sin\th)$. Note that $e_0<\m$, since
$\e(\vkk)> 0$ for $\vkk\not=0$ and $\e(\V 0)=0$; moreover
$u(\th,e)\ge c>0$ and, if $r(\th,e)$ is the curvature radius,
\be r(\th,e)^{-1}\ge c> 0\;.\lb{2.8a}\ee

\item $e_0$ is chosen so that, if $\vkk\in \Sigma(e)$ and $|e|\le
e_0$, then $\bar C_0^{-1}(\vec k)=1$.

\item If $|e|\le e_0$, then
\be 0<c_1\le \grad\e(\vpp)\cdot\vee_r(\th) \le c_2\;.\lb{2.8b}\ee

\item The following symmetry relation is satisfied
\be \e(\vpp)=\e(-\vpp)\;,\lb{2.8c}\ee
implying that the curves $\Si(e)$ are symmetric by reflection with
respect to the origin.

\end{enumerate}

\* \0 We will call $\Si_F\=\Si(0)$ the {\it Fermi surface} and we
will put $u(\th,0)\vee_r(\th)={\vec p}_F(\th)$ and
$u(\th)\=u(\th,0)=| {\vec p}_F(\th)|$.

\*\0 {\bf Remarks - } The Grassmanian functional integrals
\pref{2.9} are equal, in the limit $M\to\io$, to the Schwinger
functions of an {\it Hamiltonian} model of fermions in two
dimensions, expressed in terms of fermionic creation or
annihilation operators. Among the dispersion relations which are
in the class we are considering is that of the {\it Hubbard
model}, defined in a lattice with local interaction
$v_{\s,\s'}(\vxx-\vyy)\= \d_{\s,-\s'}\d_{\vxx,\vyy}$ (without the
counterterm) and $\e(\vec k)=2-\cos k_1-\cos k_2$, and that of the
{\it jellium model}, defined in the continuum with $\e(\vec
k)=|\vec k|^2/ 2m$. The index $\s$ is the {\it spin index}; {\it
in the following it will be not play any role and it will be
omitted to shorten the notation.}

\*

We are mainly interested in the two point Schwinger function
$S(\xx-\yy)\equiv S(\xx,-; \yy,+)$, with $S(\xx,-;\yy,+)$ given by
\pref{2.9}. For $\l=0$ and $\hat\n(\vkk,\l)=0$, $S(\xx-\yy)$ is
equal to the propagator \pref{2.3}, hence its Fourier transform is
singular at $k_0=0$ (which is not an allowed value at finite
temperature) and $\e(\vkk)=\m$. As we said in \S\ref{ss1.1}, we
want to fix $\hat\n(\vkk,\l)$ so that the location of this
singularity does not change for $\l\not=0$; this allows to study
the model as a perturbation of the model with $\l=0$.

Our main result is the following theorem \*

\begin{theorem}\lb{th2.1}
There exist two positive constants $\e$ and
$c_0$, the last one only depending on first and second order terms
in the perturbative expansion, and a continuous function
$\hat\n(\vec k,\l)=O(\l)$, such that, for all $|\l|\le\e$ and
$T\ge \exp\{- (c_0 |\l|)^{-1}\}$,
\be \hat S(\kk)=\hat g(\kk)(1+\l \hat S_1(\kk))\;,\lb{2.11}\ee
where $\hat g(\kk)$ is the free propagator at finite $\b$ (\ie it
is equal to the Fourier transform of
$\lim_{L\to\io}g^{L;\b}(\xx-\yy)$, see \pref{2.3}) and $|\hat
S_1(\kk)|\le c$, for some constant $c$. In the continuum case with
$\e(\vec k)=|\vec k|^2/ 2m$ and $v(\vrr)=\tilde v(|\vrr|)$, there
exists another constant $c_1$ such that, if $|\l|\le\e$ and $T\ge
\exp\{- (c_1 |\l|)^{-1}\}$, $\hat\n(\vec k,\l)=\nu(\l)$ is a
constant.
\end{theorem}

This theorem says that the two point Schwinger function of the
interacting system is close to the free one, for weak interactions
and up to exponentially small temperatures; the condition on the
temperature is not technical, as at temperatures low enough phase
transitions are expected and a result like \pref{2.11} cannot
hold. The theorem is proved by an expansion similar to the one in
\cite{BG}, in which the {\it relevant} and the {\it marginal}
interactions are renormalized at any iteration of the
Renormalization Group. One writes $\hat S(\kk)$ in terms of a set
of {\it running coupling functions}, which obey recursive
equations, the {\it beta function} of the model. We prove that the
expansions of $\hat S(\kk)$ and of the beta function are
convergent, if the running coupling functions are small in a
suitable norm; the convergence proof is based on the {\it tree
expansion} and the {\it determinant bounds} used for instance in
\cite{BM} and on a suitable generalization to the present problem
of the {\it sector counting lemma} of \cite{FMRT}. Finally we
show, by choosing properly the counterterm $\hat\nu(\vec k,\l)$
and by solving iteratively the beta function, that the running
coupling functions are small up to temperatures exponentially
small $T\ge \exp{-(c_0|\l|)^{-1}}$; $c_0$ is expressed in terms of
a few terms of first and second order, so much closer to the
expected value for the onset of superconductivity. Our non
perturbative definition of the beta function is interesting by
itself, as it could be used to detect the main instabilities of
the model at lower temperatures.

In order to complete our program, we should prove that
$\hat\nu(\vec k,\l)$ and $\e(\vkk)$ can be chosen in a space of
functions with the same differentiability properties and that the
relation $\e_0(\vkk)= \e(\vkk) + \hat\nu(\vec k,\l)$ can be solved
with respect to $\e(\vkk)$, given $\e_0(\vkk)$ and $\l$. This
would imply that the introduction of the counterterm is only a
technical trick, but does not restrict the class of allowed
dispersion relations; for example one could consider the Hubbard
model away from half-filling.

We did not yet get this result, mainly because our bounds can only
show that $\hat\nu(\vec k,\l)$ is a continuous function of compact
support, whose Fourier transform is summable, while $\e(\vkk)$ has
to be a bit more regular than a twice differentiable function. A
similar problem appears in [FKT] in which a result similar to
Theorem \ref{th2.1} above is proved in a class of asymmetric Fermi
surfaces (the asymmetry makes an equation like \pref{2.11} valid
up to $T=0$). It is likely that an improvement in the
differentiability properties of the counterterm could be obtained
by applying the more detailed analysis on the derivatives of the
self energy introduced in \cite{DR}.

This problem is not present in the Jellium model, where, by using
rotational invariance (so that both the free and the interacting
Fermi surfaces are circles), one can choose $\hat\nu(\vec k,\l)$
as a constant with respect to $\vkk$; this is the last statement
in the Theorem, already proved in \cite{DR}. In order to get this
result in a simple way, we chose to give up the ``close to
optimal'' upper bound on the critical temperature; in fact the
constant $c_1$ depends on an all order bound, like in \cite{DR}.
However, even in this case, our result has some interest, since we
get it without being involved in the delicate one particle
irreducibility analysis of \cite{DR}.

\section{Renormalization Group analysis}\lb{ss3}
\setcounter{equation}{0}
\subsection{The scale decomposition}\lb{ss3.1}

It is convenient, for clarity reasons, to start by studying the
"free energy" of the model, defined as
\bea && E_{L,\b} = -{1\over L^2\b}\log\int P(d\psi^{(\le
1)})e^{-\VV^{(1)}},\lb{3.1}\\
&& P(d\psi^{(\le 1)})\=P(d\psi),\qquad
\VV^{(1)}\=\VV+\NN\;.\nn\eea
Note that our model has an ultraviolet cut-off in the $\vkk$
momentum, but the $k_0$ variable is unbounded in the limit
$M\to\io$. Hence, it is convenient to decompose the field as
$\psi^{(\le 1)}= \psi^{(+1)} + \psi^{(\le 0)}$, where
$\psi^{(+1)}$ and $\psi^{(\le 0)}$ are independent fields whose
covariances have Fourier transforms with support, respectively, in
the {\it ultraviolet region} and the {\it infrared region},
defined in the following way.

Item 1) in the list of properties of the dispersion relation given
in \S\ref{ss1.2} implies that, if $H_0(t)$ is a smooth function of
$t\in\RRR^1$, such that
\be H_0(t) = \cases{
1 & if $t <e_0/\g \;,$ \cr 0 & if $t >e_0\;,$\cr}\lb{3.1a}\ee
$\g>1$ being a parameter to be fixed below, then, since $\bar
C_0^{-1}(\vec k)=1$ if $|\e(\vkk)-\m|\le e_0$,
\bea \bar C_0^{-1}(\vec k) &=& C_0^{-1}(\kk) +f_1(\kk)\;,\nn\\
C_0^{-1}(\kk) &=& H_0
\left[\sqrt{k_0^2+[\e(\vkk)-\m]^2}\right]\;,\lb{3.1b}\\
f_1(\kk) &=& \bar C_0^{-1}(\vkk) \left\{1-H_0 \left[
\sqrt{k_0^2+[\e(\vkk)-\m]^2}\right]\right\}\;.\nn \eea
The covariances $g^{(+1)}$ and $g^{(\le 0)}$ of the fields
$\psi^{(+1)}$ and $\psi^{(\le 0)}$ are defined as in \pref{2.8},
with $f_1(\kk)$ and $C_0^{-1}(\kk)$ in place of $\bar
C_0^{-1}(\vec k)$.

If we perform the integration of the ultraviolet field variables
$\psi^{(+1)}$, we get
\be e^{- L^2\b E_{L,\b}} =e^{- L^2\b E_0} \int P(d\psi^{(\le 0)})e^{-\VV^0
(\psi^{\le 0})}\;,\lb{3.2}\ee
where $\VV^{(0)}(\psi^{(\le 0)})$, the {\it effective potential on
scale $0$}, is given by an expression like \pref{3.8} below and
$E_0$ is defined by the condition $\VV^{(0)}(0)=0$.

The analysis of the ultraviolet integration is far easier than the
infrared one. It can be done by the same procedure applied below
for the infrared problem, by making a multiscale expansion of the
u.v. propagator $g^{(1)}(\xx)$, based on an obvious smooth
partition of the interval $\{|k_0|>1\}$. In this way, one can
build a tree expansion for $\VV^{(0)}$, with endpoints on scale
$M>0$, similar to the infrared tree expansion, to be described
below, see Fig. 1 and following items 1)-6). It is easy to see
that there is no relevant or marginal term on any scale $>0$,
except those which are obtained by contracting two fields
associated with the same space-time point in a vertex located
between an endpoint and the first non-trivial vertex following it
(\ie the {\it tadpoles}). However the sum over the scales of this
type of terms, which is not absolutely convergent for $M\to +\io$,
can be controlled by using the explicit expression of the single
scale propagator, since there is indeed no divergence, but only a
discontinuity at $x_0=0$ for $\vec x=0$. We shall omit the
details, which are of the same type of those used below for the
infrared part of the model.

\* Let us now consider the infrared integration; it will be
performed, as usual, by an iterative procedure. Note first that we
can write
\be H_0(t)=\sum_{h=-\io}^0  \tilde f_h(t)\;,\lb{3.4}\ee
where $\tilde f_h(t)=H_0(\g^{-h}t)-H_0(\g^{-h+1}t)$ is a smooth
function, with support in the interval $[\g^{h-2}e_0,\g^h e_0]$,
and $\g>1$ is the scaling parameter. In order to simplify some
calculations, we will put in the following $\g=4$, but this choice
is not essential.

Since $|k_0|\ge \p/\b$, $\forall \kk\in \DD$, if we define
\be h_\b= \max \{h\le 0:\g^{h-1} e_0 < \p/\b\}\;,\lb{3.4a}\ee
we have the identity
\be C_0^{-1}(\kk) =\sum_{h=h_\b}^0 f_h(\kk)\virg
f_h(\kk)\=\tilde
f_h\left(\sqrt{k_0^2+[\e(\vkk)-\m]^2}\right)\;.\lb{3.4b}\ee

We associate with the decomposition \pref{3.4b} a sequence of
constants $E_h$, $h=h_{\b},...,0$, and a sequence of {\it
effective potentials} $\VV^{(h)}(\psi)$ such that $\VV^{(h)}(0)=0$
and
\be e^{- L^2\b E_{L,\b}} =e^{-L^2\b E_h}
\int P(d\psi^{(\le h)})e^{-\VV^{(h)}(\psi^{(\le h)})}\;,\lb{3.5}\ee
where $P(d\psi^{\le h})$ is the fermionic integration with
propagator
\be g^{(\le h)}(\xx) = {1\over L^2\b}\sum_{\kk\in\DD}
{C_{h}^{-1}(\kk)\over -ik_0+\e({\vec k})-\mu}
e^{-i\kk\xx}\;,\lb{3.6}\ee
with
\be C_{h}^{-1}(\kk)=\sum_{j=h_\b}^h f_j(\kk) =
C_{h-1}^{-1}(\kk) + f_h(\kk)\;.\lb{3.7}\ee
The definition \pref{3.5} implies that
\be E_{L,\b} = E_{h_\b} -{1\over L^2\b}\log\int P(d\psi^{(\le h_\b)})
e^{-\VV^{(h_\b)}(\psi^{(\le h_\b)})}\;.\lb{3.7a}\ee

If we neglect the spin indices and we put $\e_1=\ldots=\e_n=+$,
$\e_{n+1}=\ldots=\e_{2n}=-$, we can write the effective potentials
in the form
\be \VV^{(h)}(\psi^{(\le h)})=\sum_{n=1}^\io
\int d\xx_1...d\xx_{2n} \left[\prod_{i=1}^{2n} \psi^{(\le
h)\e_i}_{\xx_i}\right]W^{(h)}_{2n}(\xx_1,...,\xx_{2n})\;.\lb{3.8}\ee

\* {\bf Remark - } The terms in the r.h.s. of \pref{3.8} are well
defined at finite $M$ and $L$, as elements of a finite Grassmanian
algebra, but have only a formal meaning for $M=L=\io$. However,
one can prove that the kernels, as well as $E_{L,\b}$, have well
defined limits as $M$ and $L$ go to infinity. Such result is
achieved by studying a suitable perturbative expansion of these
quantities and by proving that they are uniformly (in $M$ and $L$)
convergent and, in the case of the kernels, that they have fast
decaying properties in the $\xx$ variables; see \cite{BM} for a
complete analysis of this type in the one dimensional case.
However, since this procedure is cumbersome and difficult to
describe rigorously without making obscure the main ideas, which
have nothing to do with the details related with the finite values
of $M$ and $L$, we shall discuss in the following only the formal
limit of our expansions and we shall prove that the kernels as
well as the free energy constants $E_h$ are well defined. For
similar reasons, we shall also consider $k_0$ as a continuous
variable and we shall take into account the essential infrared
cut-off related with the finite temperature value, by preserving
the definition \pref{3.7} of the cut-off functions. This means, in
particular that, from now on
\be {1\over L^2\b}\sum_{\kk\in\DD} \rightarrow {1\over (2\p)^3}\int_\DD
d\kk\;.\lb{3.8a}\ee
Moreover, we shall suppose that the space coordinates are
continuous variables, both in the continuum and lattice models.
This means that, from now on, $\int d\xx$ will denote the integral
over $\RRR^3$. Finally, we shall still use the symbol $L^2\b$ to
denote the formally infinite space-time volume in the extensive
quantities like $L^2\b E_h$.

\subsection{The localization procedure}\lb{ss3.2}

Let us now describe our expansion, which is produced by using an
inductive procedure. First of all, we define an $\LL$ operator
acting on the kernels in the following way:

\begin{enumerate}

\item $\LL W^{(h)}_{2n}=0$ if $n\ge 3$.

\item If $n=2$ and we put $\xxx=(\xx_1,\ldots,\xx_4)$,
$\xx_i=(x_{i,0},\vxx_i)$, $\tilde \xx_i=(\tilde x_{i,0},\vxx_i)$,
$\d(\ux_0)= \d(x_{1,0}-x_{2,0}) \d(x_{1,0}-x_{3,0})
\d(x_{1,0}-x_{4,0})$
\be \LL W^{(h)}_{4}(\xxx) = \d(\ux_0) \int d(\tilde\ux_0\bs \tilde x_{1,0})
W^{(h)}_{4}(\tilde\xxx)\;.\lb{3.9b}\ee
Note that, because of translation invariance, this definition is
independent of the choice of the {\it localization point}, that is
the point whose time coordinate is not integrated ($\xx_1$ in
\pref{3.9b}).

\item If $n=1$ and we put (by using translation invariance)
$W^{(h)}_{2}(\xx_1,\xx_2) = \tilde W^{(h)}_{2}(\xx_1-\xx_2)$,
\bea%
&&\LL W^{(h)}_{2}(\xx_1,\xx_2)= \d(x_{1,0}-x_{2,0})\int dt\;
\tilde W^{(h)}_2(t,\vec x_1-\vxx_2)+\nn\\
&&+\partial_{x_{2,0}} \d(x_{1,0}-x_{2,0})\int d t\; t\; \tilde
W^{(h)}_2(t,\vec x_1-\vxx_2)\;.\lb{3.9}
\eea

\end{enumerate}

The definition of $\LL$ is extended by linearity to $\VV^{(h)}$,
so that we can write
\bea \LL \VV^{(h)}(\psi^{(\le h)}) &=& \int d\xx_1 d\xx_2
\d(x_{1,0}-x_{2,0})\g^h \n_h(\vec{x_1}-\vec{x_2}) \psi^{(\le
h)+}_{\xx_1} \psi^{(\le h)-}_{\xx_2}+\nn\\
&+& \int d\xx_1 d\xx_2 \d(x_{1,0}-x_{2,0})
z_h(\vec{x_1}-\vec{x_2})\psi^{(\le h)+}_{\xx_1}
\dpr_{x_{2,0}}\psi^{(\le h)-}_{\xx_2}+\nn\\
&+& \int d\xxx \l_h(\vxxx) \d(\ux_0) \psi^{(\le
h)+}_{\xx_1}\psi^{(\le h)+}_{\xx_2} \psi^{(\le
h)-}_{\xx_3}\psi^{(\le h)-}_{\xx_4}\;,\lb{3.28} \eea
where $\l_h(\vxxx) = \int d(\ux_0\bs x_{1,0}) W^{(h)}_4(\xxx)$,
$\g^h \n_h(\vec{x_1}-\vec{x_2})=\int dt \tilde W^{(h)}_2(t,\vec
x_1-\vxx_2)$ and $z_h(\vec{x_1}-\vec{x_2}) = -\int d t\; t\;
\tilde W^{(h)}_2(t,\vec x_1-\vxx_2)$.

\0 Note that, in the term containing $z_h(\vxx_1-\vxx_2)$, we can
substitute $\psi^{(\le h)+}_{\xx_1}$ $\dpr_{x_{2,0}}\psi^{(\le
h)-}_{\xx_2}$ with $-[\dpr_{x_{1,0}}\psi^{(\le h)+}_{\xx_1}]
\psi^{(\le h)-}_{\xx_2}$.

The functions $\l_h$, $\n_h$ and $z_h$ will be called the {\it
running coupling functions of scale $h$} or simply the coupling
functions.

It is useful to consider also the representation of $\LL
V^{(h)}(\psi^{(\le h)})$ in terms of the Fourier transforms,
defined so that, for example,
\be W^{(h)}_{2}(\xx_1,\xx_2) = \int{d \kk\over (2\pi)^3}
e^{-i\kk(\xx_1-\xx_2)} \hat W^{(h)}_{2}(\kk)\;,\lb{3.9a}\ee
\be W^{(h)}_{4}(\xx_1,\xx_2,\xx_3,\xx_4)
=\int \prod_{i=1}^3\left[{d\kk_i\over(2\pi)^3}
e^{-i\e_i\kk_i(\xx_i-\xx_4)} \right] \hat
W^{(h)}_{4}(\kk_1,\kk_2,\kk_3)\;.\lb{3.26}\ee
We can write
\bea \LL\VV^{(h)} (\psi^{(\le h)}) &=& \int{d \kk\over
(2\pi)^3}[\g^h \hat\n_h(\vec k )-i k_0 \hat
z_h(\vec{k})]\psi^{(\le h)+}_{\kk}\psi^{(\le h)-}_{\kk}+\nn\\
&+& \int\prod_{i=1}^4 {d \kk_i\over (2\pi)^3} \psi^{(\le
h)+}_{\kk_1}\psi^{(\le h)+}_{\kk_2} \psi^{(\le
h)-}_{\kk_3}\psi^{(\le h)-}_{\kk_4}\;\cdot\lb{3.10}\\
&\cdot\;& \hat\l_h(\vkk_1,\vkk_2,\vkk_3)
\d(\kk_1+\kk_2-\kk_3-\kk_4)\;,\nn\eea
where $\hat\l_h(\vkk_1,\vkk_2,\vkk_3)=\hat W^{(h)}_4((0,\vkk_1),
(0,\vkk_2), (0,\vkk_3))$, $\g^h\hat\nu_h(\vec k)=\hat
W^{(h)}_2(0,\vec{k})$, $\hat z_h(\vec k)=$ $i\partial_{k_0}\hat
W^{(h)}_2(0,\vec{k})$.

We also define $\RR\=1-\LL$; by using \pref{3.9}, we get:
\bea \RR W_{2}^{(h)}(\xx_1,\xx_2)&=& \tilde
W_{2}^{(h)}(\xx_1-\xx_2) -\d(x_{1,0}-x_{2,0})\bar W_2^{(h)}(0,
{\vec x}_1-{\vec x}_2)-\nn\\
&-& i\dpr_{x_{1,0}}\d(x_{1,0}-x_{2,0})\partial_{k_0} \bar
W_2^{(h)}(0, {\vec x}_1-{\vec x}_2)\;,\lb{3.16}\eea
where $\bar W_2^{(h)}(k_0, {\vec x})=\int dt\, e^{ik_0t} \tilde
W^{(h)}_2(t,\vec x)$. Furthermore
\be \RR W_4^{(h)}(\xxx)= W_4^{(h)}(\xxx) - \d(\ux_0) \bar W_4^{(h)}
(\underline{0},\vxxx)\;,\lb{3.17}\ee
where $\bar W_4^{(h)}(\uk_0,\vxxx)$ is the Fourier transform of
$W_4^{(h)}(\xxx)$ with respect to the time coordinates.

\subsection{The sector decomposition}\lb{ss3.2a}

We now further decompose the field $\psi^{(\le h)}$, by slicing
the support of $C_h^{-1}(\kk)$ as in \cite{FMRT}. Let $H_2(t)$ be
a smooth function on the interval $[-1,+1]$, such that
\be H_2(t) = \cases{
1 & if $|t| < 1/4 \;$ \cr 0 & if $|t| >3/4; $\cr} \virg H_2(t)+H_2
(1-t)=1 \ \ \hbox{\rm if}\ \ 1/4 <t<3/4,\lb{3.11}\ee
and let us define, if $\o$ is an integer in the set
$O_h\=\{0,1,\ldots,\g^{-(h-1)/2}-1\}$ (recall that $\g=4$) and
$h\le 0$,
\be \bar\z_{h,\o}(t)= H_2\left({\g^{-{h\over 2}}\over \pi}(t-\th_{h,\o})\right)
\virg \th_{h,\o}=\pi (\o+{1\over 2}) \g^{{h\over
2}}\;.\lb{3.14a}\ee
It is easy to see that $\bar\z_{h,\o}(t)$ can be extended to the
real axis as a periodic function of period $2\p$, that we can use
to define a smooth function on the one-dimensional torus $\TTT^1$,
to be called $\z_{h,\o}(\th)$; moreover
\be \sum_{\o\in O_h} \z_{h,\o}(\th)=1 \virg \forall  \th\in \TTT^1\;.\lb{3.14}\ee
On the other hand, the properties of $\e(\vkk)$ assumed in
\S\ref{ss1.2} imply that, if $C_h^{-1}(\kk)\not=0$, $\vkk=u(\th,e)
\vee_r(\th)$ with $e=\e(\vkk)-\m$. Hence, we can write
\be \psi_{\xx}^{(\le h)\pm}\=\sum_{\o\in O_h}
e^{\pm i {\vec p_F}(\th_{h,\o})\vec x}\psi_{\xx,\o}^{(\le h)\pm}
\virg P(d\psi^{(\le h)})=\prod_{\o\in O_h} P(d\psi^{(\le
h)}_{\o})\;,\lb{3.12}\ee
where $P(d\psi^{(\le h)}_\o)$ is the Grassmanian integration with
propagator
\be  g^{(\le h)}_\o(\xx-\yy)={1\over (2\p)^3} \int d\kk
e^{-i\left[\kk(\xx-\yy)-{\vpp_F}(\th_{h,\o})(\vec x-\vec
y)\right]} {C_{h}^{-1}(\kk)\z_{h,\o}(\th) \over -i k_0+\e(\vec
k)-\mu}\;.\lb{3.13}\ee

If we insert the l.h.s. of \pref{3.12} in \pref{3.8}, we get
\bea &&\VV^{(h)}\left( \sum_{\o\in O_h} e^{\e i
\vpp_F(\th_{h,\o})\vec x} \psi_{\o}^{(\le h))\e} \right) =
\sum_{n=1}^\io\sum_{\o_1,\ldots,\o_{2n}\in O_h} \cdot\nn\\
&&\cdot \int d\xx_1...d\xx_{2n} \left[\prod_{i=1}^{2n} e^{\e_i
i{\vec p_F}(\th_{\o_{h,i}})\vec x_i}\psi^{(\le
h)\e_i}_{\xx_i,\o_i}
\right]W^{(h)}_{2n}(\xx_1,...,\xx_{2n}).\lb{3.15}\eea

By using \pref{3.7}, we can write
\bea &&\int \prod_{\o\in O_h} P(d\psi^{(\le
h)}_{\o})e^{-(\LL+\RR)\VV^{(h)} \left( \sum_{\o\in O_h} e^{\e i
{\vpp_F}(\th_{h,\o}) \vec x}\psi_{\o}^{(\le h)\e} \right)}=\nn\\
&&=\int P(d\psi^{(\le h-1)})\int \prod_{\o\in O_h}
P(d\psi_{\o}^{(h)})\cdot\lb{3.19}\\
&&\cdot e^{-(\LL+\RR)\VV^{(h)}\left(\psi_{\xx}^{(\le h-1)\e}+
\sum_{\o\in O_h} e^{\e i {\vpp_F}(\th_{h,\o}) \vec x}\psi_{\xx,
\o}^{(h)\e}\right)}\;,\nn\eea
where $P(d\psi_{\o}^{(h)})$ is the integration with propagator
\be  g_{\o}^{(h)}(\xx)\={1\over (2\p)^3} \int d\kk
e^{-i\left(\kk\xx-{\vec p_F}(\th_{h,\o})\vec x\right)}
{F_{h,\o}(\kk)\over -ik_0+\e (\vec k)-\mu}\;,\lb{3.20}\ee
\be F_{h,\o}(\kk) = f_h(\kk)\z_{h,\o}(\th)\;.\lb{3.20a}\ee
The support of $F_{h,\o}(\kk)$ will be called the {\it sector of
scale $h$ and sector index $\o$}.

In order to compute the asymptotic behavior of $g_{\o}^{(h)}(\xx)$
it is convenient to introduce a coordinate frame adapted to the
Fermi surface in the point ${\vec p_F}(\th_{h,\o})$. By using the
definitions of \S\ref{ss1.2} and putting
$\vec{e}_t(\th)=(-\sin\th,\cos\th)$, we define
\bea \vec{\t}(\th) &=& {d \vec{p}_F(\th)\over d\th}\left|{d
\vec{p}_F(\th)\over d\th}\right|^{-1}=
{u'(\th)\vec{e}_r(\th)+u(\th)\vec{e}_t(\th)
\over\sqrt{u'(\th)^2+u(\th)^2}}\;,\nn\\
{\vec n}(\th) &=& {u(\th)\vec{e}_r(\th)-u'(\th)\vec{e}_t(\th)
\over\sqrt{u'(\th)^2+u(\th)^2}}\;.\lb{3.21}\eea
Moreover, given any $\kk$ belonging to the support of
$F_{h,\o}(\kk)$, we put
\be \vec{k}=\vec p_F(\th_{h,\o})+k'_1 \vec{n}(\th_{h,\o})+
k'_2 \vec{\t}(\th_{h,\o}) =\vec p_F(\th_{h,\o}) +
\vkk'\;;\lb{3.21a}\ee
it is easy to verify that $|k'_1|\le C\g^h$, $|k'_2|\le
C\g^{h\over 2}$, see Lemma \ref{lmA1.3} in \S\ref{ssA1} for
details. By using \pref{3.21a}, we can rewrite \pref{3.20} as
\be  g_{\o}^{(h)}(\xx)\={1\over (2\p)^3} \int dk_0 d\vkk'
e^{-i\left(k_0 x_0 +\vkk' \vxx \right)}
{F_{h,\o}(k_0,\vpp_F(\th_{h,\o})+\vkk')\over -ik_0+
\e(\vpp_F(\th_{h,\o})+\vkk')-\mu}\;.\lb{3.20c}\ee
Let us now put
\be \vxx= x'_1 \vec{n}(\th_{h,\o})+ x'_2 \vec{\t}(\th_{h,\o})\;;\lb{3.21b}\ee
the following lemma gives a bound on the asymptotic behavior of
$g_{\o}^{(h)}(\xx)$, which is very important in our analysis, as
in \cite{FMRT}. It will be proved in \S\ref{ssA1}.

\begin{lemma}\lb{lm3.1}
Given the integers $N,m,n_0,n_1,n_2\ge 0$, with
$m=n_0+n_1+n_2$, there exists a constant $C_{N,m}$ such that
\be |\dpr_{x_0}^{n_0} \dpr_{x'_1}^{n_1} \dpr_{x'_2}^{n_2} g_{\o}^{(h)}(\xx)|
\le {C_{N,m} \g^{{3\over 2}h} \g^{(n_0+n_1+{1\over 2} n_2)h}\over
1+(\g^h|x_0| +\g^h|x'_1|+\g^{{1\over 2}h}|x'_2|)^N}\;.\lb{3.22}\ee
\end{lemma}

\* {\bf Remark} Lemma \ref{lm3.1} holds also for non $C^{\io}$ Fermi
surfaces: it is sufficient the condition that the derivatives of
$\e(\vkk)$ diverge ``not too fast'' (\ie that $\dpr^n/\dpr
{k'}_1^{n_1}\dpr {k'}_2^{n_2} [\e(\vec k) -\m]
=O(\g^{-h(n_1+{1\over 2}n_2-1)})$).

\subsection{The tree expansion}\lb{ss3.3}

Our expansion of $\VV^{(h)}$, $0\ge h\ge h_\b$ is obtained by
integrating iteratively the field variables of scale $j\ge h+1$
and sector index $\o=1,\ldots,\g^{-h/2}$ and by applying at each
step the {\it localization procedure} described above, which has
the purpose of summing together the relevant contributions of the
same type. The result can be expressed in terms of a {\it tree
expansion}, similar to that described, for example, in \cite{BM}.

\insertplot{300}{150}%
{\ins{30pt}{85pt}{$r$}\ins{50pt}{85pt}{$v_0$}\ins{130pt}{100pt}{$v$}%
\ins{35pt}{-2pt}{$h$}\ins{55pt}{-2pt}{$h+1$}\ins{135pt}{-2pt}{$h_v$}%
\ins{215pt}{-2pt}{$0$}\ins{235pt}{-2pt}{$+1$}\ins{255pt}{-2pt}{$+2$}}%
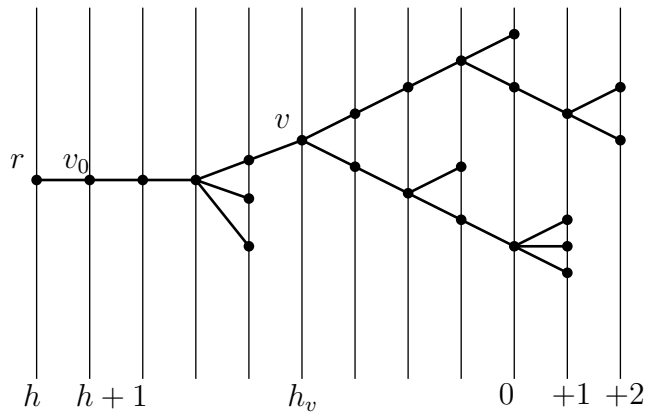
{fig51}{A possible tree of the expansion for the effective
potentials.\lb{f1}}{0}

We need some definitions and notations.

\0 1) Let us consider the family of all trees which can be
constructed by joining a point $r$, the {\it root}, with an
ordered set of $n\ge 1$ points, the {\it endpoints} of the {\it
unlabeled tree} (see Fig. 1), so that $r$ is not a branching
point. $n$ will be called the {\it order} of the unlabeled tree
and the branching points will be called the {\it non trivial
vertices}. The unlabeled trees are partially ordered from the root
to the endpoints in the natural way; we shall use the symbol $<$
to denote the partial order.

Two unlabelled trees are identified if they can be superposed by a
suitable continuous deformation, so that the endpoints with the
same index coincide. It is then easy to see that the number of
unlabeled trees with $n$ end-points is bounded by $4^n$.

We shall consider also the {\it labelled trees} (to be called
simply trees in the following); they are defined by associating
some labels with the unlabeled trees, as explained in the
following items.

\0 2) We associate a label $h\le 0$ with the root and we denote
$\TT_{h,n}$ the corresponding set of labelled trees with $n$
endpoints. Moreover, we introduce a family of vertical lines,
labelled by an integer taking values in $[h,2]$, and we represent
any tree $\t\in\TT_{h,n}$ so that, if $v$ is an endpoint or a non
trivial vertex, it is contained in a vertical line with index
$h_v>h$, to be called the {\it scale} of $v$, while the root is on
the line with index $h$. There is the constraint that, if $v$ is
an endpoint, $h_v>h+1$.

The tree will intersect in general the vertical lines in set of
points different from the root, the endpoints and the non trivial
vertices; these points will be called {\it trivial vertices}. The
set of the {\it vertices} of $\t$ will be the union of the
endpoints, the trivial vertices and the non trivial vertices. Note
that, if $v_1$ and $v_2$ are two vertices and $v_1<v_2$, then
$h_{v_1}<h_{v_2}$.

Moreover, there is only one vertex immediately following the root,
which will be denoted $v_0$ and can not be an endpoint (see
above); its scale is $h+1$.

Finally, if there is only one endpoint, its scale must be equal to
$h+2$.

\0 3) With each endpoint $v$ of scale $h_v=+2$ we associate one of
the two contributions to $\VV^{(1)}(\psi^{(\le 1)})$, and a set
$\xx_v$ of space-time points (the two corresponding integration
variables); we shall say that the endpoint is of type $\l$ or
$\n$, respectively. With each endpoint $v$ of scale $h_v\le 1$ we
associate one of the three terms appearing in \pref{3.28} and the
set $\xx_v$ of the corresponding integration variables; we shall
say that the endpoint is of type $\n$, $z$ or $\l$, respectively.

Given a vertex $v$, which is not an endpoint, $\xx_v$ will denote
the family of all space-time points associated with one of the
endpoints following $v$.

Moreover, we impose the constraint that, if $v$ is an endpoint,
$h_v=h_{v'}+1$, if $v'$ is the non trivial vertex immediately
preceding $v$.

\0 4) If $v$ is not an endpoint, the {\it cluster } $L_v$ with
scale $h_v$ is the set of endpoints following the vertex $v$; if
$v$ is an endpoint, it is itself a ({\it trivial}) cluster. The
tree provides an organization of endpoints into a hierarchy of
clusters.

\0 5) The trees containing only the root and an endpoint of scale
$h+1$ will be called the {\it trivial trees}; note that they do
not belong to $\TT_{h,1}$, if $h\le 0$ (see the end of item 3
above), and can be associated with the three terms in the local
part of $\VV^{(h)}$.

\0 6) We introduce a {\it field label} $f$ to distinguish the
field variables appearing in the terms associated with the
endpoints as in item 3); the set of field labels associated with
the endpoint $v$ will be called $I_v$. Analogously, if $v$ is not
an endpoint, we shall call $I_v$ the set of field labels
associated with the endpoints following the vertex $v$; $\xx(f)$
and $\e(f)$ will denote the space-time point and the $\e$ index,
respectively, of the field variable with label $f$.

If $h_v\le +1$, one of the field variables belonging to $I_v$
carries also a time derivative $\dpr_0$ if the corresponding local
term is of type $z$, see \pref{3.28}. Hence we can associate with
each field label $f$ an integer $m(f)\in\{0,1\}$, denoting the
order of the time derivative. Note that $m(f)$ is not uniquely
determined, since we are free to choose on which field exiting
from a vertex of type $z$ the derivative falls, see comment after
\pref{3.28}; we shall use this freedom in the following.

\* If $h\le 0$, the effective potential can be written in the
following way:
\be \VV^{(h)}(\psi^{(\le h)}) + L\b \tilde E_{h+1}=
\sum_{n=1}^\io\sum_{\t\in\TT_{h,n}} \VV^{(h)}(\t,\psi^{(\le
h)})\;,\lb{3.29}\ee
where, if $v_0$ is the first vertex of $\t$ and $\t_1,..,\t_s$
($s=s_{v_0}$) are the subtrees of $\t$ with root $v_0$,
$\VV^{(h)}(\t,\psi^{(\le h)})$ is defined inductively by the
relation
\be \VV^{(h)}(\t,\psi^{(\le h)})= {(-1)^{s+1}\over s!}
\EE^T_{h+1}[ \bar\VV^{(h+1)}(\t_1,\psi^{(\le h+1)});\ldots;
\bar\VV^{(h+1)}(\t_{s},\psi^{(\le h+1)})]\;,\lb{3.30}\ee
and $\bar\VV^{(h+1)}(\t_i,\psi^{(\le h+1)})$

\0 a) is equal to $\RR\VV^{(h+1)}(\t_i,\psi^{(\le h+1)})$ if the
subtree $\t_i$ is not trivial;

\0 b) if $\t_i$ is trivial and $h\le -1$, it is equal to one of
the three terms in $\LL\VV^{(h+1)}(\psi^{(\le h+1)})$ or, if
$h=0$, to one of the two terms contributing to
$\VV^{(1)}(\psi^{\le 1})$.

\0 $\EE^T_{h+1}$ denotes the truncated expectation with respect to
the measure\\ $\prod_{\o} P(d \psi_\o^{(h+1)})$, that is
\bea
&&\EE^T_{h+1}(X_1;\ldots;X_p)\=\nn\\
&&\={\dpr^p\over\dpr\l_1\ldots\dpr\l_p} \left.\log\int\prod_{\o}
P(d \psi_\o^{(h+1)}) e^{\l_1 X_1+
\cdots\l_pX_p}\right|_{\l_i=0}.\lb{3.31}\eea
This means, in particular, that, in \pref{3.30}, one has to use for
the field variables the sector decomposition \pref{3.12}.

We can write \pref{3.30} in a more explicit way, by a procedure
very similar to that described, for example, in \cite{BM}. Note
first that, if $h=0$, the r.h.s. of \pref{3.30} can be written
more explicitly in the following way. Given $\t\in\TT_{0,n}$,
there are $n$ endpoints of scale $2$ and only another one vertex,
$v_0$, of scale $1$; let us call $v_1,\ldots, v_n$ the endpoints.
We choose, in any set $I_{v_i}$, a subset $Q_{v_i}$ and we define
$P_{v_0}=\cup_i Q_{v_i}$; then we associate a sector index
$\o(f)\in O_0$ with any $f\in P_{v_0}$ and we put
$\O_{v_0}=\{\o(f): f\in P_{v_0}\}$. We have
\be \VV^{(0)}(\t, \psi^{(\le 0)})=\sum_{P_{v_0},\O_{v_0}}
\VV^{(0)}(\t,P_{v_0},\O_{v_0})\;,\lb{3.31a}\ee
\be \VV^{(0)}(\t,P_{v_0},\O_{v_0})= \int d\xx_{v_0}
\tilde\psi^{\le 0}_{\O_{v_0}} (P_{v_0})
K_{\t,P_{v_0}}^{(1)}(\xx_{v_0})\;, \lb{3.31b}\ee
\be K_{\t,P_{v_0}}^{(1)}(\xx_{v_0})={1\over n!} \EE^T_{1}[
\bar\psi^{(1)}(P_{v_1}\bs Q_{v_1}),\ldots,
\bar\psi^{(1)}(P_{v_n}\bs Q_{v_n})] \prod_{i=1}^n
K^{(2)}_{v_i}(\xx_{v_i})\;,\lb{3.31c}\ee
where we use the definitions ($\dpr_0$ is from now on the time
derivative)
\be \tilde\psi^{(\le h)}_{\O_v}(P_v)= \prod_{f\in P_v}
e^{i\e(f)\vpp_F(\th_{h,\o(f)})\vxx(f)} \dpr_0^{m(f)}\psi^{(\le
h)\e(f)}_{\xx(f),\o(f)}\virg h\le 0\;,\lb{3.34}\ee
\be \bar\psi^{(1)}(P_v) =\prod_{f\in P_v} \psi^{(1)\e(f)}_{\xx(f)}
\;,\lb{3.31d}\ee
\be K^{(2)}_{v_i}(\xx_{v_i})= \cases{ \l v(\vxx-\vyy) \d(x_0-y_0)
& if $v_i$ is of type $\l$ and $\xx_{v_i}=(\xx,\yy)$,\cr
\n(\vxx-\vyy)\d(x_0-y_0) & if $v_i$ is of type $\n$,}\lb{3.31e}\ee
and we suppose that the order of the (anticommuting) field
variables in \pref{3.31d} is suitable chosen in order to fix the
sign as in \pref{3.31c}. Note that the terms with
$P_{v_0}=\emptyset$ in the r.h.s. of \pref{3.31a} contribute to
$L\b \tilde E_1$, while the others contribute to
$\VV^{(0)}(\psi^{(\le 0)})$.

We now write $\VV^{(0)}$ as $\LL \VV^{(0)}+\RR \VV^{(0)}$, with
$\LL\VV^{(0)}$ defined as in \S\ref{ss3.2} (it represent, in the usual
RG language, the relevant and marginal contributions to
$\VV^{(0)}(\psi^{(\le 0)})$), and we write for $\RR\VV^{(0)}$ a
decomposition similar to the previous one, with
$\RR\VV^{(0)}(\t,P_{v_0},\O_{v_0})$ in place of
$\VV^{(0)}(\t,P_{v_0},\O_{v_0})$; this means that we modify,
according to the representation \pref{3.16}, \pref{3.17} of the
$\RR$ operation, the kernels
\be W_{\t,P_{v_0}}^{(0)}(\xx_{P_{v_0}})=\int d(\xx_{v_0}\bs\xx_{P_{v_0}})
K_{\t,P_{v_0}}^{(1)}(\xx_{v_0})\;,\lb{3.31f}\ee
where $\xx_{P_{v_0}}=\cup_{f\in P_{v_0}} \xx(f)$. In order to
remember this choice, we write
\be \RR \VV^{(0)}(\t,P_{v_0},\O_{v_0})= \int d\xx_{v_0}
\tilde\psi^{(\le 0)}_{\O_{v_0}} (P_{v_0})[\RR
K_{\t,P_{v_0}}^{(1)}(\xx_{v_0})] \;.\lb{3.31g}\ee
It is not hard to see that, by iterating the previous procedure,
one gets for $\VV^{(h)}(\t,\psi^{(\le h)})$, for any
$\t\in\TT_{h,n}$, the representation described below.

\*

We associate with any vertex $v$ of the tree a subset $P_v$ of
$I_v$, the {\it external fields} of $v$. These subsets must
satisfy various constraints. First of all, if $v$ is not an
endpoint and $v_1,\ldots,v_{s_v}$ are the vertices immediately
following it, then $P_v \subset \cup_i P_{v_i}$; if $v$ is an
endpoint, $P_v=I_v$. We shall denote $Q_{v_i}$ the intersection of
$P_v$ and $P_{v_i}$; this definition implies that $P_v=\cup_i
Q_{v_i}$. The subsets $P_{v_i}\bs Q_{v_i}$, whose union ${\cal
I}_v$ will be made, by definition, of the {\it internal fields} of
$v$, have to be non empty, if $s_v>1$.

Moreover, we associate with any $f\in {\cal I}_v$ a scale label
$h(f)=h_v$ and, if $h(f)\le 0$, an index $\o(f)\in O_{h(f)}$,
while, if $h(f)=+1$, we put $\o(f)=0$. Note that, if $h(f)\le 0$,
$h(f)$ and $\o(f)$ single out a sector of scale $h(f)$ and sector
index $\o(f)$ associated with the field variable of index $f$. In
this way we assign $h(f)$ and $\o(f)$ to each field label $f$,
except those which correspond to the set $P_{v_0}$; we associate
with any $f\in P_{v_0}$ the scale label $h(f)=h$ and a sector
index $\o(f)\in O_h$. We shall also put, for any $v\in\t$, $\O_v=
\{\o(f),f\in P_v\}$.

Given $\t\in\TT_{h,n}$, there are many possible choices of the
subsets $P_v$, $v\in\t$, compatible with all the constraints; we
shall denote $\PP_\t$ the family of all these choices and $\bP$
the elements of $\PP_\t$. Analogously, we shall call ${\cal O}_\t$
the family of possible values of $\O=\{\o(f), f\in \cup_v I_v\}$.

Then we can write
\be \VV^{(h)}(\t,\psi^{(\le h)})=\sum_{\bP\in\PP_\t,\O\in {\cal O}_\t}
\VV^{(h)}(\t,\bP,\O)\;.\lb{3.32}\ee
$\VV^{(h)}(\t,\bP,\O)$ can be represented as
\be \VV^{(h)}(\t,\bP,\O)=\int d\xx_{v_0} \tilde\psi^{(\le h)}_{\O_{v_0}}
(P_{v_0}) K_{\t,\bP,\O}^{(h+1)}(\xx_{v_0})\;,\lb{3.33}\ee
with $K_{\t,\bP,\O}^{(h+1)}(\xx_{v_0})$ defined inductively
(recall that $h_{v_0}=h+1$) by the equation, valid for any
$v\in\t$ which is not an endpoint,
\be K_{\t,\bP,\O}^{(h_v)}(\xx_v)={1\over s_v !}
\prod_{i=1}^{s_v} [K^{(h_v+1)}_{v_i}(\xx_{v_i})]\; \;\EE^T_{h_v}[
\tilde\psi^{(h_v)}_{\O_1}(P_{v_1}\bs Q_{v_1}),\ldots,
\tilde\psi^{(h_v)}_{\O_{s_v}}(P_{v_{s_v}}\bs
Q_{v_{s_v}})]\;,\lb{3.35}\ee
where $\O_i=\{\o(f), f\in P_{v_i}\bs Q_{v_i}\}$ and
$\tilde\psi^{(h_v)}_{\O_i}(P_{v_i}\bs Q_{v_i})$ has a definition
similar to \pref{3.34}, if $h_v\le 0$, while, if $h_v=+1$, is
defined as in \pref{3.31d}.

Moreover, if $v$ is an endpoint, $K^{(2)}_{v}(\xx_{v})$ is defined
as in \pref{3.31e} if $h_v=2$, otherwise
\be K^{(h_v)}_{v}(\xx_{v})= \cases{ \l_{h_v-1}(\vxxx) \d(\ux_0)
& if $v$ is of type $\l$,\cr
\g^{h_v-1}\n_{h_v-1}(\vec{x}-\vec{y})\d(x_0-y_0) & if $v$ is of
type $\n$,\cr z_{h_v-1}(\vec{x}-\vec{y})\d(x_0-y_0) & if $v$ is of
type $z$,}\lb{3.37}\ee
where $\xx_v=(\xx_1,\xx_2,\xx_3,\xx_4)$ if $v$ is of type $\l$,
and $\xx_v=(\xx,\yy)$ in the other two cases. If $v_i$ is not an
endpoint,
\be K^{(h_v+1)}_{v_i}(\xx_{v_i}) = \RR K_{\t_i,\bP^{(i)},\O^{(i)}}^{(h_v+1)}
(\xx_{v_i})\;,\lb{3.37a}\ee
where $\t_i$ is the subtree of $\t$ starting from $v$ and passing
through $v_i$ (hence with root the vertex immediately preceding
$v$), $\bP^{(i)}$ and $\O^{(i)}$ are the restrictions to $\t_i$ of
$\bP$ and $\O$. The action of $\RR$ is defined using the
representations \pref{3.16}, \pref{3.17} of the regularization
operation, as in \pref{3.31f}, \pref{3.31g}.

{\bf Remark} - In order to simplify \pref{3.34} and the following
discussion, we now decide to use the freedom in the choice of the
field that carries the $\dpr_0$ derivative in the endpoints of
type $z$, so that, given any vertex $v$, which is not an endpoint
of type $z$, $m(f)=0$ for all $f\in P_v$.

\*

\pref{3.32} is not the final form of our expansion, since we
further decompose $\VV^{(h)}(\t,\bP,\O)$, by using the following
representation of the truncated expectation in the r.h.s. of
\pref{3.35}. Let us put $s=s_v$, $P_i\=P_{v_i}\bs Q_{v_i}$;
moreover we order in an arbitrary way the sets $P_i^\pm\=\{f\in
P_i,\e(f)=\pm\}$, we call $f_{ij}^\pm$ their elements and we
define $\xx^{(i)}=\cup_{f\in P_i^-}\xx(f)$, $\yy^{(i)}=\cup_{f\in
P_i^+}\xx(f)$, $\xx_{ij}=\xx(f^-_{i,j})$,
$\yy_{ij}=\xx(f^+_{i,j})$. Note that $\sum_{i=1}^s
|P_i^-|=\sum_{i=1}^s |P_i^+|\=n$, otherwise the truncated
expectation vanishes. A couple
$l\=(f^-_{ij},f^+_{i'j'})\=(f^-_l,f^+_l)$ will be called a line
joining the fields with labels $f^-_{ij},f^+_{i'j'}$ and sector
indices $\o^-_l=\o(f^-_l)$, $\o^+_l=\o(f^+_l)$ and connecting the
points $\xx_l\=\xx_{i,j}$ and $\yy_l\=\yy_{i'j'}$, the {\it
endpoints} of $l$. Moreover, we shall put $m_l=m(f^-_l)+m(f^+_l)$
and, if $\o^-_l=\o^+_l$, $\o_l\=\o^-_l=\o^+_l$. Then, it is well
known (see \cite{Le}, \cite{BGPS}, \cite{GM} for example) that, up
to a sign, if $s>1$,
\bea && \EE^T_{h}(\tilde\psi^{(h)}_{\O_1}(P_1),...,
\tilde\psi^{(h)}_{\O_s}(P_s))=\nn\\
&& =\sum_{T}\prod_{l\in T} \dpr_0^{m_l} \tilde
g^{(h)}_{\o_l}(\xx_l-\yy_l) \d_{\o^-_l,\o^+_l}\int dP_{T}(\tt)
\det G^{h,T}(\tt)\;,\lb{3.38}\eea
where
\be \tilde g^{(h)}_\o(\xx) = e^{-i\vpp_F(\th_{h,\o})\vxx}
g^{(h)}_\o(\xx)\;,\lb{3.37b}\ee
$T$ is a set of lines forming an {\it anchored tree graph} between
the clusters of points $\xx^{(i)}\cup\yy^{(i)}$, that is $T$ is a
set of lines, which becomes a tree graph if one identifies all the
points in the same cluster. Moreover $\tt=\{t_{i,i'}\in [0,1],
1\le i,i' \le s\}$, $dP_{T}(\tt)$ is a probability measure with
support on a set of $\tt$ such that $t_{i,i'}=\uu_i\cdot\uu_{i'}$
for some family of vectors $\uu_i\in \RRR^s$ of unit norm. Finally
$G^{h,T}(\tt)$ is a $(n-s+1)\times (n-s+1)$ matrix, whose elements
are given by $G^{h,T}_{ij,i'j'}=t_{i,i'} \dpr_0^{m(f^-_{ij})+
m(f^+_{i'j'})} \tilde
g^{(h)}_{\o_l}(\xx_{ij}-\yy_{i'j'})\d_{\o^-_l,\o^+_l}$ with
$(f^-_{ij}, f^+_{i'j'})$ not belonging to $T$.

In the following we shall use \pref{3.38} even for $s=1$, when $T$
is empty, by interpreting the r.h.s. as equal to $1$, if
$|P_1|=0$, otherwise as equal to $\det
G^{h}=\EE^T_{h}(\tilde\psi^{(h)}(P_1))$. \*

If we apply the expansion \pref{3.38} in each non trivial vertex of
$\t$, we get an expression of the form
\bea \VV^{(h)}(\t,\bP,\O) &=& \sum_{T\in {\bf T}} \int d\xx_{v_0}
\tilde\psi^{(\le h)}_{\O_{v_0}}(P_{v_0}) W_{\t,\bP,\O\bs
\O_{v_0},T}^{(h)}(\xx_{v_0})\nn\\
&\=& \sum_{T\in {\bf T}} \VV^{(h)}(\t,\bP,\O,T)\;,\lb{3.38a}\eea
where ${\bf T}$ is a special family of graphs on the set of points
$\xx_{v_0}$, obtained by putting together an anchored tree graph
$T_v$ for each non trivial vertex $v$. Note that any graph $T\in
{\bf T}$ becomes a tree graph on $\xx_{v_0}$, if one identifies
all the points in the sets $x_v$, for any vertex $v$ which is also
an endpoint.

\* {\bf Remarks} - An important role in this paper, as in
\cite{FMRT}, will have the remark that, thanks to momentum
conservation and compact support properties of propagator Fourier
transforms, $\VV^{(h)}(\t,\bP,\O)$ vanishes for some choices of
$\O$. This constraint will be made explicit below in a suitable
way, see \pref{3.48}.

\0 Note also that $W_{\t,\bP,\O\bs \O_{v_0},T}^{(h)}(\xx_{v_0})$,
as underlined in the notation, is independent of $\O_{v_0}$, so
that $\VV^{(h)}(\t,\bP,\O,T)$ depends on $\O_{v_0}$ only through
the external fields sector indices.

\subsection{Detailed analysis of the $\RR$ operation}\lb{ss3.4}

The kernels $W_{\t,\bP,\O\bs \O_{v_0},T}^{(h)}(\xx_{v_0})$ in
\pref{3.38a} have a rather complicated expression, because of the
presence of the operators $\RR$ acting on the tree vertices, which
are not endpoints. In order to clarify their structure, we have to
further expand each term in the r.h.s. of \pref{3.38a}, by a
procedure similar to that explained in \cite{BM}.

We start this analysis by supposing that $|P_{v_0}|>0$ (otherwise
there is no $\RR$ operation acting on $v_0$) and by considering
the action of $\RR$ on a single contribution to the sum in the
r.h.s. of \pref{3.38a}. This action is {\it trivial}, that is
$\RR=I$, by definition, if $|P_{v_0}|>4$ or, since $\RR^2=\RR$, if
$v_0$ is a trivial vertex ($s_{v_0}=1$) and $|P_{v_0}|$ is equal
to $|P_{\bar v}|$, $\bar v$ being the vertex (of scale $h+2$)
immediately following $v_0$. Hence there is nothing to discuss in
these cases.

Let us consider first the case $|P_{v_0}|=4$ and note that, by the
remark following \pref{3.37a}, $m(f)=0$ for all $f\in P_{v_0}$. If
$P_{v_0}=(f_1,f_2,f_3,f_4)$, with $\e(f_1) = \e(f_2) =+ = -\e(f_3)
= -\e(f_4)$, and we put $\xx(f_i)=\xx_i$,
$\tilde\xx_i=(x_{1,0},\vxx_i)$, $\o(f_i)=\o_i$,
$\vpp_{F,i}=\vpp_F(\th_{h,\o_i})$, we can write, by using
\pref{3.17},
\bea &&\RR \VV^{(h)}(\t,\bP,\O,T) = \int d\xxx \; e^{ \sum_{i=1}^4
\e_i \vpp_{F,i} \vxx_i} W_4(\xxx)\;\cdot\nn\\
&&\cdot\; \left\{ (x_{2,0}- x_{1,0}) \psi^{(\le h)+}_{\xx_1,\o_1}
[\hat\dpr^1(x_{1,0}) \psi^{(\le h)+}_{\xx_2,\o_2}] \psi^{(\le
h)-}_{\xx_3,\o_3}\psi^{(\le h)-}_{\xx_4,\o_4}+ \right.\nn\\
&&+ (x_{3,0}- x_{1,0}) \psi^{(\le h)+}_{\xx_1,\o_1} \psi^{(\le
h)+}_{\tilde\xx_2,\o_2} [\hat\dpr^1(x_{1,0}) \psi^{(\le
h)-}_{\xx_3,\o_3}] \psi^{(\le h)-}_{\xx_4,\o_4}+\nn\\
&&+ \left. (x_{4,0}- x_{1,0}) \psi^{(\le h)+}_{\xx_1,\o_1}
\psi^{(\le h)+}_{\tilde\xx_2,\o_2} \psi^{(\le
h)-}_{\tilde\xx_3,\o_3} [\hat\dpr^1(x_{1,0}) \psi^{(\le
h)-}_{\xx_4,\o_4}] \right\}\;,\lb{3.38ab}\eea
where $W_4(\xxx)$ is the integral of
$W_{\t,\bP,\O,T}^{(h)}(\xx_{v_0})$ over the variables $\xx_{v_0}
\bs \xxx$, up to a sign, and $\hat\dpr^1(x_0)$ is an operator
defined by
\be \hat\dpr^1(x_0) F(\yy) = \int_0^1 ds
\dpr_0 F(\xi_0(s),\vyy)\virg \xi_0(s)=x_0+
s(y_0-x_0)\;.\lb{3.38n}\ee
Similar expressions are obtained, if the localization point (see
comment after \pref{3.9b}) is changed.

Let us now consider the case $|P_{v_0}|=2$. If only one of the
external fields of $v_0$ carries a $\dpr_0$ derivative, the action
of $\RR$ would not be trivial. However, we can limit this
possibility to the contribution corresponding to the tree with
$n=1$, whose only endpoint is of type $z$, which gives no
contribution to $\RR\VV^{(h)}$. In fact, if there is more than one
endpoint, at most one of the fields of any endpoint of type $z$
can belong to $P_{v_0}$, so that we can use the freedom in the
choice of the field which carries the derivative so that $m(f)=0$
for both $f\in P_{v_0}$ (see remark after \pref{3.37a}).

Hence, we have to discuss only the case $m(f)=0$ for both $f\in
P_{v_0}$; if we put $P_{v_0}=(f_1,f_2)$, $\xx(f_i)=\xx_i$,
$\o(f_i)=\o_i$, $\vpp_{F,i}=\vpp_F(\th_{h,\o_i})$, we can write
\bea &&\RR \VV^{(h)}(\t,\bP,\O,T) =\lb{3.38b}\\
&&= \int d\xx d\yy e^{i(\vpp_{F,1}\vxx- \vpp_{F,2}\vyy)}
(y_0-x_0)^2 W(\xx-\yy) \psi^{(\le h)+}_{\xx,\o_1} [\hat\dpr^2(x_0)
\psi^{(\le h)-}_{\yy,\o_2}]\;,\nn\eea
where $W(\xx_1-\xx_2)$ is the integral of
$W_{\t,\bP,\O,T}^{(h)}(\xx_{v_0})$ over the variables\\
$\xx_{v_0} \bs (\xx_1,\xx_2)$, up to a sign, and $\hat\dpr^2(x_0)$
is an operator defined by
\be \hat\dpr^2(x_0) \psi^{(\le h)\e}_{\yy,\o_2} = \int_0^1 ds(1-s)
\dpr_0^2 \psi^{(\le h)\e}_{\xi_0(s),\vyy,\o}\virg \xi_0(s)=x_0+
s(y_0-x_0)\;. \lb{3.38c}\ee
Instead of \pref{3.38b}, one could also use a similar expression
with $[\hat\dpr^2(y_0)\psi^{(\le h)+}_{\xx,\o_1}]$ $\psi^{(\le
h)-}_{\yy,\o_2}$ in place of $\psi^{(\le h)+}_{\xx,\o_1}
[\hat\dpr^2(x_0) \psi^{(\le h)-}_{\yy,\o_2}]$. We shall
distinguish these two different choices by saying that we have
taken $\xx$, in the case of \pref{3.38b}, or $\yy$, in the other
case, as the {\it localization point}.

By using \pref{3.35} and \pref{3.38}, we can also write
\bea &&\RR \VV^{(h)}(\t,\bP,\O,T)=\lb{3.38e}\\
&&= {1\over s_{v_0}!} \sum_{\a\in A} \int d\xx_{v_0} \int
dP_{T_{v_0}}(\tt) \;\RR[\tilde\psi^{(\le h)}_{\O_{v_0},\a}
(P_{v_0})] \;(y_{\a,0}-x_{\a,0})^{b(|P_{v_0}|)}
\;\cdot\nn\\
&&\cdot\; \Big[ \prod_{l\in T_{v_0}} \dpr_0^{m_l} \tilde
g^{(h+1)}_{\o_l}(\xx_l-\yy_l) \d_{\o^-_l,\o^+_l} \Big] \det
G^{h+1,T_{v_0}}(\tt) \prod_{i=1}^{s_{v_0}}
[K^{(h+2)}_{v_i}(\xx_{v_i})]\;,\nn\eea
where $A$ is a set of indices containing only one element, except
in the case $|P_{v_0}|=4$, when $|A|=3$, and $\xx_\a, \yy_\a$ are
two points of $\xx_{v_0}$. Moreover, $\RR[\tilde\psi^{(\le
h)}_{\O_{v_0},\a}(P_{v_0})] = \tilde\psi^{(\le
h)}_{\O_{v_0}}(P_{v_0})$, except if $|P_{v_0}|=4$ or $|P_{v_0}|=2$
and $m(f)=0$ for both $f\in P_{v_0}$; in these case, its
expression can be easily deduced from \pref{3.38ab} and
\pref{3.38b}. Finally, $b(p)$ is an integer, equal to $1$, if
$p=4$, equal to 2, if $p=2$, and equal to $0$ otherwise.

We would like to apply iteratively equation \pref{3.38ab} and
\pref{3.38b}, starting from $v_0$ and following the partial order
of the tree $\t$, in all the $\t$ vertices with $|P_v|=4$ or
$|P_v|=2$ and $m(f)=0$ for $f\in P_v$. However, in order to
control the combinatorics, it is convenient to decompose the
factor $(y_{\a,0}-x_{\a,0})^{b(|P_{v_0}|)}$ in the following way.
Let us consider the unique subset $(l_1,\ldots,l_m)$ of $T_{v_0}$,
which selects a path joining the cluster containing $\xx_\a$ with
the cluster containing $\yy_\a$, if one identifies all the points
in the same cluster; if this subset is empty (since $\xx_\a$ and
$\yy_\a$ belong to the same cluster), we put $m=0$. If $m>0$, we
call $(\bar v_{i-1},\bar v_i)$, $i=1,m$, the couple of vertices
whose clusters of points are joined by $l_i$. We shall put
$\xx_{2i-1}$, $i=1,m$, equal to the endpoint of $l_i$ belonging to
$\xx_{\bar v_{i-1}}$, $\xx_{2i}$ equal to the endpoint of $l_i$
belonging to $\xx_{\bar v_i}$, $\xx_0=\xx_\a$ and
$\xx_{2m+1}=\yy_\a$. These definitions imply that there are two
points of the sequence $\xx_r$, $r=0,\ldots,\bar m=2m+1$, possibly
coinciding, in any set $\xx_{\bar v_i}$, $i=0,\ldots,m$; these two
points are the space-time points of two different fields belonging
to $P_{\bar v_i}$. Then, we can write
\be y_{\a,0}-x_{\a,0}= \sum_{r=1}^{\bar m} (x_{r,0}-x_{r-1,0})\;.\lb{3.38d}\ee
If we insert \pref{3.38d} in \pref{3.38e}, the r.h.s. can be written
as the sum over a set $B_{v_0}$ of different terms, that we shall
distinguish with a label $\a_{v_0}$; note that $|B_{v_0}| \le
3(2s_{v_0}-1)^2$ . We get an expression of the form
\bea &&\RR \VV^{(h)}(\t,\bP,\O,T)= {1\over s_{v_0}!}
\sum_{\a_{v_0}\in B_{v_0}} \int d\xx_{v_0} \int dP_{T_{v_0}}(\tt)
\RR[\tilde\psi^{(\le h)}_{\O_{v_0},\a} (P_{v_0})] \;\cdot\nn\\
&&\cdot\; \Big[ \prod_{l\in T_{v_0}}
(x_{l,0}-y_{l,0})^{b_l(\a_{v_0})} \dpr_0^{m_l} \tilde
g^{(h+1)}_{\o_l}(\xx_l-\yy_l) \d_{\o^-_l,\o^+_l} \Big]\;\cdot\nn\\
&&\cdot\; \det G^{h+1,T_{v_0}}(\tt) \prod_{i=1}^{s_{v_0}}
[(x^{(i)}_0- y^{(i)}_0)^{b_{v_i}(\a_{v_0})}
K^{(h+2)}_{v_i}(\xx_{v_i})]\;,\lb{3.38f}\eea
where we called $(\xx^{(i)}, \yy^{(i)})$ the couple of points
which, in the previous argument, belong to $\xx_{v_i}$ and
$b_l(\a_{v_0})$, $b_{v_i}(\a_{v_0})$ are integers with values in
$\{0,1,2\}$, such that their sum is equal to $b(|P_{v_0}|)$. \*

Let us now see what happens, if we iterate the argument leading to
\pref{3.38f}. Let us suppose, for example, that $|P_{v_1}|=2$, that
the action of $\RR$ is not trivial on $v_1$ and that
$b\=b_{v_1}(\a_{v_0})>0$. In this case, if we exploit the action
of $\RR$ in the form of \pref{3.38b}, we have an overall factor
$(x^{(1)}_0- y^{(1)}_0)^m$, $m=2+b$, which multiplies
$K^{(h+2)}_{v_1}(\xx_{v_1})$. Hence, if we expand this factor, by
using an equation similar to \pref{3.38d}, we get terms with some
propagator multiplied by a factor $(x_{l,0}-y_{l,0})^{b_l}$, with
$b_l>2$. If we further iterate this procedure, we can end up with
an expansion, where some propagator is multiplied by a factor
$(x_{l,0}-y_{l,0})^{b_l}$ with $b_l$ of order $|h|$, which would
produce bad bounds. However, we can avoid very simply this
difficulty, by noticing that, if we insert \pref{3.16} in an
expression like
\be J_b =\int d\xx d\yy F_1(\xx) F_2(\yy) (y_0-x_0)^b
\RR W(\xx-\yy)\;,\lb{3.38g}\ee
we get, by a simple integration by part, if $b=2$,
\be J_2 =\int d\xx d\yy F_1(\xx) F_2(\yy) (y_0-x_0)^2
W(\xx-\yy)\;,\lb{3.38h}\ee
that is the $\RR$ operation can be substituted by the identity,
while, if $b=1$, we get
\be J_1 =\int d\xx d\yy F_1(\xx) [\hat\dpr^1(x_0) F_2(\yy)] (y_0-x_0)^2
W(\xx-\yy)\;,\lb{3.38m}\ee
where $\hat\dpr^1(x_0)$ is the operator defined by \pref{3.38n}.
This means that, if $b=1$, the action of $\RR$ only increases the
power of $(y_0-x_0)$ by one unit. Note that, in \pref{3.38m} one
could substitute $F_1(\xx) [\hat\dpr^1(x_0) F_2(\yy)]$ with
$-[\hat\dpr^1(y_0) F_1(\xx)] F_2(\yy)$; we shall again distinguish
these two different choices by saying that we have taken $\xx$, in
the case of \pref{3.38m}, or $\yy$, in the other case, as the
localization point.

Even simpler is the situation, when $|P_{v_1}|=4$. In fact, if we
insert \pref{3.17} in an expression like $\int d\xxx F(\xxx)
(y^*_0-x^*_0) \RR W_4(\xxx)$, $y^*$ and $x^*$ being two points of
$\xxx$, we get
\be \int d\xxx F(\xxx) (y^*_0-x^*_0) \RR W_4(\xxx)
= \int d\xxx F(\xxx) (y^*_0-x^*_0) W_4(\xxx)\;,\lb{3.38nn}\ee
so that, even in this case, the power of the ``zero'' can not
increase.

There are in principle two other problems. First of all, one could
worry that there is an accumulation of the operators $\hat\dpr^q$
(dimensionally equivalent to a derivative of order $q$) on a same
line, if this line is affected many times by the $\RR$ operation
in different vertices. Moreover, since the definition of the
$\hat\dpr^q(x_0)$ operators depends on the choice of the
localization point $\xx$, it could happen that there is an
``interference'' between the $\RR$ operations in two different
vertices, which would make more involved the expansion. However,
one can show, by the same arguments given in \S3.3 and \S3.4 of
\cite{BM} in the one dimensional case, that these problems can be
avoided by using the freedom in the choice of the localization
point and, mainly, the fact that some regularization operations
are not really present. Let us consider, for example, the first
term in the r.h.s. of \pref{3.38ab} and note that, if we sum it
over the sector indices, we get, in terms of Fourier transforms,
an expression of the type
\bea &&\int d\kkk \prod_{i=1}^4 \hat\psi_{\kk_i}^{(\le
h),\e_i}\d(\kk_1+\kk_2-\kk_3-\kk_4)\cdot\nn\\
&&\cdot \left[ \hat W_4(\kk_1,\kk_2,\kk_3)- \hat
W_4(\kk_1,(0,\vkk_2),\kk_3)\right] \;.\lb{3.39a}\eea
However, if $\bar f$ is the label of the field $\psi^{(\le
h),+}_{\xx_2}$, it is easy to see that $\hat W_4(\kk_1,$
$(0,\vkk_2),\kk_3)=0$, if there is a vertex $\bar v>v_0$ with four
external legs, such that $f\in P_{\bar v}$ and $f$ is affected by
the $\RR$ operation in $\bar v$. Hence, in this case, we can
substitute the first term in the braces of \pref{3.38ab} with
$\prod_i \psi_{\xx_i,\o_i}^{(\le h),\e_i}$.

We refer to \S3.3 and \S3.4 of \cite{BM} for a complete analysis
of this problem, whose final result is that the action of $\RR$ on
all the vertices of $\t$ will produce terms where the propagators
related with the lines of $T$ are multiplied by a factor
$(x_{l,0}-y_{l,0})^{b_l}$ with $b_l\le 2$ and (after that) are
possibly subject to one or two operators $\hat\dpr^q$, $q=1,2$.
Moreover, some of the external lines belonging to $P_{v_0}$ can be
affected from one operator $\hat\dpr^q$, as a consequence of the
action of $\RR$ on $v_0$ or some other vertex $v>v_0$. Finally,
the lines involved in the determinants may be affected from one
operator $\hat\dpr^q$. We introduce an index $\a$ to distinguish
these different terms and, given $\a$, we shall denote by
$\hat\dpr^{q_\a(f)}$ the differential operators acting on the
external lines of $P_{v_0}$ or the propagators belonging to $T$,
as a consequence of the regularization procedure.

All the previous considerations imply that $\RR
\VV^{(h)}(\t,\bP,\O,T)=0$, if\\
$|P_{v_0}|=4$ and $n=1$ (that is there is only an endpoint of type
$\l$ and no internal line associated with $v_0$) or
$P_{v_0}=(f_1,f_2)$ and $m(f_1)+m(f_2)=1$ (since this can happen
only if $n=1$ and the endpoint is of type $z$, as a consequence of
the freedom in the choice of the field carrying the derivative in
the endpoints of type $z$) or $m(f_1)+m(f_2)=0$ and $n=1$. In all
the other cases, we can write $\RR \VV^{(h)}(\t,\bP,\O,T)$ in the
form
\be \RR\VV^{(h)}(\t,\bP,\O,T) = \sum_{\a\in A_T}
\int d\xx_{v_0} W_{\t,\bP,\O\bs \O_{v_0},T,\a}(\xx_{v_0})
\RR[\tilde\psi^{(\le h)}_{\O_{v_0},\a} (P_{v_0})]\;,\lb{3.39}\ee
where
\be \RR[\tilde\psi^{(\le h)}_{\O_{v_0},\a}(P_{v_0})] =
\prod_{f\in P_{v_0}} e^{i\e(f)\vpp_F(\th_{h,\o(f)})\vxx(f)}
[\hat\dpr^{q_\a(f)}\psi]^{(\le h)\e(f)}_{\xx_\a(f),
\o(f)},\lb{3.39b}\ee
and, up to a sign,
\bea && W_{\t,\bP,\O\bs \O_{v_0}, T,\a}(\xx_{v_0})=\nn\\
&&=\left[\prod_{i=1}^n K_{v_i^*}^{h_i} (\xx_{v_i^*})\right]
\Bigg\{\prod_{v\,\atop\hbox{\ottorm not e.p.}}{1\over s_v!} \int
dP_{T_v}(\tt_v)\det G_\a^{h_v,T_v}(\tt_v)\cdot\lb{3.40}\\
&&\cdot\Big[\prod_{l\in T_v}\d_{\o_l^+,\o_l^-}
\hat\partial^{q_\a(f^-_l)}(x'_{l,0})
\hat\partial^{q_\a(f^+_l)}(y'_{l,0}) [(x_{l,0}- y_{l,0})^{b_\a(l)}
\dpr_{0}^{m_l} \tilde
g^{(h_v)}_{\o_l}(\xx_l-\yy_l)]\Big]\Bigg\}\;,\nn\eea
where ``e.p.'' is an abbreviation of ``endpoint'' and, together
with the definitions used before, we are using the following ones:

\begin{enumerate}

\item $A_T$ is a set of indices which allows to distinguish the
different terms produced by the non trivial $\RR$ operations and
the iterative decomposition of the zeros;

\item $v^*_1,\ldots,v^*_n$ are the endpoints of $\t$ and
$h_i=h_{v_i^*}$;

\item $b_\a(v)$, $b_\a(l)$, $q_\a(f^-_l)$ and $q_\a(f^+_l)$ are
positive integers $\le 2$;

\item if $q_\a(f^-_l)>0$, $x'_{l,0}$ denote the time coordinate of
the point involved, together with $\xx_l$, in the corresponding
$\RR$ operation, see \pref{3.38c} and \pref{3.38n}, otherwise
$\hat\dpr^0(x'_{l,0})=I$;

\item if $v$ is a non trivial vertex (so that $s_v>1$),the
elements $G^{h_v,T_v}_{\a, ij,i'j'}$ of $G_\a^{h_v,T_v}(\tt_v)$
are of the form
\bea &&G^{h_v,T_v}_{\a, ij,i'j'}= t_{i,i'} \cdot\lb{3.40a}\\
&& \cdot \hat\partial_{0}^{q_\a(f^-_{ij})}({x'}_{l,0})
\hat\partial_{0}^{q_\a(f^+_{i'j'})}({y'}_{l,0})\dpr_0^{m(f_l^-)}
\dpr_0^{m(f_l^-)} \tilde
g^{(h_v)}_{\o_l}(\xx_{ij}-\yy_{i'j'})\d_{\o_l^-,\o_l^+}\;;\nn\eea
if $v$ is trivial, $T_v$ is empty and $\int dP_{T_v}(\tt_v)\det
G_\a^{h_v,T_v}(\tt_v)$ has to be interpreted as $1$, if $|{\cal
I}_v|=0$ (${\cal I}_v$ is the set of internal fields of $v$),
otherwise it is the determinant of a matrix of the form
\pref{3.40a} with $t_{i,i'}=1$.

\end{enumerate}

\subsection{Modification of the running coupling functions}\lb{ss3.5}

We want now to introduce a different representation of the running
coupling functions $\l_h, \n_h, z_h$, which will be useful in the
following, in order to perform the bounds. This new representation
is suggested by the remark that, if we substitute \pref{3.40} in
\pref{3.39} and we express the whole integral in Fourier space, the
Fourier transform of $K_{v_i^*}^{h_i}(\xx_{v_i^*})$ is multiplied
by the factor
\be \prod_{f\in P_{v_i^*}\cap P_{v_0}} \hat\psi_{\kk(f),\o(f)}^{\le h_{v_0},\e(f)}
\prod_{f\in P_{v_i^*}\setminus
P_{v_0}}F_{h(f),\o(f)}(\kk(f))\;.\lb{3.45}\ee

In order to use this property, we define, for any $h\le 0$ and
$\o\in O_h$, the {\it s-sector} $S_{h,\o}$ (see \S\ref{ss3.1} and
\S\ref{ss3.2a} for related definitions) as
\be S_{h,\o} = \{\vkk=\r\vee_r(\th) \in\RRR^2: |\e(\vkk)-\m|\le \g^h e_0,\;
\z_{h,\o}(\th)\not=0\}\;.\lb{3.44a}\ee
Note that the  definition of s-sector has the property, to be used
extensively in the following, that the s-sector $S_{h+1,\o}$ of
scale $h+1$ contains the union of two s-sectors of scale $h$:
$S_{h+1,\o}\supseteq \left\{ S_{h,2\o}\cup S_{h,2\o+1}\right\}$,
as follows from the definition of $\z_{h,\o}$, see \pref{3.14a}.

We now observe that the field variables
$\hat\psi_{\kk(f),\o(f)}^{\le h_{v_0},\e(f)}$ have the same
supports as the functions $C^{-1}_{h_{v_0}}(\kk(f))$
$\z_{h_{v_0},\o(f)}(\th(f))$ and $h(f)\le h_i-1$, $\forall f\in
P_{v_i^*}$; hence in the expression \pref{3.40}, for any $i$, we
can freely multiply $\hat K_{v_i^*}^{h_i}(\kk_{v_i^*})$ by
$\prod_{f\in P_{v^*_i}} \tilde F_{h_i-1,\tilde\o(f)}(\vkk)$, where
$\tilde F_{h,\o}(\vkk)$ is a smooth function $=1$ on $S_{h,\o}$
and with a support slightly greater than $S_{h,\o}$, while
$\tilde\o(f)\in O_{h_i-1}$ is the unique sector index such that
$S_{h(f),\o(f)} \subseteq S_{h_i-1,\tilde\o(f)}$. In order to
formalize this statement, it is useful to introduce the following
definition.

Let $G(\vxxx)$ be a function of $2p$ variables $\vxxx=(\vxx_1,
\ldots, \vxx_{2p})$ with Fourier transform $\hat G(\vkkk)$,
defined so that $G(\vxxx)=\int d\vkkk (2\p)^{-4p}
\exp(-i\sum_{l=1}^{2p} \e_i\vkk_i\vxx_i) \hat G(\vkkk)$, where
$\e_1, \ldots, \e_p=-\e_{p+1}= \ldots = -\e_{2p}=+1$. Then, we
define, given $h\le 0$ and a family $\ss=\{\s_i\in O_h,
i=1,\ldots, 2p \}$ of sector indices,
\be (\FFF_{2p,h,\ss} * G)(\vxxx) = \int {d\vkkk\over(2\p)^{4p}}
e^{-i\sum_{l=1}^{2p} \e_i\vkk_i\vxx_i} \left[ \prod_{i=1}^{2p}
\tilde F_{h,\s_i}(\vkk_i) \right] \hat G(\vkkk)\;.\lb{3.45aa}\ee
In order to extend this definition to the case $h=1$, when the
sector index can take only the value $0$, we define $\tilde
F_{1,0}(\vkk)$ as a smooth function of compact support, equal to
$1$ on the support of $\bar C_0^{-1}(\vkk)$, defined in \S\ref{ss1.2}.

Hence, if we put $p_i=|P_{v_i^*}|$, $\tilde\O_i = \{\tilde\o(f),
f\in P_{v_i^*}\}$ and we define, for any family $\ss=\{\s(f)\in
O_{h_i-1}, f\in P_{v_i^*}\}$ of sector indices of scale $h_i-1$,
labelled by the set $P_{v_i^*}$ ($\tilde\O_i$ is a particular
example of such a family),
\be \tilde K_{v_i^*, \ss}^{h_i}(\xx_{v_i^*}) =
\left( \FFF_{p_i,h_i-1,\ss} *
K_{v_i^*}^{h_i} \right)(\xx_{v_i^*})\;,\lb{3.46}\ee
we can substitute in \pref{3.40} each
$K_{v_i^*}^{h_i}(\xx_{v_i^*})$ with $\tilde K^{h_i}_{v_i^*,
\tilde\O_i}(\xx_{v_i^*})$. If $v_i^*$ is of type $\n$, $z$ or
$\l$, $\tilde K_{v_i^*, \ss}^{h_i}(\xx_{v_i^*})$ can be written as
$\g^{h_i-1} \d(x_{0,v_i^*}) \tilde\n_{h_i-1, \ss}(\vxx_{v_i^*})$,
$\d(x_{0,v_i^*}) \tilde z_{h_i-1,\ss}(\vxx_{v_i^*})$ or
$\d(x_{0,v_i^*})$ $\tilde\l_{h_i-1,\ss} (\vxx_{v_i^*})$
respectively. $\tilde\n_{h_i-1,\ss}(\vxx_i-\vyy_i)$, $\tilde
z_{h_i-1,\ss}(\vxx_i-\vyy_i)$ and $\tilde\l_{h_i-1,\ss}
(\vxx_{v_i^*})$ will be called the {\it modified coupling
functions}.

We shall call $W^{(mod)}_{\t,\bP,\O,T,\a}(\xx_{v_0})$ the
expression we get from $W_{\t,\bP,\O\bs \O_{v_0},T,\a}(\xx_{v_0})$
by the substitution of the running coupling functions with the
modified ones. Note that $W^{(mod)}_{\t,\bP,\O,T,\a}(\xx_{v_0})$
is not independent of $\O_{v_0}$, unlike $W_{\t,\bP,\O\bs
\O_{v_0},T,\a}(\xx_{v_0})$, and that
$W^{(mod)}_{\t,\bP,\O,T,\a}(\xx_{v_0})$ is equal to
$W_{\t,\bP,\O\bs \O_{v_0},T,\a}(\xx_{v_0})$, only if
$|P_{v_0}|=0$; however, the previous considerations imply that, if
$p_0=|P_{v_0}|>0$,
\be \left( \FFF_{p_0,h,\O_{v_0}} *
W^{(mod)}_{\t,\bP,\O,T,\a} \right)(\xx_{v_0})= \left(
\FFF_{p_0,h,\O_{v_0}}
* W_{\t,\bP,\O\bs \O_{v_0},T,\a} \right)(\xx_{v_0})\;,\lb{3.46a}\ee
a trivial remark which will be important in the discussion of the
running coupling functions flow in \S\ref{ss5}.

\subsection{Bounds for the effective potentials and the
free energy}\lb{ss3.5a}

Given a vertex $v$ of a tree $\t$ and an arbitrary family
$\bar\SS=\{S_{j_f,\s_f}, f\in P_v\}$ of s-sectors labelled by
$P_v$, we define
\be \c_v(\bar\SS) = \c\left( \forall f\in P_v, \exists \vkk(f)\in S_{j_f,\s_f}:
\sum_{f\in P_v}\e(f)\vkk(f)=0\right)\;,\lb{3.44}\ee
where $\chi(condition)$ is the function $=1$ when $condition$ is
verified, and $=0$ in the opposite case. Moreover, given a set $P$
of field labels, we denote by $\SS(P)$ the special family of
s-sectors labelled by $P$, defined as
\be \SS(P)=\{ S_{h(f),\o(f)}\ ,\ f\in P \}\;.\lb{3.45a}\ee

The previous considerations imply that
\be E_{L,\b} \le \sum_{h=h_\b-1}^0 \sum_{n=1}^\io  J_{h,n}(0,0)\;,
\lb{3.47}\ee
with
\bea J_{h,n}(2 l_0,q_0)&=& \sum_{\t\in\TT_{h,n}}
\sum_{\bP\in\PP_\t: |P_{v_0}|=2 l_0, \atop \sum_{f\in P_{v_0}}
q_\a(f)=q_0} \sum_{T\in {\bf T}} \sum_{\a\in A_T}
\sum_{\O\in\OO_\t}^* \left[\prod_{v}\c_v(\SS(P_v)) \right]
\;\cdot\nn\\
&\cdot&\;\int d(\xx_{v_0}\bs \xx^*) \left|
W^{(mod)}_{\t,\bP,\O,T,\a}(\xx_{v_0}) \right|\;,\lb{3.48}\eea
where $\xx^*$ is an arbitrary point in $\xx_{v_0}$, $l_0$ is a non
negative integer and $\sum_{\O\in\OO_\t}^*$ differs from
$\sum_{\O\in\OO_\t}$ since one $\o$ index, arbitrarily chosen
among the $2 l_0$ $\o$'s in $\O_{v_0}$, is not summed over, if
$l_0>0$, otherwise it coincides with $\sum_{\O\in\OO_\t}$.

\* \0 {\bf Remarks - } Note that we could freely insert
$[\prod_{v}\c_v(\SS(P_v))]$ in \pref{3.48}, because of the
constraints following from momentum conservation and the compact
support properties of propagator's Fourier transform.

\0 Note also that, if $l_0=0$, given $\t\in \TT_{h,n}$, the number
of internal lines in the lowest vertex $v_0$ (of scale $h+1$) has
to be different from zero.

\*

Hence, in order to prove that the free energy and the effective
potentials are well defined (in the limit $L \to\io$ and $\b$ not
``too large''), we need a ``good'' bound of $J_{h,n}(2 l_0, q_0)$.

In order to get this bound, we shall extend the procedure used in
\cite{BM} for the analysis of the one dimensional Fermi systems,
which we shall refer to for some details (except for the sum over
the sector indices, which is a new problem).

An important role has the following bound for the determinants
appearing in \pref{3.40}:
\bea &&|\det G_\a^{h_v,T_v}(\tt_v)| \le
c^{\sum_{i=1}^{s_v}|P_{v_i}|-|P_v|-2(s_v-1)}\;\cdot\nn\\
&&\cdot\; \g^{{h_v}{3\over
4}\left(\sum_{i=1}^{s_v}|P_{v_i}|-|P_v|-2(s_v-1)\right)} \g^{h_v
\sum_{i=1}^{s_v}\left[ q_\a(P_{v_i}\bs Q_{v_i})+m(P_{v_i}\bs
Q_{v_i}) \right]}\;\cdot\nn\\
&&\cdot\; \g^{-h_v \sum_{l\in T_v}\left[q_\a(f^+_l)+q_\a(f^-_l)+
m(f^+_l)+m(f^-_l) \right]}\;,\lb{3.49}\eea
where, if $P\subset I_{v_0}$, we define $q_\a(P)=\sum_{f\in
P}q_\a(f)$ and $m(P)=\sum_{f\in P}m(f)$.

The proof of \pref{3.49} is based on the well known {\it
Gram-Hadamard inequality}, stating that, if $M$ is a square matrix
with elements $M_{ij}$ of the form $M_{ij}=<A_i,B_j>$, where
$A_i$, $B_j$ are vectors in a Hilbert space $\HH$ with scalar
product $<\cdot,\cdot>$ and induced norm $||\cdot||$, then
\be |\det M|\le \prod_i ||A_i||\cdot ||B_i||\;.\lb{3.50}\ee

Let $\HH=\RRR^{|O_h|}\otimes \RRR^s\otimes L^2(\RRR^3)$; it can be
shown that
\be G^{h_v,T_v}_{\a, ij,i'j'}=<\vv_{\o_l^-} \otimes \uu_i \otimes
A^{(h_v)}_{\xx(f^-_{ij}),\o_l^-}, \vv_{\o_l^+} \otimes
\uu_{i'}\otimes B^{(h_v)}_{\xx(f^+_{i'j'}),\o_l^+}>\;,\lb{3.50a}\ee
where $\vv_\o\in \RRR^{|O_h|}$, $\o\in O_h$, and $\uu_i\in
\RRR^s$, $i=1,\ldots,s$, are unit vectors such that
$\vv_\o\cdot\vv_{\o'} = \d_{\o,\o'}$, $\uu_i\cdot\uu_{i'} =
t_{i,i'}$; moreover, $A^{(h_v)}_{\xx(f^-_{ij}),\o_l}$,
$B^{(h_v)}_{\xx(f^+_{i'j'}),\o_l}$ are defined so that:
\bea &&\hat\partial_{0}^{q_\a(f^-_{ij})}({x'}_{l,0})
\hat\partial_{0}^{q_\a(f^+_{i'j'})}({y'}_{l,0})\dpr_0^{m(f_l^-
)}\dpr_0^{m(f_l^+)} \tilde
g^{(h_v)}_{\o_l}(\xx_{ij}-\yy_{i'j'})=\lb{3.51}\\
&&=<A^{(h_v)}_{\xx(f^-_{ij}),\o_l},
B^{(h_v)}_{\xx(f^+_{i'j'}),\o_l}>\= \int{d \kk\over (2\p)^3}
A^{*(h_v)}_{\xx(f^-_{ij}),\o_l}(\kk)
B^{(h_v)}_{\xx(f^+_{i'j'}),\o_l}(\kk)\;,\nn\eea
with $||A_i||\cdot ||B_i||$ satisfying the same dimensional bound
as the left side of \pref{3.51}. For example, if
$q_\a(f^-_{ij})=q_\a(f^+_{i'j'})=0$, one can put,
\bea && A^{(h_v)}_{\xx,\o_l}(\kk)= e^{i\kk\xx}
{\sqrt{F_{h_v,\o_l}} \over k_0^2+\left(\e(\vec k)-\m\right)^2}
(ik_0)^{m(f_l^-)}(ik_0)^{m(f_l^+)}\nn\\
&& B^{(h_v)}_{\xx,\o_l}(\kk)= e^{i\kk\xx} \sqrt{F_{h_v,\o_l}}
\left[ ik_0+\e(\vec k)-\m\right]\;.\lb{3.52}\eea
Using Lemma \ref{lm3.1} and \pref{3.50}, we easily get \pref{3.49}.

The next step is to bound by $1$ the integrals over the
probability measures $dP_{T_v}$ appearing in \pref{3.40}. After
that, we bound the integral
\bea &&\int d(\xx_{v_0}\bs \xx^*) \Biggl| \prod_{i=1}^n \left[
\tilde K_{v_i^*,\tilde\O_i}^{h_i} (\xx_{v_i^*})\right]
\prod_{v\,\hbox{\ottorm not e.p.}} {1\over s_v!} \;\cdot\lb{3.53}\\
&& \cdot\;\prod_{l\in T_v}\left\{
\hat\partial^{q_\a(f^-_l)}_{0}({x'}_{l,0})
\hat\partial^{q_\a(f^+_l)}_{0}({y'}_{l,0})
[(x_{l,0}-y_{l,0})^{b_\a(l)}\dpr_0^{m_l} \tilde
g^{(h_v)}_{\o_l}(\xx_l-\yy_l)]\right\}\Biggr|\;.\nn\eea
We can take from \S3.15 of \cite{BM} the identity (independent of
the dimension):
\be d  (\xx_{v_0}\bs \xx^*)=\prod_{l\in T^*} d
\rr_l\;,\lb{3.54}\ee
where $T^*$ is a tree graph obtained from $T=\cup_vT_v$, by adding
in a suitable (obvious) way, for each endpoint $v_i^*$,
$i=1,\ldots,n$, one or more lines connecting the space-time points
belonging to $\xx_{v_i^*}$. Moreover
$\rr_l=\left(\x_0(t_l)-\h_0(s_l), \vec x_l-\vec y_l\right)$ (see
\pref{3.38n}), if $l\in\cup_vT_v$, and $\rr_l=\xx_l-\yy_l$, if
$l\in T^*\setminus\cup_vT_v$.

Hence \pref{3.53} can be written as
\be J_{\t,\PP,T,\a} \int \prod_{l\in T^*\setminus\cup_vT_v} d\rr_l
\left| \prod_{i=1}^n \tilde K_{v_i^*,\tilde\O_i}^{h_i}
(\xx_{v_i^*}) \right| \;,\lb{3.55}\ee
with
\bea &&J_{\t,\PP,T,\a} = \prod_{v\,\hbox{\ottorm not e.p.}}
{1\over s_v!}\int \prod_{l\in T_v} d\rr_l \cdot\lb{3.56}\\
&&\cdot\left| \hat\partial^{q_\a(f^-_l)}_{0}({x'}_{l,0})
\hat\partial^{q_\a(f^+_l)}_{0}({y'}_{l,0})
[(x_{l,0}-y_{l,0})^{b_\a(l)}\dpr_0^{m_l} \tilde
g^{(h_v)}_{\o_l}(\xx_l-\yy_l)]\right|\nn\;.\eea
By using Lemma \ref{lm3.1}, we can bound each propagator, each
derivative and each zero by a dimensional factor, so finding
\bea &&J_{\t,\bP,T,\a}\le c^n \prod_{v\,\hbox{\ottorm not e.p.}}
\Big[{1\over s_v!} c^{2(s_v-1)} \g^{-h_v\sum_{l\in
T_v}b_\a(l)}\cdot\nn\\
&&\cdot\g^{-h_v(s_v-1)}\g^{h_v\sum_{l\in
T_v}\left[q_\a(f^+_l)+q_\a(f^-_l)+m(f^+_l)+m(f^-_l)
\right]}\Big]\;.\lb{3.57}\eea

Let us now define, for any set of field indices $P$,
$\bO_h(P)=\otimes_{f\in P} O_h$. The next step is to use the
following lemma, to be proved in \S\ref{ss4}.

\*

\begin{lemma}\lb{lm3.2}
Suppose that there exist two constants $C_1$ and
$C_\n$ such that the modified coupling functions satisfy the
following conditions:

\0 i) if $|P_v|=4$, then
\be \sum_{\ss\in \bO_{h_v-1}}^*
\int d(\vxx_v\bs\vxx^*) | \tilde\l_{h_v-1,\ss}(\vxx_v)| \le 2 C_1
|\l| \g^{-{1\over 2}(h_v-1)} \;,\lb{3.58}\ee
where $\sum^*$ means that one of the sector indices is not summed
over;

\0 ii) if $|P_v|=2$ and $\xx_v=(\xx_1,\xx_2)$, then
\be \sum_{\ss\in \bO_{h_v-1}}^* \int d\vxx_1
|\tilde\nu_{h_v-1,\ss}(\vxx_1-\vxx_2)|\le 2 C_1C_\n
|\l|\;,\lb{3.59}\ee
\be \sum_{\ss\in \bO_{h_v-1}}^* \int
d\vxx_1 |\tilde z_{h_v-1,\ss}(\vxx_1-\vxx_2)|\le
C_1|\l|\;.\lb{3.60}\ee

Consider a tree $\t\in \TT_{h,n}$, a graph $T\in {\bf T}$ and the
corresponding tree graph $T^*$, defined as after \pref{3.54}. Then
\bea &&\sum^*_{\O\in\OO_\t} \left[ \prod_{v\in\t}
\left(\c_v(\SS(P_v))\prod_{l\in T_v} \d_{\o_l^+,\o_l^-}\right)
\right] \int \prod_{l\in T^*\setminus T} d\rr_l
\left|\prod_{i=1}^n \tilde
K_{v_i^*,\tilde\O_i}^{h_i}(\xx_{v_i^*})\right| \;\le\nn\\
&&\le \quad c^n|\l|^n \g^{-{1\over 2}h
[m_4(v_0)+\c(P_{v_0}=\emptyset)]} \prod_{i=1}^n \g^{(h_i-1) \c(v\;
{\rm{is\; of\; type}}\;\n)}\;\cdot\nn\\
&&\qquad \cdot\prod_{v {\rm \ not\ e.\ p.}}\g^{\left[ -{1\over 2}
m_4(v)+{1\over 2}(|P_v|-3)\c (4\leq |P_v|\leq 8)+ {1\over
2}(|P_v|-1)\c(|P_v|\geq 10)\right]}\;,\lb{3.62}\eea
where $m_4(v)$ denotes the number of endpoints of type $\l$
following the vertex $v$.
\end{lemma}

Since $\sum_{i=1}^{s_v}|P_{v_i}|-|P_v|-2(s_v-1) \le 4n$, $\sum_v
(s_v-1)=n-1$ and $|A_T|\le c^n$, \pref{3.49}, \pref{3.56} and Lemma
\ref{lm3.2} imply that
\bea &&|J_{h,n}(2l_0,q_0)| \le (c|\l|)^n \;\cdot\lb{3.62a}\\
&& \cdot \sum_{\t\in\TT_{h,n}} \sum_{\bP\in\PP_\t: |P_{v_0}|=2
l_0, \atop \sum_{f\in P_{v_0}} q_\a(f)=q_0} \sum_{T\in {\bf T}}
\g^{-{1\over 2}h (m_4(v_0)+\c(l_0=0))} \prod_{i=1}^n \g^{(h_i-1)
\c(v\; {\rm{is\; of\; type}}\;\n)}\;\cdot\nn\\
&& \cdot\; \prod_{v {\rm \ not\ e.\ p.}} \Big[ {1\over s_v!}
\g^{{h_v}{3\over 4} \left( \sum_{i=1}^{s_v}
|P_{v_i}|-|P_v|-2(s_v-1)\right)} \g^{h_v \sum_{i=1}^{s_v}\left[
q_\a(P_{v_i}\bs Q_{v_i})+m(P_{v_i}\bs Q_{v_i})
\right]}\;\cdot\nn\\
&&\cdot\; \g^{-h_v \sum_{l\in T_v} b_\a(l)} \g^{-h_v(s_v-1)}
\g^{\left[ -{1\over 2} m_4(v)+{1\over 2}(|P_v|-3)\c (4\leq
|P_v|\leq 8)+ {1\over 2}(|P_v|-1)\c(|P_v|\geq 10)\right]} \Big]
\;.\nn\eea
Note now that the constraints on the values of $q_\a(f)$ and
$b_\a(l)$ imply, as shown in detail in \S3.11 of \cite{BM}, that
\be \sum_{v {\rm \ not\ e.\ p.}} h_v \sum_{i=1}^{s_v} q_\a(P_{v_i}\bs
Q_{v_i}) +h\ q_0 = \sum_{f\in I_{v_0}} h(f) q_\a(f)\;,\lb{3.62b}\ee
\be \Big[\prod_{f\in I_{v_0}} \g^{h(f) q_\a(f)} \Big]
\Big[\prod_{l\in T} \g^{-h_\a(l)b_\a(l)} \Big]\le \prod_{v\;{\rm
not\; e.p.}\atop} \g^{-z(v)} \lb{3.63}\ee
where
\be z(v)=\cases{2 &\  if $|P(v)|=2\;,$\cr
1 &\  if $|P(v)|=4\;,$\cr 0 &\  otherwise.\cr}\lb{3.64}\ee
Moreover, since the freedom in the choice of the field carrying
the derivative in the endpoints of type $z$ was used (see remark a
few lines before \pref{3.38b}) so that $m(P_v)=0$, if $v$ is not an
endpoint, and the field with $m(f)=1$ belonging to the endpoint
$v$ is contracted in the vertex immediately preceding $v$, whose
scale is $h_v-1$, we have the identity
\be \prod_{v {\rm \ not\ e.\ p.}} \g^{h_v
\sum_{i=1}^{s_v}\left[ m(P_{v_i}\bs Q_{v_i}) \right]}=
\prod_{i=1}^n \g^{ (h_i-1)  \c(v\;{\rm
is\;of\;type\;}z)}\;.\lb{3.65}\ee

Putting together the previous bounds and supposing that the
hypothesis \pref{3.58}, \pref{3.59}, \pref{3.60} of Lemma \ref{lm3.2}
are verified, we find that
\bea && J_{h,n}(2l_0, q_0) \le (c|\l|)^n \sum_{\t\in\TT_{h,n}}
\sum_{\bP\in\PP_\t: |P_{v_0}|=2 l_0, \atop \sum_{f\in P_{v_0}}
q_\a(f)=q_0} \sum_{T\in {\bf T}} \g^{-h[ {1\over 2}
m_4(v_0)+{1\over 2}\c(l_0=0) + q_0]} \cdot\nn\\
&&\cdot \left[ \prod_{i=1}^n \g^{(h_i-1) \c( |P(v)|=2) } \right]
\prod_{v {\rm \ not\ e.\ p.}} \Biggl[ {1\over s_v!}
\g^{h_v\left[{3\over 4}\left(\sum_{i=1}^{s_v}|P_{v_i}|
-|P_v|\right)-{5\over 2}(s_v-1) \right]} \cdot\nn\\
&& \cdot \g^{ \left[-z(v) -{1\over 2} m_4(v)+{1\over 2}(|P_v|-3)\c
(4\leq |P_v|\leq 8)+ {1\over 2}(|P_v|-1)\c(|P_v|\geq 10)\right] }
\Biggr]\;.\lb{3.66a}\eea
On the other hand, if $m_2(v)$ denotes the number of endpoints of
type $\n$ or $z$ following $v$, we have, if $\tilde v$ is not an
endpoint, the identities
\bea \sum_{v\ge \tilde v \atop v {\rm \ not\ e.\ p.}}
\left(\sum_{i=1}^{s_v}|P_{v_i}| -|P_v|\right) &=& 4m_4(\tilde v) +
2m_2(\tilde v) - |P_{\tilde v}|\;,\nn\\
\sum_{v\ge \tilde v} (s_v-1) &=& m_4(\tilde v) + m_2(\tilde
v)-1\;,\lb{3.66b}\eea
which, together with \pref{3.66a} imply that
\bea &&J_{h,n}(2l_0,q_0) \le \lb{3.66}\\
&& \le (c|\l|)^n \g^{h[ -q_0 +\d_{\rm ext}(2l_0)]}
\sum_{\t\in\TT_{h,n}} \sum_{\bP\atop |P_{v_0}|= 2l_0} \sum_{T\in
{\bf T}} \prod_{v\atop {\rm not\ e.\ p.}}{1\over s_v!}
\g^{\d(|P_v|)}\;,\nn\eea
where
\bea &&\d(p) = -\c(2\le p\le 4) +\nn\\
&& +\left(1 -{p\over 4} \right) \c (6\leq p\leq 8)
 + \left( 2 -{p\over 4}\right) \c (p\ge 10)\;,\lb{3.67}\eea
\be \d_{\rm ext}(p) = {5\over 2} - {3\over 4}p - {1\over 2}\c(p=0)\;.\lb{3.67a}\ee

Since $\d(|P_v|)<0$, for any vertex $v$, which is not an endpoint,
a standard argument, see \cite{BM} or \cite{GM}, allows to show
that
\be \sum_{\t\in\TT_{h,n}} \sum_{\bP\atop |P_{v_0}|=
2l_0} \sum_{T\in {\bf T}} \prod_{v\atop {\rm not\ e.\ p.}}{1\over
s_v!} \g^{\d(|P_v|)} \le c^n\;.\lb{3.68}\ee
The bounds \pref{3.66} and \pref{3.68} imply the following theorem.

\*

\begin{theorem}\lb{th3.1}
If conditions \pref{3.58}, \pref{3.59},
\pref{3.60} are satisfied, then
\be J_{h,n}(2l_0,q_0) \le (c|\l|)^n \g^{h[ -q_0 +\d_{\rm
ext}(2l_0)]}\;. \lb{3.69}\ee
\end{theorem}

{\bf Remark - } We will prove in \S\ref{ss5} that, if $|\l|$ is
small enough and $C_{1,2}$ $\log\b|\l|\le 1$, where $C_{1,2}$ is a
constant depending only on first and second order contributions of
perturbation theory, it is possible to choose $\tilde\n_1(\vxxx)$
so that the modified running coupling functions satisfy the
hypotheses of Lemma \ref{lm3.2}, \pref{3.58}, \pref{3.59} and
\pref{3.60}. So, in that case, we see from Theorem \ref{th3.1}
that $\lim_{L\to\io}E_{L, \b}$ does exist and is of order $\l$.

\section{Proof of Lemma \ref{lm3.2}} \lb{ss4}
\setcounter{equation}{0}

\subsection{The sector counting Lemma}\lb{ss4.1}

In order to present the proof of Lemma \ref{lm3.2}, we need to
introduce some new definitions.

\begin{enumerate}

\item Given a tree $\t$ and $\bP\in \PP_\t$, we shall call
$\c$-vertices the vertices $v$ of $\t$, such that ${\cal I}_v$
(the set of internal lines, that is the lines contracted in $v$)
is not empty. We shall also call $V_\c$ the family of all
$\c$-vertices, whose number is of order $n$.

\item Given $h\le 0$ and a set of field indices $P$, we define
$\bO_h(P)=\otimes_{f\in P} O_h$ and we shall call $\ss=\{\s_f\in
O_h,\ f\in P\}$ the elements of $\bO_h(P)$.

\item Given $h\le 0$ and $\ss\in \bO_h(P)$, we define
$\SSS_h(\ss)=\{ S_{h,\s_f}, \s_f\in\ss\}$.

\item Given a set of field indices $P$ and two families of
s-sectors labelled by $P$, $\SS^{(i)}=\{S_{j_f^{(i)},
\s_f^{(i)}}$, $f\in P\}$, $i=1,2$, we shall say that $\SS^{(1)}
\prec \SS^{(2)}$, if $S_{j_f^{(1)}, \s_f^{(1)}} \subset
S_{j_f^{(2)}, \s_f^{(2)}}$, for any $f\in P$.

\end{enumerate}

The main point in the proof is the following lemma, which is an
extension of that proved in \cite{FMRT} in the jellium case; see
\S\ref{ssA1} for a proof.

\begin{lemma}\lb{lm4.1}
Let $h',h,L$ be integers such that $h'\le h\le
0$. Let $v$ be a vertex of a tree $\t$, such that $|P_v|=L$ and
$f_1$ a fixed element of $P_v$. Then, given the sector index
$\s_{f_1}\in O_{h'}$, and a set $\ss\in \bO_h(P_v\bs f_1)$, the
following bound holds:
\be \sum_{\ss'\in\bO_{h'}(P_v\bs
f_1)\atop\SSSS_{h'}(\ss')\prec\SSSS_h(\ss)}
\c_v\left(\{S_{h',\s_{f_1}}\}\cup \SSS_{h'}(\ss')\right)\le \cases
{c^L \g^{{h-h' \over 2}(L-3)} & ,\ \ if $L\ge 4$,\cr c& ,\ \ if
$L=2$\ .\cr} \lb{4.3}\ee
\end{lemma}

\subsection{Proof of Lemma \ref{lm3.2}}\lb{ss4.2}

First of all, we note that
\be \prod_v \c_v(\SS(P_v)) = \prod_{v\in V_\c}
\c_v(\SS(P_v))\;.\lb{4.1}\ee

Let us consider first the case $P_{v_0}\not = \emptyset$ and let
$\tilde v_0$ be the first $\c$-vertex following the root (possibly
equal to $v_0$); note that $P_{\tilde v_0} = P_{v_0}$ and that
$h(f)=h$ for any $f\in P_{\tilde v_0}$. In the following it will
also very important to remember that $\O$ is the family of all
sector indices $\o(f)$ associated with the field labels $f$ and
that $\o(f)\in O_{h(f)}$, $h(f)$ being the scale of the propagator
connected to the corresponding field variable, see \S\ref{ss3.3}.
In agreement with this definition, if $\bar\O$ is a subset of
$\O$, $\sum_{\bar\O}$ will denote the sum over $\o(f)\in
O_{h(f)}$, for any $f\in \bar\O$.

Let us call $f_0$ the field whose sector index $\o(f_0)\in O_h$ is
fixed in the sum over $\O$. We rewrite the sector sum in the
l.h.s. of \pref{3.62} as:
\be \sum_{\O}^*=\sum_{\O_{\tilde v_0}}^*\sum_{\O\bs\O_{\tilde v_0}}=
\sum_{\ss_{\tilde v_0}\in\bO_{h_{\tilde v_0}}(P_{\tilde v_0} \bs
f_0)} \sum^*_{\O_{\tilde v_0}: \atop\SS(P_{\tilde v_0}\bs
f_0)\prec \SSSS_{h_{\tilde v_0}}(\ss_{\tilde v_0})}
\sum_{\O\bs\O_{\tilde v_0}}\;.\lb{4.2}\ee
Then, for any fixed $\ss_{\tilde v_0}\in \bO_{h_{\tilde
v_0}}(P_{\tilde v_0}\bs f_0)$,
 we bound the product of $\c_v$ functions as
\be \prod_{v\in V_\c }
\c_v(\SS(P_v))\le\c_{\tilde v_0}(\SS(P_{\tilde v_0})) \prod_{v\in
\{V_\c\bs{\tilde v_0}\}} \c_v(\tilde\SS_{v,\tilde v_0})
\;,\lb{4.4}\ee
where
\be \tilde\SS_{v,\tilde v_0}=\SS\Bigl(P_v\bs (P_{\tilde v_0}\bs f_0)\Bigr)
\cup\left\{S_{h_{\tilde v_0},\s_f}\in\SSS_{h_{\tilde v_0}}
(\ss_{\tilde v_0}), f\in P_v\cap (P_{\tilde v_0}\bs
f_0)\right\}\;.\lb{4.4d}\ee
In other words, for any $v\not= \tilde v_0$, we relax the sector
condition by allowing the external fields of $v$, which are also
external fields of $\tilde v_0$ and are not equal to $f_0$, to
have a momentum varying, instead than in the original sector, of
scale $h$, in that of scale $h_{\tilde v_0}$ containing it.

Let us now observe that the modified running coupling functions do
not depend on $\O_{\tilde v_0}$, if $\ss_{\tilde v_0}$ is fixed,
as it follows from definition \pref{3.46}; hence the only remaining
dependence on $\O_{\tilde v_0}$ is in $\c_{\tilde
v_0}(\SS(P_{\tilde v_0}))$. It follows, by using Lemma \ref{lm4.1}
for $|P_{\tilde v_0}|\le 8$ and the trivial bound
\be \sum^*_{\O_{\tilde v_0}\atop \SS(P_{\tilde v_0}\bs f_0)\prec
\SSSS_{h_{\tilde v_0}(\ss_{\tilde v_0})}} 1 \le c \g^{ {1\over 2}
(h_{\tilde v_0}-h) (|P_{\tilde v_0}|-1) }\;,\lb{4.4b}\ee
for $|P_{\tilde v_0}|\ge 10$, that we can bound the sum over
$\O_{\tilde v_0}$, for any $\SSS_{h_{\tilde v_0}}(\ss_{\tilde
v_0})$, as
\bea &&\sum^*_{\O_{\tilde v_0}\atop \SS(P_{\tilde v_0}\bs
f_0)\prec \SSSS_{h_{\tilde v_0}}(\ss_{\tilde v_0})} \c_{\tilde
v_0}(\SS(P_{\tilde v_0}))\le\nn\\
&&\le c \g^{(h_{\tilde v_0}-h) \left[ {1\over 2}(|P_{\tilde
v_0}|-3 ) \c (4\leq |P_{\tilde v_0}|\leq 8)+{1\over 2}(|P_{\tilde
v_0}|-1)\c(|P_{\tilde v_0}|\ge 10)\right]}\;.\lb{4.5}\eea

We are thus left with the problem of bounding a sum similar to the
initial one, but with all the external sector indices on scale
$h_{\tilde v_0}$ instead of $h$. We shall do that by iterating the
previous procedure, in a way which depends on the structure of the
tree $\t$ and of the graph $T$; the iteration stops at the
endpoints, where we can use the hypotheses \pref{3.58}, \pref{3.59}
and \pref{3.60}.

To describe this inductive procedure, we establish, for any vertex
$v\in V_\c$, a partial ordering of the $s_v$ vertices
$v_1,\ldots,v_{s_v}\in V_\c$ immediately following $v$ on $\t$, by
assigning a root to the tree graph $T^*$ and to each anchored tree
graph $T_v$. We decide that the root of $T^*$ is the space-time
point containing $f_0$; then we assign a direction to the lines of
the tree graph $T^*$, the one which goes from the root towards the
leaves. Finally we decide that the root of $T_v$ is the vertex
which the line of $T_{v'}$ enters, where $v'$ is the $\c$-vertex
immediately preceding $v\in V_\c$, if $f_0\not\in P_v$; otherwise,
the root of $T_v$ is the vertex containing the root of $T^*$, see
Fig. \ref{f2}.

\insertplot{300}{130}{%
\ins{200pt}{110pt}{$\tilde v_0$}%
\ins{70pt}{85pt}{$f_0$}\ins{170pt}{50pt}{$v_1$}%
\ins{140pt}{65pt}{$\ell_1$}\ins{130pt}{100pt}{$\ell_3$}%
\ins{100pt}{85pt}{$v_2$}\ins{170pt}{110pt}{$v_3$}}{fig53} {A
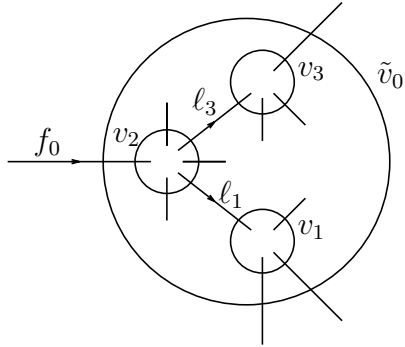
possible cluster structure corresponding to a tree $\t$ of the
expansion for the effective potentials such that $s_{\tilde
v_0}=3$. The set $T_{\tilde v_0}$ is formed by the lines $\ell_1$
and $\ell_3$. The lines different from $\ell_1$ and $\ell_3$ and
not belonging to $P_{\tilde v_0}$ have to be contracted into the
Lesniewski determinants.\lb{f2}}{0}

The l.h.s. of \pref{3.62} is bounded by the product of the r.h.s.
of \pref{4.5} and the following quantity:
\be \left[ \prod_{v >{\tilde v_0}, v\in V_\c} \;\sum_{
\tilde\O_{v,\tilde v_0}} \c_v\left(\tilde{\SS}_{v,\tilde
v_0}\right) \right] \left[ \prod_{l\in T}\d_{\o_l^+,\o_l^-}
\right] \int\prod_{l\in T^*\setminus T} d\rr_l \prod_{i=1}^n
\left|\tilde K_{v_i^*, \tilde\O_i}^{h_i}
(\xx_{v_i^*})\right|\;,\lb{4.6}\ee
where
\be \tilde\O_{v, \tilde v_0}= \{\O_v\bs\O_{\tilde
v_0}\}\cup\{\s_f\in O_{h_{\tilde v_0}}, f\in P_v\cap (P_{\tilde
v_0}\bs f_0)\}\;.\lb{4.6a}\ee
Note that there is no sector index associated with $f_0$ in
$\tilde\O_{v, \tilde v_0}$ and that, if $v$ is a $\c$-vertex
immediately following $\tilde v_0$ on $\t$, all the sector indices
included in $\tilde\O_{v, \tilde v_0}$ belong to $O_{h_{\tilde
v_0}}$, since in this case the fields associated with $P_v\bs
P_{\tilde v_0}$ are contracted on scale $h_{\tilde v_0}$.

We now consider the $s_{\tilde v_0}$ $\c$-vertices immediately
following $\tilde v_0$ and we reorder the expression \pref{4.6} in
the following way:
\bea \pref{4.6}&=& \prod_{j=1}^{s_{\tilde v_0}}\left[
\sum_{\cup_{v\ge v_j} \tilde\O_{v,\tilde v_0}}
\left(\c_{v_j}(\tilde{\SS}_{v_j,\tilde v_0})\prod_{v> v_j\atop
v\in V_\c} \c_v(\tilde{\SS}_{v,\tilde v_0})\prod_{l\in\cup_{v\ge
v_j}T_v} \d_{\o_l^+,\o_l^-}\right)\cdot\right.\nn\\
&&\left.\cdot\prod_{v_i^*\ge v_j} \int d\rr_{v_i^*} \left|\tilde
K_{v_i^*,\tilde\O_i}^{h_i} (\xx_{v_i^*})\right|\right] \prod_{l\in
T_{\tilde v_0}}\d_{\o_l^+,\o_l^-}\;,\lb{4.8}\eea
where:
\0 i) $\int d\rr_{v_i^*}$ is equal to $\int \prod_{l\in T_{v_i^*}}
d\rr_l$, where $T_{v_i^*}$ denotes the subset of the tree graph
$T^*$ connecting the set $x_{v_i^*}$;
\0 ii) if $s_v=1$, $\prod_{l\in T_v} \d_{\o_l^+,\o_l^-}$ has to be
thought as equal to $1$.

We now choose a leave of $T_{v_0}$ ($v_1$ or $v_3$ in Fig.
\ref{f2}), say $v^*$, and we consider the factor in the product
$\prod_{j=1}^{s_{\tilde v_0}}$ appearing in the r.h.s. of
\pref{4.8} corresponding to $v^*$, together with the line $l^*\in
T_{v_0}$ entering $v^*$ ($\ell_1$ or $\ell_3$ in Fig. \ref{f2}).
We can associate with $v^*$ the following quantity, which is
independent of all the other leaves and of the sector indices
associated with the lines of $T_{v_0}$:
\bea [v^*]&=& \left[ \sum_{\tilde\O_{v^*,\tilde v_0}}^*
\left(\c_{v^*}(\tilde{\SS}_{v^*,\tilde v_0}) \sum_{\cup_{v> v^*}
  \tilde\O_{v,\tilde v_0}\setminus
\tilde\O_{v^*,\tilde v_0}}^* \prod_{v> v^*\atop v\in V_\c}
\c_v(\tilde{\SS}_{v,\tilde v_0}) \prod_{l\in\cup_{v\ge v^*}T_v}
\d_{\o_l^+,\o_l^-}\right) \cdot\right.\nn\\
&&\left.\cdot \prod_{i:v_i^*\ge v^*}\int d\rr_{v_i^*} \left|\tilde
K_{v_i^*,\tilde\O_i}^{h_i} (\xx_{v_i^*})\right| \right]
\;,\lb{4.8a}\eea
where $\sum_{\tilde\O_{v^*,\tilde v_0}}^*$ means that we do not
sum over the sector index associated with $l^*$.

In order to bound the expression in the r.h.s. \pref{4.8a}, we have
to distinguish two cases
\*\0 (a) $v^*$ is an endpoint. In this case
$\sum_{\tilde\O_{v^*,\tilde v_0}}^* = \sum_{\ss\in
\O_{h_{v^*}-1}}^*$ and $\c_{v^*}(\tilde{\SS}_{v^*,\tilde v_0})=1$,
since the corresponding constraint is already included in the
definition of the modified coupling functions, so that the
expression to bound is simply:
\be \sum_{\ss\in \O_{h_{v^*}-1}}^* \int d\rr_{v^*}
\left|\tilde K_{v^*,\ss}^{h_{v^*}} (\xx_{v^*})\right|\;.\lb{4.9}\ee
Hence, conditions \pref{3.58}, \pref{3.59},
\pref{3.60} imply that
\be [v^*] \le c |\l|\g^{-{1\over 2} (h_{v^*}-1)
\c(|P_{v^*}|=4)}\g^{(h_{v^*}-1) \c(v^*\;{\rm
is\;of\;type\;}\n)}\;.\lb{4.9a}\ee

\*\0 (b) $v^*$ is not an end point. In this case, by the remark
following \pref{4.6a}, the expression in the r.h.s. of \pref{4.8a}
has exactly the same structure as the l.h.s. of \pref{3.62}, which
we started the iteration from; one has only to substitute $\tilde
v_0$ with $v^*$, $h$ with $h_{\tilde v_0}$ and $h_{\tilde v_0}$
with $h_{v^*}$. Hence we can bound the r.h.s. of \pref{4.8a} by
extracting a factor
\be c \g^{(h_{v^*}-h_{\tilde v_0})\left( {1\over 2}(|P_{v^*}|-3 )
\c (4\leq |P_{v^*}|\leq 8)+{1\over 2}(|P_{v^*}|-1)\c(|P_{v^*}|\ge
10)\right)}\lb{4.13}\ee
and we end up with an expression similar to \pref{4.6}, the line
$l^*$ acting now as an external field, since there is only one
sector sum associated with it, thanks to the factor
$\d_{\o_{l^*}^+,\o_{l^*}^-}$ present in the r.h.s. of \pref{4.8}.

\*

It is now completely obvious that we can iterate the previous
procedure, for each leave of $T_{\tilde v_0}$, ending up with a
bound of the l.h.s. of \pref{3.62} of the form
\bea &&(c|\l|)^n \left[ \prod_{v\in V_\c} \g^{(h_{v}-h_{v'})\left(
{1\over 2}(|P_{v}|-3 ) \c (4\leq |P_{v}|\leq 8)+{1\over
2}(|P_{v}|-1)\c(|P_{v}|\ge 10)\right)} \right]\;\cdot\nn\\
&&\cdot\; \left[ \prod_{i=1}^n \g^{-{1\over 2} (h_i-1)
\c(v_i^*\;{\rm is\;of\;type\;}\l)}\g^{(h_i-1)\c(v_i^*\;{\rm
is\;of\;type\;}\n)} \right]\;,\lb{4.14}\eea
where $v'$ is the $\c$-vertex immediately preceding $v$ on $\t$,
if $v>\tilde v_0$, or the root, if $v=\tilde v_0$. On the other
hand, given $v\in V_\c$, $P_{\bar v}=P_v$ if $v'<\bar v \le v$.
Moreover,
\be \prod_{i=1}^n \g^{-{1\over 2} (h_i-1) \c(v_i^*\;{\rm
is\;of\;type\;}\l)} = \g^{-{1\over 2}h m_4(v_0)} \prod_{v\;{\rm
not\;e.\;p.}} \g^{-{1\over 2}m_4(v)}\;,\lb{4.15}\ee
where $m_4(v)$ is the number of end points of type $\l$ following
vertex $v$ on $\t$. It follows that \pref{4.14} can be written in
the form

\bea &&(c|\l|)^n \g^{-{1\over 2}h m_4(v_0)}\left[\prod_{i=1}^n
\g^{(h_i-1) \c(v_i^*\; {\rm{is\; of\; type}}\;\n)}
\right]\cdot\nn\\
&&\cdot \prod_{v\;{\rm not\;e.\;p.}}\g^{\left[ -{1\over
2}m_4(v)+{1\over 2}(|P_v|-3) \c (4\leq |P_v|\leq 8)+{1\over
2}(|P_v|-1) \c(|P_v|\ge 10)\right]}\;,\lb{4.16}\eea
which proves Lemma \ref{lm3.2} in the case $|P_{v_0}|>0$.

The case $P_{v_0}=\emptyset$ is treated in a similar way. The only
real difference is that one has to sum over all sector indices.
However, since the set of internal fields ${\cal I}_{v_0}$ is
necessarily not empty (our definitions imply that, in this case,
${\tilde v_0}=v_0$), we can choose in an arbitrary way one field
$f_0\in {\cal I}_{v_0}$ and let it play the same role of the
selected external field of $v_0$ in the previous iterative
procedure. Of course, the first iteration step, which produced
before the ``scale jump'' factor in the r.h.s. of \pref{4.5}, is
now missing, but this is irrelevant, since that factor is equal to
$1$ if $|P_{v_0}|=0$. All the other steps are absolutely
identical, but, at the end of the iteration, we end up with the
sector sum related with $f_0$; this produces a factor
$\g^{-{1\over 2}h_{v_0}}=\g^{-{1\over 2}(h+1)}$. This completes
the proof.\Halmos

\section{ The flow of running coupling functions}\lb{ss5}
\setcounter{equation}{0}

\subsection{The expansion for $\LL
\VV^{(h)}(\psi^{(\le h)})$}\lb{ss5.1}

By using \pref{3.29}, \pref{3.32} and \pref{3.38a}, we get
\bea &&\LL \VV^{(h)}(\psi^{(\le h)})=
\sum_{n=1}^\io\sum_{\t\in\TT_{h,n}} \sum_{\bP\in\PP_\t:\atop
|P_{v_0}|=2,4} \sum_{\O\in {\cal O}_\t} \sum_{T\in {\bf T}}
\cdot\nn\\
&&\cdot \int d\xx_{v_0} \tilde\psi^{(\le h)}_{\O_{v_0}}(P_{v_0})
\LL W_{\t,\bP,\O\bs \O_{v_0},T}^{(h)}(\xx_{v_0})\;,\lb{5.0b}\eea
where, if $P_{v_0}=(f_1, \ldots, f_4)$ and we put $\xx(f_i)=
\xx_i= (x_{i,0}, \vxx_i)$, $\tilde \xx_i=(\tilde x_{i,0}, \vxx_i)$
and $\xx^*$ is any point in $\xx_{v_0}$,
\be \LL W_{\t,\bP,\O\bs \O_{v_0},T}^{(h)}(\xxx) = \d(\ux_0) \int
d(\tilde\ux_0\bs \tilde x_0^*) W_{\t,\bP,\O\bs
\O_{v_0},T}^{(h)}(\tilde\xxx)\;,\lb{5.0}\ee
while, if $P_{v_0}=(f_1,f_2)$ and $m(P_{v_0}) = m(f_1)+m(f_2)=0$,
\be \LL W_{\t,\bP,\O\bs \O_{v_0},T}^{(h)}(\xx_1,\xx_2)=
\d(x_{1,0}-x_{2,0})\int d\tilde x_{1,0} W_{\t,\bP,\O\bs
\O_{v_0},T}^{(h)}(\tilde\xx_1, \tilde\xx_2)\;,\lb{5.0c}\ee
and finally, if $P_{v_0}=(f_1,f_2)$, $m(P_{v_0})=1$,
\bea &&\LL W_{\t,\bP,\O\bs \O_{v_0},T}^{(h)}(\xx_1,\xx_2)=
\d(x_{1,0}-x_{2,0})\cdot\nn\\
&& \int d\tilde x_{1,0} (\tilde x_{1,0}- \tilde x_{2,0})
W_{\t,\bP,\O\bs \O_{v_0},T}^{(h)}(\tilde\xx_1,
\tilde\xx_2)\;.\lb{5.0d}\eea
Note that there is no other case to consider, since, as a
consequence of the freedom in the choice of the field carrying the
derivative in the endpoints of type $z$, there is no contribution
to the effective potential with $n\ge 2$ and a derivative acting
on the external fields of $v_0$, before the application of the
$\LL$ operator.

Let us consider first the contributions to the r.h.s. of
\pref{5.0b} coming from the trees with $n=1$. These trees have only
two vertices, $v_0$ (of scale $h+1$) and the endpoint $v^*$, whose
scale has to be equal to $h+2$. If we impose the further condition
that $P_{v^*}=P_{v_0}$, the sum of these terms is equal to $\LL
\VV^{(h+1)}(\t,\psi^{(\le h)})$. In order to control the flow of
the running coupling functions, we need a ``good bound'' of the
remaining terms.

Let us consider a contribution to the r.h.s. of \pref{5.0b}, such
that $n\ge 2$ or $n=1$ and $P_{v^*}\not=P_{v_0}$. By proceeding as
in \S\ref{ss3.4}, it is easy to show that
\be \LL W_{\t,\bP,\O\bs \O_{v_0},T}^{(h)}(\xx_{v_0})= \sum_{\a\in A_T}
W^{(L)}_{\t,\bP,\O\bs \O_{v_0},T,\a}(\xx_{v_0})\;,\lb{5.0e}\ee
where $A_T$ is a suitable set of indices and
$W^{(L)}_{\t,\bP,\O\bs \O_{v_0},T,\a}(\xx_{v_0})$ can be
represented as in \pref{3.40}. There is indeed a small difference,
because of the delta function and the integral appearing in
\pref{5.0}, \pref{5.0c} and \pref{5.0d}, but it can be treated
without any new problem. Moreover, by the considerations of
\S\ref{ss3.5}, if we insert \pref{5.0e} in the r.h.s. of \pref{5.0b},
we can substitute $W^{(L)}_{\t,\bP,\O\bs
\O_{v_0},T,\a}(\xx_{v_0})$ with $W^{(L,
mod)}_{\t,\bP,\O,T,\a}(\xx_{v_0})$, obtained by using the modified
running coupling functions in place of the original ones. As
before, these modified functions are not constant with respect to
$\O_{v_0}$. We can prove the following Theorem, analogous to
Theorem \ref{th3.1}.

\begin{theorem}\lb{th5.1}
If conditions \pref{3.58}, \pref{3.59},
\pref{3.60} are satisfied, given a couple of integers $(p,m)$ equal
to $(2,0)$, $(2,1)$ or $(4,0)$, we have:
\bea &&\sum_{\t\in\TT_{h,n}} \sum_{\bP\in \PP_\t:\atop
|P_{v_0}|=p, m(P_{v_0})=m}^{**} \sum_{T\in {\bf T}} \sum_{\a\in
A_T} \sum^*_{\O\in\OO_\t}\int d(\xx_{v_0}\bs\xx^*) \left|
W^{(L,mod)}_{\t,\bP,\O,T,\a}(\xx_{v_0}) \right| \le\nn\\
&& \le\;(c |\l|)^n \g^{h [\d_{ext}(p)-m]}\;,\lb{5.69}\eea
with $\d_{ext}(p)$ defined by \pref{3.67a} and $\sum^{**}$ means
that, if $n=1$ and $v^*$ is the endpoint, $P_{v^*}\not=P_{v_0}$.
\end{theorem}

\proof We can repeat step by step the proof of Theorem \ref{th3.1}
and use the remark that, in the identity \pref{3.62b},
$q_0=m$.\Halmos

\subsection{The beta function}\lb{ss5.2}

The discussion of \S\ref{ss5.1} and the definition of modified
running coupling functions (MRCF in the following) of
\S\ref{ss3.5} imply that
\be \tilde\l_{h,\ss}(\vxxx) = (\FFF_{4,h,\ss}*
\l_{h+1})(\vxxx) + (\FFF_{4,h,\ss}*
\b^{4,0}_{h+1})(\tilde\vv_{h+1},..,\tilde\vv_1;
\vxxx)\;,\lb{5.5a}\ee
\be \tilde\n_{h,\ss}(\vxxx) = \g\ (\FFF_{2,h,\ss}* \n_{h+1})(\vxxx) +
(\FFF_{2,h,\ss}* \b^{2,0}_{h+1})(\tilde\vv_{h+1},..,\tilde\vv_1;
\vxxx)\;,\lb{5.5b}\ee
\be \tilde z_{h,\ss}(\vxxx) = (\FFF_{2,h,\ss}* z_{h+1})(\vxxx) +
(\FFF_{2,h,\ss}* \b^{2,1}_{h+1})(\tilde\vv_{h+1},..,\tilde\vv_1;
\vxxx)\;,\lb{5.5c}\ee
where $\vv_h\= (\l_h, \n_h, z_h)$, $\tilde \vv_h$ is the set of
the corresponding MRCF, $\vxxx=(\vxx_1,\ldots,\vxx_p)$,
$\ss=(\s_1,\ldots,\s_p)$, with $\s_i\in O_{h-1}$ and $p=4$ in
\pref{5.5a}, $p=2$ in \pref{5.5b} and \pref{5.5c}. Finally, the {\it
beta function} $\b^{p,m}_{h+1}(\vv_h,..,\vv_1; \vxxx)$ is defined
by the equation
\bea &&\b^{p,m}_{h+1}(\vv_h,..,\vv_1; \vxxx)= \g^{-h\cdot
\c(p=2,m=0)} \;\cdot\lb{5.2}\\
&&\cdot\; \sum_{n=1}^\io \sum_{\t\in\TT_{h,n}} \sum_{\bP:
|P_{v_0}|=p\atop \ m(P_{v_0})=m}^{**}  \sum_{T\in {\bf T}}
\sum_{\O\bs \O_{v_0}} \sum_{\a\in A_T} \int d(\ux_0\bs x^*_0)
W^{(L)}_{\t,\bP,\O\bs \O_{v_0},T,\a}(\xxx)\;.\nn\eea

Note that, given a tree contributing to the r.h.s. of \pref{5.2},
we can substitute the RCF with the MRCF in all endpoints except
those containing one of the external fields of $v_0$. However
$(\FFF_{p,h,\ss}* \b^{p,m}_{h+1})$ is indeed a function of the
MRCF, as we made explicit in the r.h.s. of \pref{5.5a}-\pref{5.5c}
and
\bea &&(\FFF_{p,h,\ss}*
\b^{p,m}_{h+1})(\tilde\vv_{h+1},..,\tilde\vv_1; \vxxx)=
\g^{-h\cdot \c(p=2,m=0)} \sum_{n=1}^\io \sum_{\t\in\TT_{h,n}}
\sum_{\bP\atop |P_{v_0}|=p,\ m(P_{v_0})=m}^{**} \;\cdot\nn\\
&&\cdot\; \sum_{T\in {\bf T}} \sum_{\O:\atop \O_{v_0}=\ss}
\sum_{\a\in A_T} \int d(\ux_0\bs x^*_0)
\;(\FFF_{p,h,\ss}*W^{(L,mod)}_{\t,\bP,\O
,T,\a})(\xxx)\;.\lb{5.2a}\eea

Iterating \pref{5.5a}, \pref{5.5b} and \pref{5.5c} we find, for $h\le 0$,
\bea \tilde\l_{h,\ss}(\vxxx) &=& (\FFF_{4,h,\ss}*
\tilde\l_1)(\vxxx) + \sum_{j=h+1}^1(\FFF_{4,h,\ss}*
\b^{4,0}_j)(\tilde\vv_j,..,\tilde\vv_1; \vxxx)\;,\lb{5.5aa}\\
\tilde\n_{h,\ss}(\vxxx) &=& \g^{-h+1}(\FFF_{2,h,\ss}*
\tilde\n_1)(\vxxx) +\nn\\
&&+ \sum_{j=h+1}^1\g^{-h+j-1} (\FFF_{2,h,\ss}*
\b^{2,0}_j)(\tilde\vv_j,..,\tilde\vv_1; \vxxx)\;,\lb{5.5ba}\\
\tilde z_{h,\ss}(\vxxx) &=& (\FFF_{2,h,\ss}* \tilde z_1)(\vxxx) +
\sum_{j=h+1}^1(\FFF_{2,h,\ss}*
\b^{2,1}_j)(\tilde\vv_j,..,\tilde\vv_1; \vxxx)\;,\lb{5.5ca}\eea
where, ignoring in the notation the spin dependence of $v(\vxx)$,
see \pref{2.8}, $\tilde \l_1(\vxxx) = (\FFF_{4,1,{\underline 0}}*
\l_1)(\vxxx)$, with $\l_1(\vxxx)=-\l v(\vxx_1-\vxx_2)
\d(\vxx_3-\vxx_1)\d(\vxx_4-\vxx_2)$, and $\tilde\n_1(\vxxx) =
(\FFF_{2,1,(0,0)} *\n_1)(\vxxx)$. Furthermore $\tilde z_1(\vxxx) =
z_1(\vxxx)=0$ and $\n_1(\vxxx)$ must be suitably chosen.

We note that it is possible to choose the functions $\tilde
F_{h,\s}(\vkk)$ appearing in the definition of the operators
$(\FFF_{p,h,\ss}*\cdot)$, see \pref{3.45aa}, in such a way that, if
$h\le 0$,
\be |\e(\vkk)-\m|\le e_0\g^h\Rightarrow {1\over 2}
\sum_{\s\in O_h}\tilde F_{h,\s}(\vkk)=1\;.\lb{5.5d}\ee
In order to simplify the following discussion, we shall suppose
that the property \pref{5.5d} is satisfied. Moreover we define
$\bO_{h,p}=\otimes_{i=1}^pO_h$.

Theorem \ref{th5.1} implies that, given $\bar h<0$, the MRCF are
well defined for $\bar h\le h\le 1$, if $\l$ and
$\tilde\n_1(\vxxx)$ are small enough. We want to show that, given
$\l$ small enough and $\log\b\le c_0|\l|^{-1}$, it is possible to
choose $\tilde\n_1(\vxxx)$ so that the MRCF are well defined for
$h_\b\le h$, with $h_\b$ defined by \pref{3.4a}. We shall try to
fix $\tilde\n_1(\vxxx)$ in such a way that
\be \g^{-h_\b+1}\tilde\n_1(\vxxx) +
\sum_{j=h_\b+1}^1\g^{-h_\b+j-1}{1\over 4}
\sum_{\ss_j\in\bO_{j,2}}(\FFF_{2,j,\ss_j}* \b^{2,0}_j)
(\tilde\vv_j,..,\tilde\vv_1; \vxxx)=0\;,\lb{5.5e}\ee
so that \pref{5.5ba} becomes:
\bea && \tilde\n_{h,\ss}(\vxxx) =
-\sum_{j=h_\b+1}^h\g^{-h+j-1}\cdot\nn\\
&&\cdot {1\over 4} \sum_{\ss_j\in\bO_{j,2}}^* \left(
\FFF_{2,h,\ss}
* \FFF_{2,j,\ss_j}* \b^{2,0}_j \right)(\tilde\vv_j,..,\tilde\vv_1;
\vxxx)\;,\lb{5.5ee}\eea
where, given $\ss=(\s_1,\s_2)\in\bO_{h,2}$,
$\sum^*_{\ss_j\in\bO_{j,2}}$ is the sum restricted to the
$\ss_j=({\s'}_1,{\s'}_2)\in\bO_{j,2}$ such that $S_{h,\s_i}\cap
S_{j,{\s'}_i}\not=\emptyset$, $i=1,2$.

In order to present our results, we have to introduce a few other
definitions. Given $h\le 1$ and $\o\in O_h$, we denote by
$D_{h,\s}\in \RRR^2$ the support of $\tilde F_{h,\s}(\vkk)$.
Moreover, if $p=2,4$, we call $\MM_{h,p}$ the space of functions
$G_\ss(\vxxx): \bO_{h,p}\times\RRR^{2p}\to\RRR$, such that

\0 1) for any $\ss\in \bO_{h,p}$, $G_\ss(\vxxx)$ is translation
invariant;

\0 2) for any $\ss\in \bO_{h,p}$, the Fourier transform $\hat
G_\ss(\vkkk)$ of $G_\ss(\vxxx)$, defined so that,
\be G_\ss(\vxxx) = \int {d\vkkk \over (2\p)^{2p}}\;
e^{-i \vkkk \cdot \vxxx} \; \hat G_\ss(\vkkk) \; \d(\sum_{i=1}^p
\e_i \vkk_i)\;,\lb{5.5f}\ee
with $\e_1=-\e_2=+$, if $p=2$, and $\e_1=\e_2=-\e_3=-\e_4=+$, if
$p=4$, is a continuous function with support in the set
$\otimes_{i=1}^p D_{h,\s_i}$.

\0 Given $G\in \MM_{h,p}$, we shall say that $G_\ss(\vxxx)$ is the
$\ss$-component of $G$. These definitions are such that
$\tilde\n_{h,\ss}(\vxx)$ and $\tilde z_{h,\ss}(\vxx)$ are the
$\ss$-components of two functions $\tilde\n_h$ and $\tilde z_h$
belonging to $\MM_{h,2}$, while $\tilde\l_{h,\ss}(\vxx)$ is the
$\ss$-component of a function $\tilde\l_h\in \MM_{h,4}$.

\0 We shall define a norm on the set $\MM_{h,p}$ by putting
\be ||G||_{h,p}=\sup_{i,\s_i\in O_h \atop j,\vxx_j\in \rrr^2}
\sum_{\ss\bs \s_i\in\bO_{h,p-1}} \int d(\vxxx\bs\vxx_j)
|G_\ss(\vxxx)|\;.\lb{5.3}\ee

Finally we shall define $\MMM_p$ as the set of sequences
$G=\{G_h\in \MM_{h,p}, h_\b\le h\le 1\}$, such that the norm
\be ||G||_p = \max_{h_\b\le h\le 1} ||G_h||_{h,p}\lb{5.3a}\ee
is finite. We want to prove that the sequence $\tilde\l \= \{
\tilde\l_h, h_\b\le h\le 1 \}$ is well defined as an element of
$\MMM_4$, while the sequences $\tilde\n \= \{ \tilde\n_h, h_\b\le
h\le 1 \}$ and $\tilde z \= \{ \tilde z_h, h_\b\le h\le 1 \}$ are
two elements of $\MMM_2$.

We begin our analysis by ``decoupling'' equations \pref{5.5aa} and
\pref{5.5ca} from \pref{5.5ba}, that is we imagine that, in the
r.h.s. of \pref{5.5aa} and \pref{5.5ca}, $\tilde\n$ is an arbitrary
element of $\MMM_2$, acting as a parameter. We want to look for a
solution ($\tilde\l(\tilde\n)\in \MMM_4$, $\tilde z(\tilde\n)\in
\MMM_2$). We shall prove the following lemma.

\*

\begin{lemma}\lb{lm5.1}
There exist positive constants $C_1$ and $C_2$, depending only on
first and second order terms in our expansion, such that, given
two positive constants $C_3\ge C_1$ and $C_4$, there exists $\l_0$
so that, if $|\l|\le \l_0$,
\be 2 C_2 C_3 \max\{1,C_4^{-1}\} |\l| |h_\b|\le 1\ee
and $||\tilde\n||_2, ||\tilde\n'||_2 \le C_3|\l|$, then, for
$h_\b\le h\le 1$,
\be ||\tilde\l(\tilde\n)_h||_{h,4} \le 2C_1|\l|\g^{-{1\over 2}h}
\virg ||\tilde z(\tilde\n)_h||_{h,2} \le C_1|\l|\;,\lb{5.6}\ee
\bea ||\tilde\l(\tilde\n)_h - \tilde\l(\tilde\n')_h||_{h,4} &\le&
C_4 \g^{-{1\over 2}h} \max_{j> h} ||\tilde\n_h -
\tilde\n'_h||_{h,2} \;,\nn\\
||\tilde z(\tilde\n)_h - \tilde z(\tilde\n')_h||_{h,2} &\le& C_4
\max_{j> h} ||\tilde\n_h - \tilde\n'_h||_{h,2}\;.\lb{5.7}\eea
\end{lemma}

\proof Note that, if  $\bar F_{h,\o}(\vxx)$ is the Fourier
transform of $\tilde F_{h,\o}(\vkk)$, then
\be \left|\bar F_{h,\o}(\vxx)\right|\le{C_N\g^{{3\over 2}h}\over
1+\left(\g^h|x_1'|+\g^{h\over 2}|x_2'|\right)^N}\;,\lb{5.7a}\ee
so  that $\int d\vxx\left|\bar F_{h,\o}(\vxx)\right|\le c_F$ for
some constant $c_F$ independent of $h$ and $\o$. It follows that
there exists a constant $C_1$, such
\be ||\tilde\l_1||_{1,4} \le 2C_1|\l|\g^{-1/2} \virg
||\FFF_{4,h,\ss}* \tilde\l_1||_{h,4} \le
C_1|\l|\g^{-h/2}\;,\lb{5.7b}\ee
having used also Lemma \ref{lm4.1} for the second inequality.

We shall prove inductively that, if $||\tilde\n||_2 \le C_3|\l|$,
with $C_3\ge C_1$, then $||\tilde\l(\tilde\n)_h||_{h,4} \le
2C_1|\l|\g^{-{1\over 2}h}$ and $||\tilde z(\tilde\n)_h||_{h,2} \le
C_1|\l|$. This bound is satisfied for $h=1$, by the first
inequality of \pref{5.7b} and the fact that $\tilde z_1=0$; let us
suppose that it is true for any $j>h$. Then, by using \pref{5.5aa},
\pref{5.5ca}, the second inequality of \pref{5.7b}, Theorem
\ref{th5.1} and the fact that $\b_j^{(4,0)}$ and $\b_j^{(2,1)}$ do
not have first order contributions, we find
\bea ||\tilde\l(\tilde\n)_h||_{h,4} &\le& C_1|\l|\g^{-{1\over 2}h}
+ \g^{-{1\over 2}h} \sum_{j=h+1}^1\Bigl[ C_{2,\l} C_1 C_3| \l|^2 +
\sum_{n=3}^{\io} (c|\l|)^n \Bigr]\;,\nn\\
||\tilde z(\tilde\n)_h||_{h,2} &\le& \sum_{j=h+1}^1\Bigl[ C_{2,z}
C_1 C_3| \l|^2 + \sum_{n=3}^{\io} (c|\l|)^n \Bigr]\;.\lb{5.7c}\eea
Hence, if $\l$ small enough and $ 2|\l| |h_\b|C_3 \max\{C_{2,\l},
C_{2,z}\} \le 1$, then $||\tilde\l(\tilde\n)_h||_{h,4}$ $\le
2C_1|\l|\g^{-{1\over 2}h}$ and $||\tilde z(\tilde\n)_h||_{h,2} \le
C_1|\l|$, up to $h=h_\b$.

We still have to prove that, if $||\tilde\n||_2$, $||\tilde\n'||_2
\le C_3|\l|$, then the bounds \pref{5.7} are verified. We shall
again proceed by induction, by using that $\tilde\l(\tilde\n)_1 -
\tilde\l(\tilde\n')_1=0$, since $\tilde\l_1$ is independent of
$\tilde\n$, and that $\tilde z(\tilde\n)_1=0$. Then, if we suppose
that the bound is true for any $j>h$, we find
\bea ||\tilde\l(\tilde\n)_h - \tilde\l(\tilde\n')_h||_{h,4} &\le&
\g^{-{1\over 2}h} \max_{j> h} ||\tilde\n_j-
{\tilde\n'}_j||_{j,2}\cdot\nn\\
&&\cdot \sum_{j=h+1}^1 \left[ \tilde C_{2,\l} C_3\max\{1,C_4\}
|\l| +\sum_{n=3}^{\io} c^n |\l|^{n-1} \right]\nn\\
||\tilde z(\tilde\n)_h - \tilde z(\tilde\n')_h||_{h,2} &\le&
\max_{j> h} ||\tilde\n_j- {\tilde\n'}_j||_{j,2}\cdot\lb{3.7d}\\
&&\cdot \sum_{j=h+1}^1 \left[ \tilde C_{2,z} C_3 \max\{1,C_4\}
|\l| + \sum_{n=3}^{\io} c^n |\l|^{n-1} \right]\;.\nn\eea
Hence, if $\l$ small enough and $2 |\l||h_\b|C_3 \max\{1,C_4^{-1}\}
\max\{\tilde C_{2,\l}, \tilde C_{2,z}\} \le 1$, the bound is
verified up to $h=h_\b$.\Halmos

\*

We want now to show that there is indeed a solution of the full
set of equations \pref{5.5aa}-\pref{5.5ca}, satisfying condition
\pref{5.5e}.

\*

\begin{theorem}\lb{th5.2}
If $|\l|$ is small enough and $C_{1,2}\log\b|\l|\le 1$, where
$C_{1,2}$ is a constant depending only on first and second order
contributions of perturbation theory, it is possible to choose
$\tilde\n_1(\vxxx)$ so that the MRCF satisfy the hypothesis of
Lemma \ref{lm3.2}, \pref{3.58}, \pref{3.59} and \pref{3.60}.
\end{theorem}

\proof In order to prove the Theorem, it is sufficient to look for
a fixed point of the operator $\bT: \MMM_2 \to \MMM_2$, defined in
the following way, if $\tilde\n'\=\bT(\tilde\n)$:
\bea &&\tilde\n_h' = -\sum_{j=h_\b+1}^h\g^{-h+j-1}\nn\\
&& {1\over 4} \sum_{\ss_j\in\bO_{j,2}}^*
\bigl(\FFF_{2,h,\ss}*\FFF_{2,j,\ss_j}*
\b^{2,0}_j\Bigr)(\tilde\vv_j(\tilde\n),..,\tilde\vv_1(\tilde\n);
\vxxx)\;,\lb{5.8}\eea
where $\tilde\vv_j(\tilde\n)=(\tilde\l(\tilde\n), \tilde\n, \tilde
z(\tilde\n))$.

We want to prove that it is possible to choose the constant
$C_\n\ge 1$, so that, if $C_1$ is the constant defined in Lemma
\ref{lm5.1} and $|\l|$ is small enough, the set
$\FF=\{\tilde\n\in\MMM_2: ||\tilde\n||_2\le 2C_1 C_\n|\l|\}$ is
invariant under $\bT$ and that $\bT$ is a contraction on it. This
is sufficient to prove the Theorem, since $\MMM_2$ is a Banach
space, as one can easily show.

By using Theorem \ref{th5.1} and Lemma \ref{lm5.1} (with
$C_3=2C_1C_\n$), we see that, if $|\l|$ is small enough and $4 C_2
C_1 C_\n \max\{1,C_4^{-1}\}|\l| |h_\b|\le 1$ ($C_4$ will be chosen
later),
\be ||\tilde\n_h'||_{h,2}\le
\sum_{j=h_\b+1}^h\g^{-h+j-1} \g^{h-j\over 2}
\left[C_{1,\n}C_1|\l|+ \sum_{n=2}^{\io}c^n|\l|^n\right]
\;,\lb{5.11}\ee
where $\g^{h-j\over 2}$ is, up to a constant, a bound for the
number of sectors $\s'\in O_j$ with non empty intersection with a
given $\s\in O_h$, $h\ge j$ and $C_{1,\n}$ is a constant depending
on the first order contribution (\ie the {\it tadpole}). So, if
$C_\n\ge {C_{1,\n}\over \g-\sqrt\g}$ and
$\sum_{n=2}^{\io}c^n|\l|^n\le C_{1,\n}C_1|\l|$, then
$||\tilde\n_h'||\le 2C_1C_\n|\l|$.

We then show that $\bT$ is a contraction on $\FF$. In fact, given
$\tilde\n_1,\tilde\n_2\in\FF$, by using again Theorem \ref{th5.1}
and Lemma \ref{lm5.1}, we see that, under the same conditions
supposed above,
\bea &&||\tilde\n_{1,h}'-\tilde\n_{2,h}'||\le
\sum_{j=h_\b+1}^h\g^{-h+j-1}\g^{h-j\over 2}\cdot\nn\\
&& \cdot \max_{i\ge j} ||\tilde\n_{1,i}-\tilde\n_{2,i}||\left[
C_{1,\n}C_4+\sum_{n=2}^{\io}c^n|\l|^{n-1}\right]\;, \lb{5.13}\eea
so that, if $\sum_{n=2}^{\io}c^n|\l|^{n-1}\le C_{1,\n}C_4/2$ and
$C_4=(2C_\n)^{-1}$, then $||\tilde\n_1'-\tilde\n_2'||\le {3\over
4}||\tilde\n_1-\tilde\n_2||$, if $8 C_2 C_1 C_\n^2|\l| |h_\b|\le
1$.\Halmos

{\bf Remark} In \cite{DR}, where only the rotational invariant
case is considered, the localization acts only on the kernels of
the effective potential with $n=2$. The consequence of this choice
is that the effective potential is bounded at order $n$ by $(c|\l|
|h_\b|)^n$. Hence in \cite{DR} the effective potentials are found
to be convergent only for $T\ge O(e^{-{1\over c|\l|}})$, $c$ being
a ``bad'' constant, so that such value is very far from the true
critical temperature, which is supposed to be driven by the second
order contribution to the effective potential, for $\l$ small
enough. Note in fact that $c$ depend on bounds at every order in
$\l$, while $C_{1,2}$ only depends on a few lower orders.

\section{ The two point Schwinger function}\lb{ss6}
\setcounter{equation}{0}

In this section we prove Theorem \ref{th2.1}.

The Schwinger functions can be derived by the {\it generating
function} defined as
\be \WW(\phi)= \log \int P(d\psi) e^{-\VV(\psi)-\NN(\psi)+\int d\xx
\left[ \phi^+_{\xx}\psi^{-}_{\xx}+
\psi^{+}_{\xx}\phi^-_{\xx}\right]}\;,\lb{6.1}\ee
where the variables $\phi^\s_{\xx}$ are defined to be Grassmanian
variables, anticommuting with themselves and $\psi^{\s}_{\xx}$. In
particular the two point Schwinger function is given by
\be S(\xx-\yy)= \left. {\dpr^2\over\dpr\phi^+_{\xx}\dpr\phi^-_{\yy}}
\WW(\phi) \right|_{\phi=0}\;.\lb{6.2}\ee

We can get a multiscale expansion for $\WW(\phi)$, by a procedure
very similar to that used for the free energy, by taking into
account that the interaction contains a new term, linear in $\psi$
and $\phi$. This novelty has the consequence that new terms appear
in the expansion, containing one or more $\phi$ fields linked to
the corresponding graphs through a single scale propagator. In
order to study $S(\xx-\yy)$, it is sufficient to analyze the
structure of the terms with one or two $\phi$ fields.

Let us consider first the terms produced after integrating the
scales greater or equal to $h+1$ and linear in $\phi$. These terms
can be obtained by taking one of the contributions $\VV^{(h)}(\t,
\bP) \= \sum_\O \VV^{(h)}(\t, \bP, \O)$ to the effective potential
on scale $h$ and by linking one of its external lines, say $\bar
f$, with the $\phi$ field through a propagator of scale $j\ge
h+1$, to be called the {\it external propagator}. However, one has
to be careful in the choice of the localization point in the
vertices $v$ such that $\bar f\in P_v$ and $|P_v|\le 4$ (so that
the action of $\RR$ in $v$ is not trivial); we choose it as that
one which connects $\bar f$ with the $\phi$ field (hence no
derivative can act on the external propagator, when one exploits
the effect of the $\RR$ operations as in \S\ref{ss3.4}). This choice
has the aim of preserving the regularizing effect of the $\RR$
operation, based on the fact that, if a field acquires a
derivative as a consequence of the $\RR$ operation on scale $i$,
then it has to be contracted on a scale $j<i$, so producing an
improvement of order $\g^{-(i-j)}$ in the bounds. Note also that,
because of the localization operation, the scale $j$ of the
external propagator can be higher of the scale of the endpoint
$\bar v$, such that $\bar f\in P_{\bar v}$.

The situation is different in the terms with two $\phi$ fields,
connected through two external propagators of scale $j_x$ and
$j_y$ greater than $h$ and involving two $\psi$ fields, of labels
$f_x$ and $f_y$. There are two different type of contributions.
The first type is associated with trees $\t$ satisfying the
following conditions:

\begin{enumerate}

\item the root has scale $h_r\ge h$, \item ${\cal I}_{v_0}$ (the
set of internal lines in the vertex immediately following the
root) is not empty, \item{3)} there is no external line in $v_0$,
except $f_x$ and $f_y$, the lines contracted in the external
propagators.

\end{enumerate}

\0 These terms are produced, in the iterative integration
procedure, at scale $h_r+1$ and, after that scale are constant
with respect to the integration process. The other type of terms
is associated with trees such that

\begin{enumerate}

\item the root has scale $h$, \item $|P_{v_0}|>2$.

\end{enumerate}

\0 These terms depend on the integration field $\psi^{(\le h)}$,
so that are involved in the subsequent integration steps.

Given a tree $\t$ (of any type) with two $\phi$ fields, the
corresponding contributions to $\WW(\phi)$ are obtained in a way
slightly different from that described in the case of the
effective potential. Given $j_x$ and $j_y$, larger or equal to
$h+1$, select two field labels $f_x$ and $f_y$ and call $\bar v$
the higher vertex, of scale $\bar h$, such that

\begin{enumerate}

\item $\bar h\le \min\{j_x,j_y\}$, \item $f_x$ and $f_y$ belong to
$P_{\bar v}$.

\end{enumerate}

\0 Let $\cal C$ be the path on $\t$ connecting $\bar v$ with
$v_0$. Given $v\in \cal C$, we avoid to apply there the
localization procedure, because the $\RR$ operation, no matter we
choose the localization point, would give rise to terms with a
derivative acting on the external propagators (which is not
convenient, see above). In all other vertices of $\t$ the
localization procedure is defined as in the case of the free
energy expansion, by suitably choosing the localization point in
the vertices following $\bar v$ and containing $f_x$ or $f_y$, as
explained above. Then we substitute $f_x$ and $f_y$ with two
external propagators of scale $j_x$ and $j_y$, respectively. Note
that these propagators can acquire a derivative, as a consequence
of the $\RR$ operation acting on a vertex $v$, only if $h_v$ is
greater or equal to their scale ($j_x$ or $j_y$).

The previous considerations imply that $S(\xx-\yy)$ is given by
the following sum:
\be S(\xx-\yy)= g(\xx-\yy) + \sum_{\bar h=h_\b}^1
\sum_{h_r=h_\b-1}^{\bar h-1} \sum_{n=1}^\io \sum_{\t\in\TT_n^{\bar
h,h_r}} \sum_\bP S_{\t,\bP}(\xx-\yy) \;,\lb{6.3}\ee
where the family of labelled trees $\TT_n^{\bar h, h_r}$ and the
families of external lines $P_v$ can be described as in
\S\ref{ss3}, with the following modifications (see Fig. \ref{f3}).

\insertplot{300}{150}{ \ins{30pt}{85pt}{$r$}
\ins{50pt}{85pt}{$v_0$} \ins{130pt}{100pt}{$\bar v$}
\ins{265pt}{115pt}{$v_x$} \ins{205pt}{85pt}{$v_y$}
\ins{35pt}{-2pt}{$h_r$} \ins{135pt}{-2pt}{$\bar h$}
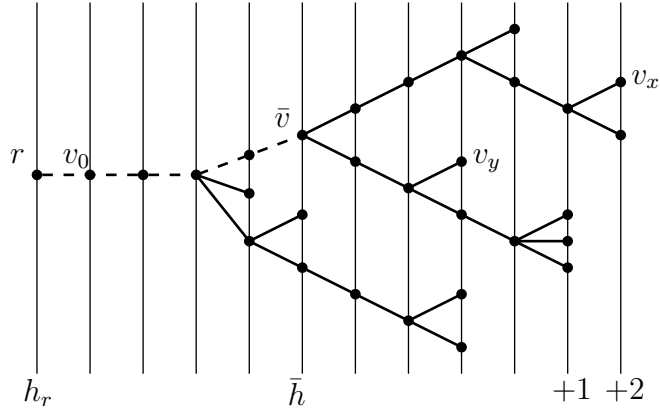
\ins{235pt}{-2pt}{$+1$} \ins{255pt}{-2pt}{$+2$}} {fig16}{An
example of tree contributing to $S(\xx-\yy)$. \lb{f3}}{0}

\0 1) There are two field labels, $f_x$ and $f_y$, two scale
labels $j_x\ge \bar h$ and $j_y\ge \bar h$, and a vertex $\bar v$
such that $h_{\bar v} = \bar h$, $f_x, f_y \in P_{\bar v}$ and
there is no other vertex $v>\bar v$ such that $h_v\le
\min\{j_x,j_y\}$ and $f_x,f_y\in P_v$; we shall call $v_x$ and
$v_y$ the endpoints (possibly coinciding) that $f_x$ and $f_y$
belong to. Note that we are not introducing the sector
decomposition for the external propagators and that the vertex
$\bar v$ can be lower than the higher vertex preceding both $v_x$
and $v_y$ (opposite to what happens in Fig. \ref{f3}).

\0 2) Given $f_x$ and $f_y$, let $\cal C$ be the path on the tree
(see dashed line in Fig. \ref{f3}), connecting $\bar v$ with the
lowest vertex $v_0$, of scale $h_r+1$. If $v\in {\cal C}$ and
$v\not= v_0$, $|P_v|\ge 4$, while $|P_{v_0}|=2$.

Given $\t\in \TT_n^{\bar h, h_r}$ and $\bP$, we have
\be S_{\t,\bP}(\xx-\yy) = \left[ g^{(j_x)} \ast
W_{\t,\bP,j_x,j_y} \ast g^{(j_y)} \right](\xx-\yy)\;,\lb{6.4}\ee
where $\ast$ means the convolution in $\xx$ space and
$W_{\t,\bP,j_x,j_y}$ differs from the kernel $K^{(h_r+1)}_{\t,\bP}
= \sum_{\O\bs \O_{v_0}} K^{(h_r+1)}_{\t,\bP,\O}$ of
$\VV^{(h_r)}(\t,\bP,\O)$ (see \pref{3.33} and note that
$K^{(h_r+1)}_{\t,\bP,\O}$ does not depend on $\O_{v_0}$) only
because no $\RR$ operation acts on the vertices of ${\cal C}$.

We now consider the Fourier transform $\hat S(\kk)$ of
$S(\xx-\yy)$, which can be written in the form:
\be \hat S(\kk)=\hat g(\kk)\left( 1+\l \hat S_1(\kk)\right)\;,\lb{6.5}\ee
where $\hat g(\kk)$ is the free propagator. In order to prove
Theorem \ref{th2.1}, we have to show that $\hat S_1(\kk)$ is a
bounded function.

Let us define $h_\kk=\max\{h: \hat g^{(h)}(\kk)\not= 0\}$. By
using \pref{6.4}, it is easy to see that
\be \l \hat g(\kk) \hat S_1(\kk)= \sum_{j_x,j_y=h_\kk-1}^{h_\kk}
\sum_{n=1}^\io \sum_{\bar h=h_\b}^{\min\{j_x,j_y\}}
\sum_{h_r=h_\b-1}^{\bar h-1} \sum_{\t\in\TT_n^{\bar h,h_r}}
\sum_\bP \hat S_{\t,\bP,j_x,j_y}(\kk)\;,\lb{6.6}\ee
implying that
\bea &&|\hat S_1(\kk)| \le c|\l|^{-1} \g^{h_\kk}\cdot\nn\\
&&\cdot \sup_{j_x,j_y=h_\kk-1,h_\kk} \sum_{n=1}^\io \sum_{\bar
h=h_\b}^{\min\{j_x,j_y\}} \sum_{h_r=h_\b-1}^{\bar h-1}
\sum_{\t\in\TT_n^{\bar h,h_r}} \sum_\bP ||
S_{\t,\bP,j_x,j_y}||_1\;,\lb{6.6a}\eea
where $||.||_1$ denotes the $L_1$ norm.

We can bound $\sum_{\t\in\TT_n^{\bar h,h_r}} \sum_\bP
||S_{\t,\bP,j_x,j_y}||_1$ by proceeding as in \S\ref{ss3.5a}. Since
the combinatorial problems are of the same nature, we can describe
in a simple way the result by dimensional arguments. We can take
as a reference the bound of $J_{h_r,n}(2,0)$, see \pref{3.48} and
\pref{3.66}, that is the bound of the $L_1$ norm of the effective
potential terms with two external lines on scale $h_r$ and no
external derivative, and
multiply it by a factor $\g^{-j_x-j_y}$, which comes from the
external propagators (the derivatives possibly acting on them are
absorbed in the ``gain factors'' $\g^{-(h-h')}$, produced by the
localization procedure, so that they do not give any contribution
to the final bound). There are two relevant differences.

\0 1) There is no regularization on the vertices with four
external lines belonging to $\cal C$. This implies that one
``looses'' a factor $\g^{-1}$, with respect to the bound
\pref{3.66}, for each vertex $v\in {\cal C}$ such that $|P_v|=4$.

\0 2) The external propagators sectors are not on the scale $h_r$,
but they are exactly fixed. Hence, we have to modify the momentum
conservation constraint \pref{3.44} in the tree vertices $v$ such
that $f_x$ or $f_y$ belong to $P_v$, in order to remember this
condition when we bound the sector sums. Then, we have to prove a
lemma similar to Lemma \ref{lm4.1}, by substituting one sector sum
with the constraint that one momentum is exactly fixed. It is not
hard to see, by using Lemma \ref{lms1.5} and by proceeding as in
\S\ref{ssA1.3}, that we get a bound of the same type of that of Lemma
\ref{lm4.1}.

The previous considerations, together with the bound \pref{3.66},
allow to prove that
\bea && \sum_{\t\in\TT_n^{\bar h,h_r}} \sum_\bP
||S_{\t,\bP,j_x,j_y}||_1 \le (c|\l|)^n \g^{h_r-2h_\kk}\cdot\nn\\
&&\cdot \sum_{\t\in\TT_n^{\bar h,h_r}} \sum_{\bP\atop |P_{v_0}|=
2} \sum_{T\in {\bf T}} \prod_{v\atop {\rm not\ e.\ p.}}{1\over
s_v!} \g^{\d_v^*}\;,\lb{6.7}\eea
where $\d_v^*=\d(|P_v|)$, if $v\notin {\cal C}$, otherwise
$\d_v^*=\d(|P_v|)+ \chi(|P_v|=4)$. By using \pref{3.67}, it is easy
to see that, if we define $\tilde\d_v = \d^*_v -1/2$, if $v\in
\cal C$ and $\tilde\d_v=\d_v^*$ otherwise, $\tilde\d_v<0$ for all
$v\in\t$. Hence, the bound \pref{3.68} is still valid, if we put
$\tilde\d_v$ in place of $\d(|P_v|)$, and we get
\bea |\hat S_1(\kk)| &\le& c \g^{-h_\kk} \sum_{\bar
h=h_\b}^{h_\kk} \sum_{h_r=h_\b-1}^{\bar h-1} \g^{h_r} \g^{(\bar
h-h_r)/2}\le\nn\\
&\le& c \sum_{\bar h=h_\b}^{h_\kk} \g^{-(h_\kk-\bar h)}
\sum_{h_r=h_\b-1}^{\bar h-1} \g^{-(\bar h-h_r)/2} \le
c\;.\lb{6.9}\eea
\Halmos

\section{ The rotation-invariant case}\lb{ss7}
\setcounter{equation}{0}

We consider now the {\it Jellium model}, which is defined in the
continuum with $\e(\vec{k})= |\vec{k}|^2/(2m)$ and $v(\vec x-\vec
y)=\tilde v(|\vec x-\vec y|)$, implying rotation invariance
symmetry. In particular, $p_F= |\vpp_F(\th)|$ does not depend on
$\th$ and the two point contribution to the effective potential
$\hat W^{(h)}_2(k_0,\vec k)=\int d\xx \tilde W^{(h)}_2(\xx)
\exp(i\kk\xx)$, see \pref{3.9a} and the line before \pref{3.9}, is
of the form $\WW^{(h)}_2(k_0,|\vec k|)$, where
$\WW^{(h)}_2(k_0,\r)$ is a function of two variables. We show that
in such a case we can choose the counterterm $\hat\nu(\vec k)$ as
a constant $\nu$, if the temperature $T$ is big enough, \ie $T\ge
e^{-{1\over c_1|\l|}}$, where $c_1$ is a constant depending on a
bound to all orders of multiscale perturbation theory.

In order to get this result, we must change the localization
definition, so that

\begin{enumerate}

\item $\LL W^{(h)}_{2n}=0$ if $n\ge 2$;

\item if $n=1$, $\LL \hat W^{(h)}_2(k_0,\vec k)= \WW^{(h)}_2(0,
p_F)\=\g^h\n_h$.

\end{enumerate}

We want now to analyze the properties of the $\RR$ operator. If we
put, as in \pref{3.21a}, for any $\vkk\in S_{h,\o}$, $\o\in O_h$,
$\vkk=\vkk'+\vpp_F(\th_{h,\o})$, we can write
\bea \RR \hat W^{(h)}_2(k_0,\vkk) &=& \int_0^1d  t{d \over d
t}\WW^{(h)}_2\left(t k_0, |t\vec k'+\vpp_F(\th_{h,\o})|\right)=\lb{7.2}\\
&=& \int_0^1 d  t \Bigl[ k_0\dpr_{k_0}\WW^{(h)}_2(tk_0,\r (t)\
)+ \nn\\
&+& {(tk_1'+ p_F)k_1'+ t(k_2')^2\over \r (t)} \dpr_\r\WW^{(h)}_2(t
k_0,\r (t)\ )\Bigr]\;,\nn\eea
where, for any vector $\vec v$, we are defining $v_1\=\vec v\cdot
\vnn(\th_{h,\o})$, $v_2\=\vec v \cdot \vt(\th_{h,\o})$ (see
\pref{3.21}, recalling that now $\vec e_r(\th)=\vnn(\th)$ and $\vec
e_t(\th)=\vt(\th)$) and $\r (t)\= \sqrt{(tk_1'+|\vpp_F|)^2 +
(tk_2')^2}$.

It is easy to see that the term $\dpr_\r\WW^{(h)}_2 (t k_0, \r
(t)\ )$ in \pref{7.2} can be rewritten in the following form:
\bea &&\dpr_\r\WW^{(h)}_2 (t k_0, \r (t)\ )= \cos\th(t)\dpr_{k_1}
\hat W^{(h)}_2 (t k_0, t\vkk'+\vpp_F(\th_{h,\o})) +\nn\\
&&+ \sin\th(t)\dpr_{k_2} \hat W^{(h)}_2(t k_0, t
\vkk'+\vpp_F(\th_{h,\o}) )=\lb{7.3}\\
&&=\int d\yy \left(iy_1{tk_1' + p_F\over \r (t)}+ iy_2{tk_2'\over
\r (t)}\right) \tilde W_{2}^{(h)}(\yy)e^{itk_0y_0+i(t\vec
k'+\vpp_F(\th_{h,\o}))\vec y}\;,\nn\eea
where $\th (t)$ is the angle between $\vnn(\th_{h,\o})$ and
$t\vkk'+\vpp_F(\th_{h,\o})$. Substituting \pref{7.3} in \pref{7.2}
we get, if $\pp_\o=(0, \vpp_F(\th_{h,\o}))$,
\bea &&\RR\VV^{(h)}(\psi^{(\le h)}) = \sum_{\s,\o\in O_h} \int
{d\kk\over (2\p)^3} \hat\psi^{(\le h)+}_{\kk-\pp_\s, \s}
\hat\psi^{(\le h)-}_{\kk-\pp_\o, \o} \RR \hat W^{(h)}_2(\kk)
=\nn\\
&&= \sum_{\s,\o\in O_h} \int_0^1 dt \int {d\kk'\over (2\p)^3}
\hat\psi^{(\le h)+}_{\kk' + \pp_\o -\pp_\s, \s} \hat\psi^{(\le
h)-}_{\kk', \o} \int d\yy \tilde W_2^{(h)}(\yy)  e^{i (t \kk'
+\pp_\o) \yy} \cdot\nn\\
&&\cdot \left[ik_0y_0+{(tk_1'+ p_F)k_1'+t(k_2')^2\over \r (t)}
\left( {tk_1'+ p_F\over \r (t)}iy_1+{tk_2'\over \r (t)}
iy_2\right)\right]\;.\lb{7.4}\eea

Let us define the operators $D_i(t)$, $i=1,2$, so that
\bea D_1(t) \psi_{\xx,\o}^{(\le h)\e} &=& i\e \int {d\kk' \over
(2\p)^3} e^{i\e\kk'\xx} {tk_1'+ p_F\over \r (t)} {(tk_1'+ p_F)
k_1'+ t(k_2')^2\over \r (t)} \hat\psi_{\kk',\o}^{(\le h)\e}\;,
\nn\\
D_2(t) \psi_{\xx,\o}^{(\le h)\e} &=& i\e \int {d\kk' \over
(2\p)^3} e^{i\e\kk'\xx} {tk_2'\over \r(t)} {(tk_1'+ p_F)k_1'+
t(k_2')^2 \over \r (t)} \hat \psi_{\kk',\o}^{(\le
h)\e}\;.\lb{7.5}\eea
Hence, \pref{7.4} can be written as
\bea &&\RR\VV^{(h)}(\psi^{(\le h)}) = -\sum_{\s,\o\in O_h}
\int_0^1 dt \int d\xx \int d\yy \; e^{i\pp_\s\yy -i\pp_\o\xx}
\tilde W_2^{(h)}(\yy-\xx) \psi_{\yy,\s}^{(\le h) +} \cdot\nn\\
&&\cdot \left[(y_0-x_0)\dpr_0 +(y_1-x_1) D_1(t) +
(y_2-x_2)D_2(t)\right] \ps_{\x(t), \o}^{(\le h) -}=\nn\\
&&=\sum_{\s,\o\in O_h} \int_0^1 dt \int d\xx \int d\yy \;
e^{i\pp_\s\yy -i\pp_\o\xx} \tilde W_2^{(h)}(\yy-\xx)
\cdot\lb{7.6}\\
&&\cdot \left[ (y_0-x_0)\dpr_0+(y_1-x_1)D_1(t) +(y_2-x_2)D_2(t)
\right] \ps_{{\bf\h}(t),\s}^{(\le h) +} \ps_{\xx,\o}^{(\le h)
-}\;,\nn\eea
where
\bea {\bf\h}(t)&\=& \yy+t(\xx-\yy)\nn\\
{\bf\x}(t) &\=& \xx+t(\yy-\xx)\;.\lb{7.7}\eea

\* It is easy to prove the following dimensional bound.

\begin{lemma}\lb{lm7.1}
Given non negative integers $N,
n_0, n_1, n_2$, $m=n_0+n_1+n_2$, there exists a constant
$C_{N,m}$, such that
\be |\dpr_0^{n_0} D_1^{n_1} D_2^{n_2}
g_{\o}^{(h)}(\xx)| \leq C_{N,m}{\g^{h({3\over 2} + n_0+n_1 + {3\over
2}n_2)} \over 1+\left[ (\g^h x_0)^2+(\g^h x_1)^2+(\g^{h\over 2}
x_2)^2\right]^N}\;,\lb{7.11}\ee
where $D_i^n$ denotes the product of $n$ factors $D_i(t_j)$,
$j=1,\ldots,n$.
\end{lemma}

{\bf Remark - } Each operator $D_i(t)$ improves the bound of the
covariance by a factor at least $\g^h$; this is what we need to
obtain the right dimensional gain from renormalization operations,
which also produce a factor $\g^{-h'}$ on a scale $h'>h$. This is
a consequence of rotational invariance; in fact a naive Taylor
expansion would apparently produce a term of the form
$(y_2-x_2)\dpr_{x_2}$, which would give rise to a ``bad factor''
$\g^{-h'+h/2}$ in the bounds. \*

We can now repeat the analysis of the previous sections, in a much
more simple context. In fact it is easy to see that it is possible
to fix $\n_1$ in such a way that $\n_h$ stay bounded for $h_\b\le
h\le 1$. Furthermore we can easily perform the bounds for the
$n^{th}$-order contributions to the kernel of the effective
potentials or to the two point Schwinger functions. In both cases
we find that, unless for the external dimensional factors, the
$n^{th}$-order contributions are bounded by
$(c|\l|)^n(\log\b)^{n-1}$, where the diverging factor
$(\log\b)^{n-1}$ is due to the choice of not localizing the
four-legs clusters and of localizing the two-legs clusters only at
the first order. So the result of Theorem \ref{th2.1} in the
rotational invariant case easily follows.

\section{ Some technical lemmata.}\lb{ssA1}
\setcounter{equation}{0}

\subsection{Geometrical properties of the dispersion relation}\lb{ssA1.1}

Let ${\cal B}=\{\vpp\in\RRR^2: |\e(\vpp)-\m|\le e_0\}$; the
hypotheses on $\e(\vpp)$ described in \S\ref{ss1.2} imply that there
is a $C^\io$ diffeomorphism between $\cal B$ and the compact set
${\cal A}= \TTT^1\times[-e_0, e_0]$, defined by
\be \vpp = \vqq(\th,e)=u(\th,e) \vee_r(\th)\virg (\th,e)\in{\cal A}
\;.\lb{A1.1}\ee
Moreover, the symmetry property \pref{2.8c} implies that
\be \vqq(\th+\p,e)=-\vqq(\th,e)\;,\lb{A1.4}\ee
a property that will have an important role in the following.

Let us now introduce some more geometrical definitions, which we
shall need in the following. For any fixed $e$, we can locally
define the arc length $s(\th,e)$ on $\Si(e)$; we shall denote
$\dpr/\dpr s$ the partial derivative with respect to $s$, at fixed
$e$, and we shall sometime use the prime to denote the partial
derivative with respect to $\th$. If $\vt(\th,e)=\dpr \vpp(\th,e)/
\dpr s$ is the unit tangent vector at $\Si(e)$ in $\vqq(\th,e)$,
we have
\bea s'(\th,e) \vt(\th,e) &=& {\dpr \vpp\over \dpr\th}(\th,e) =
u'(\th,e)\vee_r(\th) + u(\th,e)\vee_t(\th)\;,\nn\\
s'(\th,e) &=& \sqrt{ u'(\th,e)^2 + u(\th,e)^2}\;,\lb{A1.6}\eea
where $\vee_t(\th)=(-\sin\th,\cos\th)$.

Analogously, if $\vnn(\th,e)$ is the outgoing unit normal vector
at $\Si(e)$ in $\vqq(\th,e)$ and $1/r(\th,e)$ is the curvature
(which satisfies the convexity condition \pref{2.8a}), we have
\bea s'(\th,e) \vnn(\th,e) &=& u(\th,e)\vee_r(\th) -
u'(\th,e)\vee_t(\th)\;,\nn\\
{\dpr^2 \vpp\over \dpr\th^2}(\th,e) &=& s''(\th,e) \vt(\th,e) -
{s'(\th,e)^2\over r(\th,e)} \vnn(\th,e)\;.\lb{A1.7}\eea

\begin{lemma}\lb{lmA1.1}
The angle $\a(\th,e)$ between
$\vnn(\th,e)$ and $\vee_r(0)$ is a monotone increasing function of
$\th$, such that, if $||\th_1-\th_2||$ denotes the distance on
$\TTT^1$.
\be c_1||\th_2-\th_1|| \le ||\a(\th_2,e) - \a(\th_1,e)|| \le c_2||\th_2-\th_1||
\;;\lb{A1.9}\ee
moreover, $\a(\th+\p,e)-\a(\th,e)=\p$.
\end{lemma}

\proof By using \pref{A1.6} and \pref{A1.7} and Taylor expansion,
one can easily prove that, if $\a_i=\a(\th_i,e)$,
\bea \sin(\a_2-\a_1) &=& \vnn(\th_2,e)\cdot \vt(\th_1,e)=
(\th_2-\th_1){s'(\th_1,e)\over r(\th_1,e)} +
O(\th_2-\th_1)^2\;,\lb{A1.10}\\
\cos(\a_2-\a_1) &=& \vnn(\th_2,e)\cdot \vnn(\th_1,e)=1-
{(\th_2-\th_1)^2\over 2} {s'^2(\th_1,e)\over r^2(\th_1,e)} +
O(\th_2-\th_1)^3\;,\nn\eea
which implies \pref{A1.9} for $|\th_2-\th_1|$ small, hence even for
any value of $\th_2-\th_1$, together with the monotonicity
property. The fact that $\a(\th+\p,e)-\a(\th,e)=\p$ is a trivial
consequence of \pref{A1.4}.\Halmos

\*

We denote by $\vpp_F(\th)=\vqq(\th,0)$ the generic point of the
Fermi surface $\Si_F\=\Si(0)$. Moreover, to simplify the notation,
from now on we shall in general suppress the variable $e$ when it
is equal to $0$; for example, we shall put
$\vpp_F(\th)=u(\th)\vee_r(\th)$. Let us consider an s-sector
$S_{h,\o}$, see \pref{3.44a}.

\begin{lemma}\lb{lms1.1}
If $\vpp=\r\vee_r(\th)\in
S_{h,\o}$, $h\le 0$, $\o\in O_h$, then
\be |\r-u(\th)| \le c\g^h\virg |\th-\th_{h,\o}|\le \p \g^{h/2}\;.
\lb{A1.10a}\ee
\end{lemma}

\proof The bound on $\th$ follows directly from the definition of
$S_{h,\o}$. On the other hand, the identity
\be \e(\vpp)-\m = \e(\r\vee_r(\th))-\e(u(\th)\vee_r(\th))=
\int_\r^{u(\th)} d\r'\; \vee_r(\th)\grad\e \left(\r'\vee_r(\th)
\right)\;,\lb{A1.10b}\ee
and the property \pref{2.8b} of $\e(\vpp)$, easily imply the bound
on $\r-u(\th,0)$.\Halmos

\*

The following lemma shows that, if $\vpp\in S_{h,\o}$, the
difference between $\vpp$ and $\vpp_F(\th_{h,\o})$ is of order
$\g^h$ in the direction normal to $\Si_F$ in the point
$\vpp_F(\th_{h,\o})$, while it is of order $\g^{h/2}$ in the
tangent direction.

\begin{lemma}\lb{lmA1.3}
If $\vpp\in S_{h,\o}$, $h\le 0$,
$\o\in O_h$, then
\be \vpp = \vpp_F(\th_{h,\o}) + k_1\vnn(\th_{h,\o}) + k_2\vt(\th_{h,\o})\virg
|k_1|\le c\g^h\virg |k_2|\le c\g^{h/2}\;.\lb{A1.13}\ee
Moreover, \be \left| {\dpr \over \dpr k_2} \e(\vpp_F(\th_{h,\o}) +
k_1\vnn(\th_{h,\o}) + k_2\vt(\th_{h,\o})) \right| \le
c\g^{h/2}\;.\lb{A1.14}\ee
\end{lemma}

\proof If $\vpp=\r\vee_r(\th)$, by Lemma \ref{lms1.1}
$|\vpp-\vpp_F(\th)|\le c\g^h$. Hence, to prove \pref{A1.13}, it is
sufficient to prove that $|[\vpp_F(\th) -\vpp_F(\th_{h,\o})]
\vnn(\th_{h,\o})|\le c\g^h$ and $|[\vpp_F(\th)
-\vpp_F(\th_{h,\o})] \vt(\th_{h,\o})|\le c\g^{h/2}$. These bounds
immediately follows from the the following ones, which can be
easily proved, by using \pref{A1.6}, \pref{A1.7} and some Taylor
expansions:

\be [\vpp_F(\th_1) -\vpp_F(\th_2)]\cdot \vnn(\th_2) =
O(\th_1-\th_2)^2\;,\lb{A1.15}\ee

\be [\vpp_F(\th_1) -\vpp_F(\th_2)]\cdot \vt(\th_2) =
O(\th_1-\th_2)\;.\lb{A1.16}\ee
It is sufficient to put here $\th_1=\th$, $\th_2=\th_{h,\o}$ and
to recall that $\th-\th_{h,\o}=O(\g^{h/2})$.

Let us now observe that, if we derive with respect to $\th$ the
identity $\e(u(\th,e) \vee_r(\th))=e$, we get, for any
$\vpp\in{\cal B}$,
\be [\grad\e(\vpp) \vt(\th,e)] s'(\th,e)=0
\quad \Rightarrow \quad
\grad\e(\vpp)=a(\th,e)\vnn(\th,e)\;,\lb{A1.17}\ee
$a(\th,e)$ being a smooth function, strictly positive by
\pref{2.8b}. Hence, if $\vpp\in S_{h,\o}$, by using the first line
of \pref{A1.10}, \pref{A1.17} and the fact that $|\e(\vpp)-\m|\le
c\g^h$, $|\th-\th_{h,\o}|\le c\g^{h/2}$,
\bea {\dpr \e(\vpp)\over \dpr k_2} &=& \grad \e(\vpp)\cdot
\vt(\th_{h,\o})= a(\th,e)\vnn(\th,e)\cdot \vt(\th_{h,\o}) =\nn\\
&=& a(\th,e)\vnn(\th)\cdot \vt(\th_{h,\o}) + O(\g^h)=
O(\g^{h/2})\;,\lb{A1.18}\eea
which proves \pref{A1.14}.\Halmos

\*

Given $\vpp\in S_{h,\o}$, we shall also consider the projection on
the Fermi surface, defined as
\be \vpp_\bot = \vpp_F(\th_\bot) = \vpp- x\,\vnn(\th_\bot)\;.\lb{A1.11}\ee
Note that \pref{A1.11} has to be thought as an equation for
$\th_\bot$ and $x$, given $\vpp\ $; it is easy to prove that, as a
consequence of the condition \pref{2.8a}, this equation has a
smooth unique solution, if $e_0$ is small enough, what we shall
suppose from now on.

\begin{lemma}\lb{lmA1.2}
If $\vpp=\r\vee_r(\th) \in S_{h,\o}$, $h\le 0$, and $x$ and
$\th_\bot$ are defined as in \pref{A1.11}, then $|x|\le c\g^h$ and
$|\th_\bot-\th_{h,\o}|\le c\g^{h/2}$.
\end{lemma}

\proof \pref{A1.13} and \pref{A1.11} imply that
\bea k_1 &=& [\vpp_F(\th_\bot) -\vpp_F(\th_{h,\o})]\cdot
\vnn(\th_{h,\o})+ x \vnn(\th_\bot)\cdot \vnn(\th_{h,\o})\;,
\lb{s1.12}\\
k_2 &=& [\vpp_F(\th_\bot) -\vpp_F(\th_{h,\o})]\cdot
\vt(\th_{h,\o})+ x \vnn(\th_\bot)\cdot \vt(\th_{h,\o})\;.
\lb{s1.13}\eea
By using the \pref{A1.15}, \pref{A1.16} and \pref{A1.10}, one can
easily complete the proof of the lemma.\Halmos

\subsection{Proof of Lemma \ref{lm3.1}}\lb{ssA1.0}

The bounds on $k_1$ and $k_2$ in \pref{A1.13} imply that $\int d\pp
F_{h,\o}(\pp) \le c\g^{5h/2}$. On the other hand, if
$F_{h,\o}(\pp) \not=0$, $|-ip_0+\e(\vpp)-\m| \ge c\g^h$, so that
\pref{3.20c} implies the bound $|g^{(h)}_\o(\xx)|\le c\g^{3h/2}$.
It is also very easy to prove that $|\dpr^n \hat
g^{(h)}_\o(\pp)/\dpr p_0^n|$ and $|\dpr^n \hat
g^{(h)}_\o(\pp)/\dpr k_1^n|$ are bounded by $c\g^{-h(n+1)}$.
Hence, using simple integration by parts arguments, one can show
that $|x_0^n g^{(h)}_\o(\xx)|\le c\g^{h(3/2-n)}$ and $|{x_1'}^n
g^{(h)}_\o(\xx)|\le c\g^{h(3/2-n)}$. Moreover, it is easy to prove
that
\be |\dpr^n \hat g^{(h)}_\o(\pp)/\dpr k_2^n| \le c \g^{-h} \Big[\g^{-h}
\sup_{\vpp\in S_{h,\o}} |{\dpr \e(\vpp)\over \dpr k_2}|
\Big]^n\;,\lb{A1.11a}\ee
which implies the bound $|{x_2'}^n g^{(h)}_\o(\xx)|\le
c\g^{h(3/2-n/2)}$. Finally, by using Lemma \ref{lmA1.3}, it is easy
to prove that the previous bounds have to be multiplied by
$\g^{mh}$, if one substitutes $g^{(h)}_\o(\xx)$ with $\dpr^m
g^{(h)}_\o(\xx)/ \dpr x_0^m$ or $\dpr^m g^{(h)}_\o(\xx)/ \dpr
{x'_1}^m$, while they have to be multiplied by $\g^{mh/2}$ if
$g^{(h)}_\o(\xx)$ is changed in $\dpr^m g^{(h)}_\o(\xx)/ \dpr
{x'_2}^m$.

The bound \pref{3.22} is a simple consequence of the previous
considerations.\Halmos

\subsection{The parallelogram lemma}\lb{ssA1.2}

Let us consider the map F, defined on the two dimensional torus
$\TTT^2$, with values in $\RRR^2$, such that, if
$(\th_1,\th_2)\in\TTT^2$ and $\V b=F(\th_1,\th_2)$, then
\be \V b= \vpp_F(\th_1) + \vpp_F(\th_2)\;.\lb{A1.19}\ee
The differential $J(\th_1,\th_2)$ of F is a matrix, whose columns
coincide with $s'(\th_1)\vt(\th_1)$ and $s'(\th_2)\vt(\th_2)$.
Then Lemma \ref{lmA1.1} implies that $\det J\not=0$, hence F is
invertible, around any point $(\th_1,\th_2)\in {\cal T}$, where
\be {\cal T}= \{ (\th_1,\th_2)\in \TTT^2: \sin(\th_1-\th_2)\not=0\}
\;.\lb{A1.20}\ee
Moreover, if $||\th_1-\th_2||=\p$, $\V b=0$, while, if
$\th_1=\th_2=\th$, $\V b=2 u(\th)\vee_r(\th)$. Finally, $\cal T$
is the union of two disjoint subsets, which are obtained one from
the other by exchanging $\th_1$ with $\th_2$, and each one of them
is in a one to one correspondence through $F$ with the open set
\be {\cal D} = \{\vpp=\r\vee_r(\th):\,0 < \r < 2u(\th),\, \th\in\TTT^1\}\;.
\lb{A1.21}\ee

The following Lemma will have an important role in the following.

\begin{lemma}\lb{lms1.5}
Let $(\bar\th_1,\bar\th_2)\in
{\cal T}$, $\V b= \vpp_F(\bar\th_1) + \vpp_F(\bar\th_2)$,
\be \phi=\min\{||\bar\th_1-\bar\th_2||,\p-||\bar\th_1-\bar\th_2|||\}
> 0\;,\lb{s1.16}\ee
\be \vrr=r_1\vnn(\bar\th_1) + r_2\vt(\bar\th_1)\virg |r_1|\le c_1\h\phi
\virg |r_2|\le\h\le c_2\phi\;.\lb{s1.17}\ee
Then there exist $c_0$, $\bar c_2$ and $\h_0$, such that, if
$c_2\le \bar c_2$ and $\h\le\h_0$, then $\vbb+\vrr\in {\cal D}$
and
\be \vbb+\vrr = \vpp_F(\th_1) + \vpp_F(\th_2)\virg
|\th_i-\bar\th_i|\le c_0\h\;. \lb{s1.18}\ee
\end{lemma}

\proof We shall consider only the case
$\phi=||\bar\th_1-\bar\th_2||$; the case
$\phi=\p-||\bar\th_1-\bar\th_2||$ can be easily reduced to this
one, by using the symmetry property \pref{A1.4}. We shall also
choose the sign of $\bar\th_1-\bar\th_2$, so that
$\phi=\bar\th_2-\bar\th_1$.

Let us define $\d_i=\th_i-\bar\th_i$, $\d=\sqrt{\d_1^2+\d_2^2}$;
then we can write, by using \pref{A1.19}, \pref{A1.6} and
\pref{A1.7}, if $\vbb+\vrr\in {\cal D}$ (which is certainly true,
if $\vrr$ is small enough),
\bea \vrr &=& {d\vpp_F(\bar\th_1)\over d\th} \d_1 +
{d\vpp_F(\bar\th_2)\over d\th} \d_2 +O(\d^2)=\nn\\
&=& \d_1 s'(\bar\th_1)\vt(\bar\th_1) + \d_2
s'(\bar\th_2)\vt(\bar\th_2) +O(\d^2)\;.\lb{s1.19}\eea
Let us now put $\d_i=\h x_i$, $r_1=\h\phi\tilde r_1$,
$r_2=\h\tilde r_2$; condition \pref{s1.17} takes the form $|\tilde
r_1|\le c_1$ and $|\tilde r_2|\le 1$. Since, by hypothesis, $\h\le
c_2\phi$, the condition $\vbb+\vrr\in {\cal D}$ is satisfied,
together with \pref{s1.18}, if and only if the following system of
two equations in the unknowns $x_1, x_2$ has a unique solution:
\bea x_2 &=& {\tilde r_1\phi\over s'(\bar\th_2) \sin[\a(\bar\th_1)
- \a(\bar\th_2)]} +O(c_2)\;,\nn\\
x_1 &=& {\tilde r_2\over s'(\bar\th_1)} -{\tilde r_1\phi
\cos[\a(\bar\th_1) - \a(\bar\th_2)]\over s'(\bar\th_1)
\sin[\a(\bar\th_1) - \a(\bar\th_2)]} +O(\h)+O(c_2)\;,
\lb{s1.19a}\eea
where $\a(\th)$ is defined as in Lemma \ref{lmA1.1} and $O(c_2)$,
$O(\h)$ are of second order as functions of the $x_i$'s.

By using Lemma \ref{lmA1.1}, we see that the right sides of
\pref{s1.19a} are bounded for $\phi\to 0$. Hence, by Dini Theorem,
\pref{s1.19a} allow to uniquely determine $x_1$ and $x_2$ for any
$\phi>0$, given $\vrr$, if $\h$ and $c_2$ are small enough, and
$|\d_i|\le c_0\h$, with $c_0$ independent of $c_2$.\Halmos

\subsection{Proof of Lemma \ref{lm4.1} (sectors counting lemma)}
\lb{ssA1.3}

Let $h',h,L$ be integers such that $h'\le h\le 0$. Given $\o_1\in
O_{h'}$ and $\tilde\o_i\in O_h$, $i=2,\ldots L$, let
$A_{h,h'}(\o_1;\tilde\o_2,\ldots,\tilde\o_L)$ be the set of the
sequences $(\o_2,\ldots,\o_L)$, such that
\0 i) $S_{h',\o_i}\subset S_{h, \tilde\o_i}$ for $i=2,\ldots,L$;
\0 ii) there exists, for $i=1,\ldots,L$, a vector $\vkk^{(i)}\in
S_{h',\o_i}$, so that $\sum_{i=1}^L \vkk^{(i)}=0$.

\0 If $L=2$, the momentum conservation ii) and the symmetry
property \pref{2.8c} immediately imply that
$|A_{h,h'}(\o_1;\tilde\o_2,\ldots,\tilde\o_L)|=1$. Hence, in order
to prove Lemma \ref{lm4.1} it is sufficient to consider the case
$L\ge 4$; we have to prove that
\be |A_{h,h'}(\o_1;\tilde\o_2,\ldots,\tilde\o_L)| \le c^L \g^{{h-h'
\over 2}(L-3)}\;.\lb{4.3app}\ee

Let $\th_i\= \th_{h',\o_i}$, so that $\th_i$ is the center of the
$\th$-interval, which the polar angle of $\vpp$ has to belong to,
if $\vpp\in S_{h',\o_i}$. For any pair $(i,j)$, we define
\be \phi_{i,j}=\min\{||\th_i-\th_j||, \p-||\th_i-\th_j||\}\;.\lb{s1.21}\ee
By a reordering of the sectors, which is unimportant since we are
looking for a bound proportional to $c^L$, we can get the
condition (recall that $L\ge 4)$:
\be \phi\=\phi_{L-1,L}\ge \phi_{i,j}\virg \forall i,j\in [2,L]\;.\lb{s1.22}\ee

Note that, given $\tilde\o\in O_h$,
\be \left| \o\in O_{h'}: S_{h',\o}\subset S_{h, \tilde\o} \right|
=\g^{h-h'\over 2}\;.\lb{s1.23}\ee
Hence, given any positive constant $c_0$, if we define
\be {\cal A}_< = \{(\o_2,\ldots,\o_L) \in A_{h,h'}(\o_1,\tilde\o_2,\ldots,
\tilde\o_L) \,:\, \phi\le L c_0^{-1}\g^{h'/2}\}\;,\lb{s1.24}\ee
we have:
\be |{\cal A}_<| \le \g^{{h-h'\over 2}(L-3)} (c L c_0^{-1})^2\;,\lb{s1.25}\ee
where $(c L c_0^{-1})^2$ is a bound on the number of possible
choices of $\o_{L-1}$ and $\o_L$, given $\o_1,\ldots,\o_{L-2}$.
Hence, in order to prove \pref{4.3app}, it is sufficient to prove
that, if $c_0$ is small enough, a similar bound is valid for the
set
\be {\cal A}_> = \{(\o_2,\ldots,\o_L) \in A_{h,h'}(\o_1,\tilde\o_2,\ldots,
\tilde\o_L) \,:\, \phi \ge L c_0^{-1}\g^{h'/2}\}\;.\lb{s1.26}\ee

We have
\be |{\cal A}_>| \le m_L \g^{{h-h'\over 2}(L-3)} \;,\lb{s1.25a}\ee
where $m_L$ is a bound on the number of choices of $\o_{L-1}$ and
$\o_L$, given $\o_1,\ldots, \o_{L-2}$.

In order to get $m_L$, we consider a particular choice of
$\o_2,\ldots,\o_{L-2}\in O_{h'}$ and we suppose that the set
${\cal E}=\{(\o_{L-1},\o_L): (\o_2,\ldots,$ $\o_L) \in {\cal
A}_>\}$ is not empty. Moreover, we define
\be \phi_0=\max_{(\o_{L-1},\o_L)\in\cal E} \phi_{L-1,L}
\;,\lb{s1.32}\ee

By definition, for any choice of $(\o_{L-1},\o_L)\in {\cal E}$, we
can find $L$ vectors $\vkk^{(1)},\ldots,\vkk^{(L)}$, such that
$\vkk^{(i)}\in S_{h',\o_i}$ and
\be \sum_{i=1}^L \vkk^{(i)}=0\;.\lb{s1.31}\ee
Moreover, by Lemma \ref{lmA1.3}, for $i=1,\ldots,L$, we can write
\be \vkk^{(i)} = \vpp_F(\th_i) + x_i\vnn(\th_i) + y_i\vt(\th_i)\virg
|x_i|\le c \g^{h'} \virg |y_i|\le c \g^{h'/2} \;. \lb{s1.34a}\ee
Hence, since $\phi_0\ge L c_0^{-1}\g^{h'/2}$, we get
\be
|\vkk^{(i)}\wedge\vpp_F(\th_2)|=|\vpp_F(\th_i)|\,|\vpp_F(\th_2)|
\sin\phi_{i,2} + O(\g^{h'\over 2}) \le c\phi_0 \;,\lb{s1.33}\ee
for $i=2,\ldots,L$, and, by using \pref{s1.31},
\be |\vkk^{(1)}\wedge\vpp_F(\th_2)|= \left| \sum_{i=2}^L\vkk^{(i)}
\wedge\vpp_F(\th_2) \right| \le cL\phi_0\;,\lb{s1.34}\ee
so that $\phi_{1,2}\le cL\phi_0$.

Lemma \ref{lmA1.1}, \pref{s1.34a} and \pref{s1.34} easily imply that
\bea && \vkk^{(i)} = \vpp_F(\th_i) + \bar x_i\vnn(\th_2) + \bar
y_i\vt(\th_2), \nn\\
&& |\bar x_i|\le\cases{ c\phi_0 \g^{h'/2}& if $i>1$\cr cL\phi_0
\g^{h'/2}& if $i=1$\cr} \virg |\bar y_i|\le \cases{ c \g^{h'/2}&
if $i>1$\cr cL \g^{h'/2}& if $i=1$\cr} \;.
\lb{s1.38}\eea

Let us now define
\be \V a= -\sum_{i=1}^{L-2} \vpp_F(\th_i) \virg \V b=\vkk^{(L-1)}_\bot
+\vkk^{(L)}_\bot \virg \vrr=\V b-\V a\;,\lb{s1.39}\ee
where $\vkk_\bot$ denotes the projection on the Fermi surface, see
\pref{A1.11}. By using Lemma \ref{lmA1.2}, the momentum conservation
\pref{s1.31} and \pref{s1.38}, we get
\be \vrr= r_1\vnn(\th_2) + r_2\vt(\th_2) \virg |r_1|\le cL\phi_0 \g^{h'/2}
\virg |r_2|\le cL\g^{h'/2}\;.\lb{s1.40}\ee

Note now that the vector $\V a$ defined in \pref{s1.39} is fixed,
if the indices $\o_1,\ldots,\o_{L-2}$ are fixed. Hence, if we put
$\vpp_F(\bar\th_i)=\vkk^{(i)}_\bot$, $i=L-1,L$, $m_L$ can be
calculated by studying the possible solutions of the equation
\be \vpp_F(\bar\th_{L-1}) + \vpp_F(\bar\th_L) = \V a+\vrr\;,\lb{s1.42}\ee
as $\vrr$ varies satisfying \pref{s1.40}. Let
$(\bar\th_{L-1}^{(0)}, \bar\th_L^{(0)})$ be a particular solution
of \pref{s1.42}, such that $\vkk^{(i)}\in S_{h',\o_i}$, $i=L-1,L$,
with $\phi_{L-1,L}=\phi_0$, and put $\V b_0=
\vpp_F(\bar\th_{L-1}^{(0)})+ \vpp_F(\bar\th_L^{(0)})= \V
a+\vrr_0$, so that \pref{s1.42} can be written as
$\vpp_F(\bar\th_{L-1}) + \vpp_F(\bar\th_L)= \V b_0+(\vrr-\vrr_0)$.
The definition of $\phi_0$ implies that $\vrr-\vrr_0$ can be
represented as $\vrr-\vrr_0= r'_1\vnn(\bar\th_L^{(0)}) +
r'_2\vt(\bar\th_L^{(0)})$, with $|r'_1|\le cL\phi_0 \g^{h'/2}$ and
$|r'_2|\le cL\g^{h'/2}$. Hence a simple application of Lemma
\ref{lms1.5} shows that the solutions of \pref{s1.42} belong, up to a
exchange between $\bar\th_{L-1}$ and $\bar\th_L$, to a connected
set and that $m_L\le cL^2$, if $c_0\le \bar c_2/c_1$, where $c_1$
is the constant $c$ of \pref{s1.40} and $\bar c_2$ is defined in
Lemma \ref{lms1.5}.\Halmos

\vskip1.5truecm

\0 {\bf\large Acknowledgements} - We are indebted to M. Disertori
and V. Rivasseau for their enlightening explanation of their work
\cite{DR}.

\end{document}